\documentclass[10pt]{iopart}
\usepackage[abbr]{harvard}

\usepackage{iopams}
\expandafter\let\csname equation*\endcsname\relax
\expandafter\let\csname endequation*\endcsname\relax
\usepackage{amsmath}

\usepackage{moreverb,url}
\usepackage[colorlinks,bookmarksopen,bookmarksnumbered,citecolor=red,urlcolor=red]{hyperref}

\usepackage{amssymb,amsfonts,algorithm,algorithmic,subcaption,graphicx,color,flushend,mysymbol,mathrsfs,lipsum,dsfont,multirow,textgreek}
\graphicspath{{figures/}}
\AtEndDocument{\par\leavevmode}

\makeatletter
\long\def\@makefntext#1{\parindent 1em\noindent 
 \makebox[1em][l]{\footnotesize\rm$\m@th{\arabic{footnote}}$}%
 \footnotesize\rm #1}
\def\@makefnmark{\textsuperscript{\arabic{footnote}}}
\def\@thefnmark{\arabic{footnote}}
\makeatother

\begin{document}

% Remove page numbering
%\pagenumbering{gobble}

\title[Plane Wave Elastography]{Plane Wave Elastography: A Frequency-Domain Ultrasound Shear Wave Elastography Approach}

\author{Reza Khodayi-mehr$^1$, Matthew W. Urban$^2$,\\Michael M. Zavlanos$^1$, and Wilkins Aquino$^1$}
\vspace{10pt}

\address{$^1$ Department of Mechanical Engineering and Materials Science, Duke University, Durham, NC 27708, USA.}
\ead{\{reza.khodayi.mehr, michael.zavlanos, wilkins.aquino\}@duke.edu}
\vspace{10pt}

\address{$^2$ Department of Radiology, Mayo Clinic, Rochester, MN 55905, USA.}
\ead{urban.matthew@mayo.edu}
\vspace{10pt}

\begin{indented}
\item[]December 2020, Patent Pending.
\end{indented}

% make the title area
%\maketitle

\begin{abstract}
In this paper, we propose Plane Wave Elastography (PWE), a novel ultrasound shear wave elastography (SWE) approach. Currently, commercial methods for SWE rely on directional filtering based on the prior knowledge of the wave propagation direction, to remove complicated wave patterns formed due to reflection and refraction. The result is a set of decomposed directional waves that are separately analyzed to construct shear modulus fields that are then combined through compounding. Instead, PWE relies on a rigorous representation of the wave propagation using the frequency-domain scalar wave equation to automatically select appropriate propagation directions and simultaneously reconstruct shear modulus fields. Specifically, assuming a homogeneous, isotropic, incompressible, linear-elastic medium, we represent the solution of the wave equation using a linear combination of plane waves propagating in arbitrary directions. Given this closed-form solution, we formulate the SWE problem as a nonlinear least-squares optimization problem which can be solved very efficiently. Through numerous phantom studies, we show that PWE can handle complicated waveforms without prior filtering and is competitive with state-of-the-art that requires prior filtering based on the knowledge of propagation directions.
\end{abstract}

\vspace{2pc}

%\submitto{\PMB}

% ------------------------------------------------------------------------------------------------------------------------------ %
\section{Introduction} \label{sec:intro}
%
% Theme: motivation and evidence of correlation
%
Soft tissue pathology is known to be correlated with change in tissue stiffness \cite{WGRCUE2015SNPH}. For instance, an increase in cellular density caused by malignant tumors, leads to an increased tissue stiffness \cite{CPUSEETN2014MSMF,SWEBVQC2015BZ,ESWECFLL2017GFSW}.
Ultrasound elastography methods utilize this correlation by mechanically exciting the tissue and analyzing the subsequent motion to quantify its stiffness over a region of interest (ROI).
Various clinical studies of such methods indicate their relevance for non-invasive monitoring and diagnosis of breast and liver diseases among others \cite{BLQESSI2010ATTS,USRTCA2017SLEC}.
In the case of breast cancer for instance, \citeasnoun{BDCAUD2006IUTKT} used the quantitative ratio of lesion to background stiffness to determine the probability of malignancy, with higher contrasts indicating higher probability of the lesion being malignant.

% Theme: shear wave elastography
%
Different ultrasound elastography techniques exist in the literature that can broadly be categorized into strain imaging and shear wave elastograohy (SWE) methods \cite{USRTCA2017SLEC}.
Particularly, SWE methods generate and track shear waves that travel in the tissue with wave-speeds considerably smaller than their compressional counterparts (typically below $10$\,m/s) and have a low frequency content (below $1500$\,Hz). In these methods, the tissue excitation is achieved by a vertical displacement ($1-20$\,\textmu m) using an acoustic radiation force (ARF) impulse from an ultrasound push beam, while high frame-rate ultrasound imaging techniques are used to track the induced shear wave \cite{USWEISDP2016DRPN}.
%
%The SWE methods can be categorized into methods that measure displacement along the ARF axis \cite{ARFII2002NSNT} and methods that track the shear wave displacement in a ROI that excludes the ARF push beam \cite{SSI2004BTF, CPUSE2012SZMU}. In this paper, we consider the latter approaches.
The supersonic shear imaging (SSI) method proposed by \citeasnoun{SSI2004BTF}, is one of the early works to outline the SWE approach and demonstrate its performance.
On the other hand, the comb-push ultrasound shear elastography (CUSE) of \citeasnoun{CPUSE2012SZMU}, is a more recent work that relies on simultaneous, parallel ARF beams and compounding to improve the quality of shear modulus\footnote{Shear modulus is one of the several quantities used to measure the stiffness of materials.}estimations.

% Theme: signal processing algorithms - BME methods
%
A number of different SWE algorithms have been proposed to estimate the shear modulus field from wave data. For instance, \citeasnoun{SSI2004BTF} utilized the scalar wave equation along with the Fourier-transformed second-order derivatives of the scalar displacement field, while \citeasnoun{LSIATRSWS2006MR} proposed the use of a level-set function of wave-front arrival time.
However, most currently used SWE methods, including the commercialized versions of SSI \cite{BLQESSI2010ATTS} and CUSE \cite{CUSEBMIR2015DMSM} systems, utilize the time-of-flight (ToF) technique to estimate the wave-speed and corresponding shear modulus field \cite{QABLV2008TBAD,QHSMVARF2008PWDF,PARAQSWI2012RWPN,CPUSE2012SZMU,FSCSWSC2014SMZU,ISWGVEMSPTM2017CCMGU}.
For instance, \cite{QABLV2008TBAD,FSCSWSC2014SMZU} utilize cross-correlation of the signal at nearby locations along the propagation direction (typically the lateral axis) to estimate the ToF, i.e., the time that it takes the wave-front to reach the second point from the first point.
Alternatively, the time-to-peak of shear waveform can also be used to estimate the time-of-flight \cite{QHSMVARF2008PWDF,PARAQSWI2012RWPN,ISWGVEMSPTM2017CCMGU}.
These methods assume that the medium is locally homogeneous, isotropic, incompressible, and linear-elastic \cite{GFEMARFSW2016PQSU} and rely on the prior knowledge of the propagation direction which is assumed to be perpendicular to the ARF push axis \cite{USWEISDP2016DRPN}. This assumption is reasonable for `directional' shear waves with flat wave-fronts; see \cite{USRTCA2017SLEC} for details. We define a directional shear wave as a wave that has a clear dominant propagation direction. %A plane wave is the extreme form of a directional wave with a single propagation direction. In general, a directional wave can be expressed as a linear combination of plane waves.

% Theme: directional filtering
%
In the presence of inhomogeneities like tumors, shear waves scatter due to reflection and refraction, violating the directionality assumption of ToF methods, which can lead to artifacts in the estimated shear modulus field \cite{ERWTSWE2011DGBT,RSWFETP2017POZC}. Directional filters are then necessary to remove the reflections and obtain directional waves \cite{SDFIIMR2003MLKE,ERWTSWE2011DGBT,EISWSEMDF2016LRPN}.
In their simplest form, directional filters are frequency-domain projections that only retain the wave components aligned with the given propagation direction; see \cite{ERWTSWE2011DGBT} for details. They decompose the shear wave into a set of directional waves that can be separately analyzed via the ToF method.
In addition to reflection and refraction, noise in the wave data can create artifacts in the estimated fields and a second radial filter is often necessary to improve the signal-to-noise ratio; see \ref{app:directionalFilter} for details.
Ultimately, the shear modulus fields corresponding to different directions are combined through compounding (averaging) to obtain a final shear modulus field.
When the direction of propagation is not aligned with the lateral axis, one dimensional (1D) analysis can overestimate the wave-speed. The fast shear compounding (FSC) method proposed by \citeasnoun{FSCSWSC2014SMZU} utilizes a 2D analysis of wave-speed along with prior filtering for robust shear modulus estimation in inhomogeneous mediums. It extends the CUSE method \cite{CPUSE2012SZMU} to include multiple simultaneous ARF beams with arbitrary \textit{known} directions. In Section \ref{sec:exp}, we compare our proposed approach to this method.

% Theme: frequency-domain methods
%
Another set of SWE methods exist that rely on similar assumptions but perform the wave-speed (phase velocity) analysis in the frequency domain \cite{QEVMSWS2004CFG,SDUVMTE2009CUPK,DCSLMIURF2016BBBDT,VMILSWD2017VWWM,LPVI2018KU}.
For instance, the local phase velocity imaging method proposed by \citeasnoun{LPVI2018KU}, relies on short space 2D Fourier transform analysis of the wave data to extract the most dominant wavenumber within a homogeneous window. Assuming a directional propagation within the window, the phase velocity is estimated as the ratio of angular frequency to wavenumber. %This method achieves a comparable performance to state-of-the-art time-domain methods \cite{LPVI2018KU}.
%
% Theme: PDE-constrained methods
%
As discussed earlier, the assumption of directional propagation breaks down in the presence of inhomogeneities.
Another family of frequency-domain methods utilize the elastodynamic partial differential equation (PDE) to rigorously model the shear wave propagation in this general form, automatically accounting for reflection and refraction.
These methods can be classified into direct \cite{SMRDETHC2006PM} and iterative \cite{VCSTDFEM2008ESRO,MECEUE2017GZBA,AECETE2019AB} approaches. Direct methods, although more efficient, are sensitive to noise.
%The elastodynamic PDE rigorously describes the propagation of the shear wave through an inhomogeneous medium and can even model the viscoelastic behavior of the soft tissue \cite{MECEFD2015DAB}.
The iterative methods on the other hand, despite versatility, are computationally demanding which has limited their practical utility.

% Theme: PWE and placement in the literature
%
It is known that for shear waves induced by ARF beams, the displacement component parallel to the push axis is dominant \cite{GFEMARFSW2016PQSU}. Thus, we can use a scalar wave equation, instead of the elastodynamic PDE, to capture only the dominant shear wave component. Relying on this observation, we develop a novel frequency-domain SWE approach, called Plane Wave Elastography (PWE), that does not require the prior knowledge of propagation direction and is considerably more computationally efficient than the PDE-constrained approaches.
More specifically, given a homogeneous subdomain within the ROI, we represent the solution of the scalar wave equation as a linear combination of plane waves with arbitrary propagation directions. Using this representation, we formulate the SWE problem as a nonlinear least-squares problem that can be efficiently solved for the constant wave-speed within the subdomain. The PWE method relaxes the need for prior denoising by relying on a regularized least-squares formulation and directional filtering through automatic selection of dominant plane waves and does not require the post-processing step of compounding.
Moreover, the optimal mean squared error (MSE) is an indicator of how closely the data conforms to the scalar wave equation. This can be used to provide feedback on the quality of reconstruction since higher levels of MSE indicate deviation from the wave model due possibly to noise.
Finally, when the geometry of inclusions is known, e.g., from B-mode images, the PWE method can reconstruct the shear modulus field by a single solve for each homogeneous subdomain.

% Theme: contributions
%\blue{A schematic representation of the literature discussed above is given in Fig. \ref{fig:litReview}.}
%%
%\begin{figure}[t!]
%  \centering
%    \includegraphics[width=0.7\textwidth]{Sundry/Diagram/main.pdf}
%  \caption{A schematic representation of the available SWE approaches. The horizontal axis depicts the complexity of the model used to describe the shear wave propagation while the vertical axis shows the assumptions that each model relies on. The commonly used time-of-flight approaches are the most computationally efficient but make the most restrictive assumptions about the wave propagation. On the other hand, PDE-constrained optimization methods are the most versatile but computationally demanding. The proposed PWE method attempts to strike a balance between these two families of approaches by relying on a closed-form solution of the scalar wave equation to describe the propagation of the dominant shear wave component in the ROI.} \label{fig:litReview}
%\end{figure}
%
In summary, compared to the common SWE methods, our approach (i) has competitive reconstruction performance without relying on prior denoising, directional filtering and the prior knowledge of the propagation direction, or the need for compounding, (ii) provides feedback on quality of reconstructions using the MSE, and (iii) can take advantage of the prior knowledge of the inclusion geometry, if available, to speed up computations and improve the estimation contrast.

% Theme: traveling wave elastography (TWE)
%
%\red{A closely related approach to PWE was proposed by \citeasnoun{TWEMFAIP2011BSHS} for magnetic resonance elastography (MRE). This method utilizes the elastodynamic PDE, as opposed to the scalar wave equation, and relies on a set of plane wave expansions to represent the solution of the PDE. Using the vector form of the wave equation increases the computational cost and is unnecessary for ultrasound SWE, as we demonstrate in this paper. This also requires measurements of all displacement components which unlike MRE, are unavailable for ultrasound SWE. A major advantage of PWE compared to this method, is the systematic treatment of noise through regularization which enables more accurate reconstructions without prior filtering. Finally, due to the nature of MRE acquisitions and to limit the computational cost, this method utilizes data from a single frequency for reconstructions. As we demonstrate in Section \ref{sec:exp}, the use of multiple frequencies is imperative for accurate ultrasound SWE reconstructions.
%A comparison of this approach to other popular MRE techniques is given by \citeasnoun{CFEIALF2017HSRS}.
%%
%Finally, \citeasnoun{RSWFETP2017POZC} also utilize a plane-wave expansion but to model narrow-band reverberant wave fields in which, no propagation direction is dominant. This does not apply to the ARF-based SWE problem considered here which often has a handful of dominant directions.}
%
A closely related approach to PWE was proposed by \citeasnoun{TWEMFAIP2011BSHS} for magnetic resonance elastography (MRE). In their approach, the solution of an elastodynamics PDE was approximated using a set of plane wave expansions in a similar vein as in our PWE approach. However, their work considers a vector-valued problem in which all components of displacements are required. This is not directly applicable to an ultrasound modality where only one component of displacement is usually available. Our approach improves upon this previously proposed method in several key aspects by: i) adapting the plane-wave expansion to the scalar wave equation, thus enabling the use of ultrasound tracking, ii) decreasing the ensuing computational cost, iii) adding a multi-frequency treatment and, iv) incorporating the systematic and consistent treatment of noise through regularization which enables more accurate reconstructions without prior filtering. A comparison of the approach proposed by \citeasnoun{TWEMFAIP2011BSHS} to other popular MRE techniques is given by \citeasnoun{CFEIALF2017HSRS}. Finally, \citeasnoun{RSWFETP2017POZC} also utilized a plane-wave expansion, but to model narrow-band reverberant wave fields in which, no propagation direction is dominant. This does not apply to the ARF-based SWE problem considered here which often has a few dominant directions.

% Theme: structure of the paper
The remainder of this paper is organized as follows. In Section \ref{sec:PF}, we formulate the SWE problem and in Section \ref{sec:PWE}, we present the PWE approach to solve it. In Section \ref{sec:exp}, we present various phantom studies demonstrating the performance of the PWE method. Section \ref{sec:discussion} is dedicated to discussing various aspects of our method highlighting its strengths and weaknesses, and Section \ref{sec:concl} concludes the paper.

% ------------------------------------------------------------------------------------------------------------------------------ %
\section{Problem Formulation} \label{sec:PF}

% ------------------------------------------------------ %
\subsection{Scalar Wave Equation} \label{sec:scalarWE}
Let $\Omega_{\text{ROI}} \subset \reals^2$ denote the region of interest (ROI) and consider a shear wave propagating in $\Omega_{\text{ROI}}$ in response to one or a set of acoustic radiation force (ARF) push beams applied outside $\Omega_{\text{ROI}}$; let $\hhatbbu(t, \bbx): [0, T] \times \Omega_{\text{ROI}} \to \reals^2$ denote the in-plane displacement at time $t$ and point $\bbx$ due to this shear wave.
Ultrasound transducers often only measure the dominant component of the displacement parallel to the push directions; let $\hhatu(t, \bbx) = \hhatbbu_d(t, \bbx)$ denote this dominant component.

Assuming that the medium is isotropic, incompressible, and linear-elastic \cite{GFEMARFSW2016PQSU}, and assuming that the ARF push beams are applied outside the ROI, we can represent the propagation of the shear wave in $\Omega_{\text{ROI}}$ using the scalar wave equation
$$ \rho \, \ddot{\hhatu} = \nabla \cdot (\mu_f \nabla \hhatu) , $$
where $\rho = 1000$\,kg/m$^3$ is the mass density of the soft tissue and $\mu_f : \Omega_{\text{ROI}} \to \reals_{++}$ is the shear modulus field; $\reals_{++}$ denotes the positive real numbers.
Note that the shear modulus is related to the shear wave-speed $c_f : \Omega_{\text{ROI}} \to \reals_{++}$ as
\begin{equation} \label{eq:shearM}
\mu_f(\bbx) = \rho \, c_f^2(\bbx).
\end{equation}
Let $u: \Omega_{\text{ROI}} \to \mbC$ denote the Fourier transformed signal with respect to the temporal coordinate at a frequency $\omega$, i.e., $u(\bbx; \omega) = \ccalF\set{\hhatu(t, \bbx)}$. Then, given frequency $\omega$, the scalar wave equation can be written in the frequency-domain as
\begin{equation} \label{eq:scalarWE}
\rho \, \omega^2 u + \nabla \cdot (\mu_f \nabla u) = 0 .
\end{equation}

\begin{figure}
	\centering
	\begin{subfigure}[b]{0.36\textwidth}
		\includegraphics[width=\textwidth]{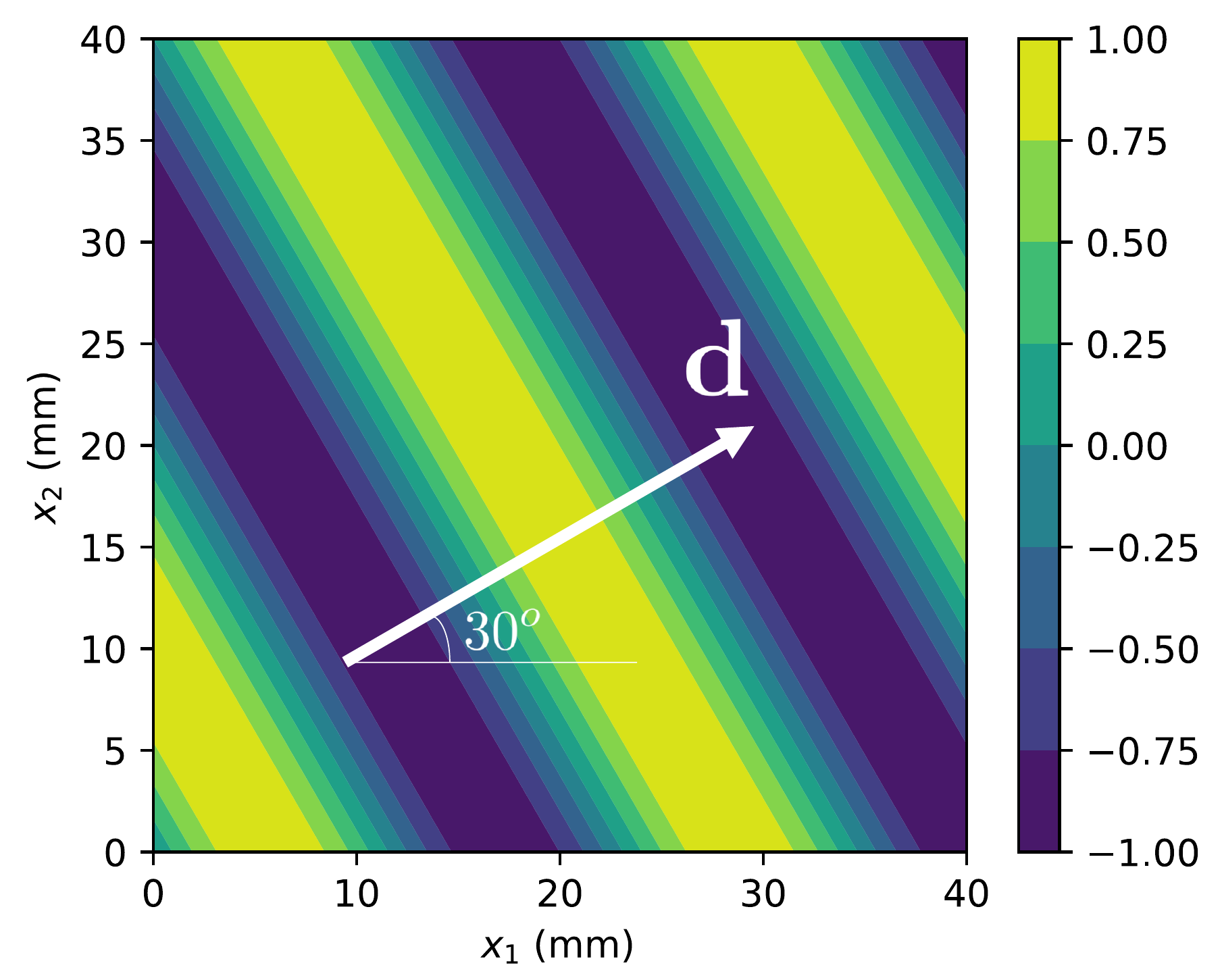}
	\caption{} \label{fig:plane_wave}
	\end{subfigure}
	\quad
	\begin{subfigure}[b]{0.35\textwidth}
		\includegraphics[width=\textwidth]{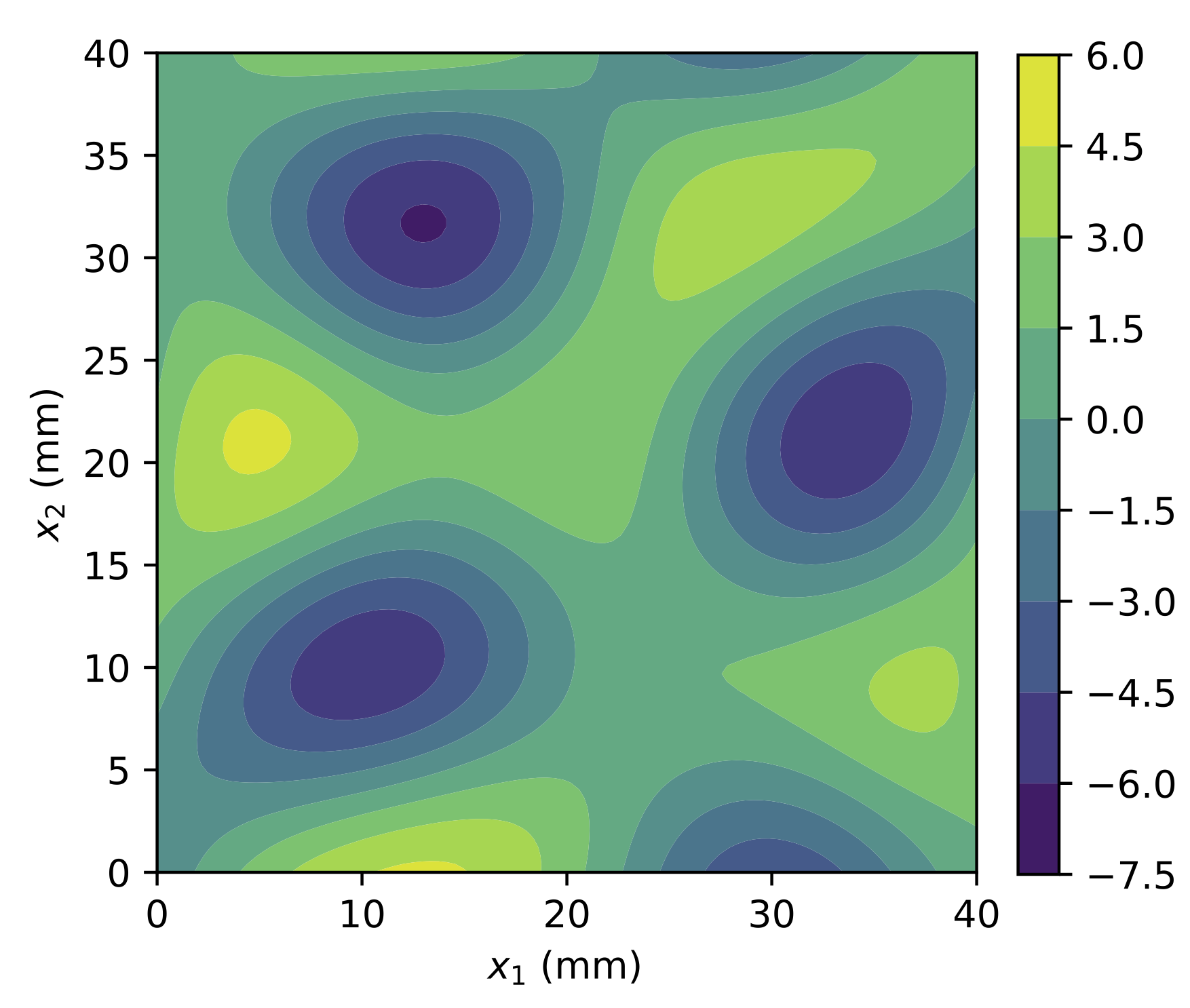}
	\caption{} \label{fig:helmholtz}
	\end{subfigure}
\caption{A plane wave and a shear wave field obtained by superposing plane waves within a homogeneous subdomain $\Omega = [0, 40] \times [0, 40]$\,mm$^2$ with wave-speed $c = 5$\,m/s at frequency $\omega = 500\pi$\,rad/s. (a) Fig. \ref{fig:plane_wave} shows the imaginary component of the plane wave \eqref{eq:basisFun} with $\bbd = [\cos(30^o), \sin(30^o)]$. (b) Fig. \ref{fig:helmholtz} depicts the imaginary component of the shear wave obtained from \eqref{eq:basisExpansion} by superposing $n_b = 12$ plane waves with $30^o$ angular spacing and random coefficients with arbitrary units.}	\label{fig:plane_waves}
\end{figure}

Consider a \textit{homogeneous} subdomain $\Omega \subset \Omega_{\text{ROI}}$ with \textit{constant} shear modulus $\mu = \mu_f(\bbx \in \Omega)$. Then, we can write solutions to the scalar wave equation \eqref{eq:scalarWE} in $\Omega$ as a linear combination of basis functions $\phi_j(\bbx) : \Omega \to \mbC$, i.e.,
\begin{equation} \label{eq:basisExpansion}
u(\bbx) = \sum\nolimits_{j=1}^{n_b} a_j \, \phi_j(\bbx) ,
\end{equation}
where $n_b$ is the number of basis functions, $a_j \in \mbC$, and the basis functions $\phi_j(\bbx)$ are plane waves explicitly given by
\begin{equation} \label{eq:basisFun}
\phi_j(\bbx) = \exp \left( \bbi \, \frac{\omega}{c} \, \bbd_j \cdot \bbx \right) .
\end{equation}
In this expression, $\bbi$ denotes the unit imaginary number, $c = c_f(\bbx \in \Omega)$ is the constant wave-speed within the homogeneous subdomain $\Omega$, and $\bbd_j \in \reals^2$ are unit direction vectors of propagation, i.e., $\norm{\bbd_j} = 1$, where $\norm{\cdot}$ denotes the Euclidean $\ell_2$-norm.
Fig. \ref{fig:plane_waves} depicts a plane wave of form \eqref{eq:basisFun} and a shear wave obtained by superposing $n_b=12$ plane wave bases according to \eqref{eq:basisExpansion}.
By increasing $n_b$ and appropriate selection of coefficients $a_j$ in \eqref{eq:basisExpansion}, we can approximate any solution to the scalar wave equation \eqref{eq:scalarWE} arbitrarily close in a normed sense \cite{ODHWFEHP2001CK}.

% ------------------------------------------------------ %
\subsection{Shear Wave Elastography Problem} \label{sec:elastProb}
Let $\hhatbby (t) \in \reals^m$ denote the temporal signal measured at $m$ points $\bbx_i \in \Omega$ for $i \in \set{1, \dots, m}$ and $\bby \in \mbC^m$ denote the corresponding (discrete) Fourier transformed signal at a frequency $\omega$.
The objective of the elastography problem then is to find the constant shear modulus $\mu$ or equivalently the wave-speed $c$ within the homogeneous subdomain $\Omega$. Given the basis expansion \eqref{eq:basisExpansion}, we can formulate this problem as a nonlinear least-squares optimization problem as follows:
%
%$$ \min_{c, \, \bba} \frac{1}{m} \sum_{i=1}^m \abs{ \sum_{j=1}^{n_b} a_j \, \phi_j(\bbx_i) - y_i }^2 , $$
%%
%where $\bba = [a_1, \dots, a_{n_b}]$ is the vector of coefficients.
%We can rewrite this optimization problem more compactly as
%
\begin{equation} \label{eq:opt0}
\min_{c, \, \bba} \frac{1}{m}  \norm{\bbPhi(c) \, \bba - \bby}^2 , 
\end{equation}
where $\bba = [a_1, \dots, a_{n_b}]$ is the vector of coefficients and $\bbPhi \in \mbC^{m \times n_b}$ is the design matrix with
\begin{equation} \label{eq:designMat}
\bbPhi_{ij} = \phi_j(\bbx_i)
\end{equation}
for $i \in \set{1, \dots, m} \and j \in \set{1, \dots, n_b}$.

Whenever the number of measurements $m$ is less than the number of bases $n_b$, the optimization problem \eqref{eq:opt0} is ill-posed. Furthermore, often the measured signal is contaminated with noise and we do not want the solution \eqref{eq:basisExpansion} of the wave equation \eqref{eq:scalarWE} to perfectly match the measurements $\bby$. To address these challenges, we add a regularization term to \eqref{eq:opt0} that improves stability and allows us to control how closely we fit the data:% More specifically, we rewrite the optimization problem as
\begin{equation*} \label{eq:opt1}
\min_{c, \, \bba} \frac{1}{m} \norm{\bbPhi(c) \, \bba - \bby}^2 + \tau \norm{\bba}^2,
\end{equation*}
where $\tau \in \reals_{++}$ is the regularization parameter. Note that the use of regularization encourages selection of a subset of plane waves and helps distinguish dominant propagation directions by penalizing nonzero coefficients $a_j$.

So far we have utilized the data at a single frequency $\omega$. It is often necessary to consider a set of dominant frequencies $\set{\omega_1, \dots, \omega_{n_{\omega}}}$, where $n_{\omega}$ denotes the number of frequencies. Given measurements $\bby_k \in \mbC^m$ for $k \in \set{1, \dots, n_{\omega}}$, we use the linear expansion \eqref{eq:basisExpansion} with coefficients $\bba_k$ to represent the solution of the scalar wave equation \eqref{eq:scalarWE} at frequency $\omega_k$. Then, the corresponding elastography problem is given by
\begin{equation} \label{eq:elastProb}
\min_c \sum\nolimits_{k=1}^{n_{\omega}} \min_{\bba_k} \frac{1}{m} \norm{\bbPhi_k(c) \, \bba_k - \bby_k}^2 + \tau \norm{\bba_k}^2 .
\end{equation}

%\begin{rem}
Shear wave data are often calculated from the in-phase-quadrature data using an autocorrelation algorithm \cite{RTBFIAT1985KNKO} and are given as particle \textit{velocity} and not displacement. Noting that $\ccalF\set{\dot{\hhatu}} = \bbi \, \omega \ccalF\set{\hhatu} = \bbi \, \omega \, u$, we can represent the (discrete) Fourier transformed velocity data with an expansion similar to \eqref{eq:basisExpansion} and scaled coefficients. Therefore, regardless of whether displacement or velocity data are used, the elastography problem is formulated as \eqref{eq:elastProb}.
%\end{rem}
%
In the next section, we discuss an efficient approach to solve this optimization problem.

% ------------------------------------------------------------------------------------------------------------------------------ %
\section{Plane Wave Elastography} \label{sec:PWE}

% ------------------------------------------------------ %
\subsection{Solution to the Elastography Problem in a Homogeneous Subdomain} \label{sec:solution}
Solving the optimization problem \eqref{eq:elastProb} can be challenging due to nonlinearity; see \ref{app:homogeneous}. However, for a fixed wave-speed $c$, \eqref{eq:elastProb} is a standard $\ell_2$-regularized least-squares problem whose solution for each frequency, is given in closed-form by
\begin{equation} \label{eq:closed-form}
\bba^*_k(c) = \left( \frac{1}{m} \bbPhi_k(c)^H \bbPhi_k(c) + \tau \, \bbI \right)^{-1} \bbPhi_k(c)^H \bby_k ,
\end{equation}
where $\bbI \in \reals^{n_b \times n_b}$ is the identity matrix and the superscript $H$ denotes the conjugate transpose operator.
Given this closed-form expression and since within the homogeneous subdomain $\Omega$, the wave-speed $c$ is a constant scalar, we can utilize a global search algorithm or a simple discretization method to find the optimal wave-speed as
\begin{equation} \label{eq:PWEsol}
c^* = \argmin_{c \in [c_{\min}, c_{\max}]} \sum\nolimits_{k=1}^{n_{\omega}} \frac{1}{m} \norm{\bbPhi_k(c) \, \bba^*_k(c) - \bby_k}^2 + \tau \norm{\bba^*_k(c)}^2 ,
\end{equation}
where $c_{\min}, c_{\max} \in \reals_{++}$ are the lower-bound and upper-bound on the wave-speed and $\bba^*_k (c)$ is the optimal regularized least-squares solution \eqref{eq:closed-form} for a given wave-speed $c$.

Assume that we use a limited number of basis functions $n_b$ and frequencies $n_{\omega}$ along with appropriate regularization to prevent overfitting the noise. Then, the value of the least-squares term corresponding to the optimal wave-speed $c^*$ can be used as a measure of conformity of the data to the scalar wave equation \eqref{eq:scalarWE}; the smaller the least-squares error, the closer the data is to the underlying physics. Thus, we can use this feedback to evaluate the quality of reconstruction. Particularly, we use the mean squared error (MSE), given by
\begin{equation} \label{eq:MSE}
\text{MSE} = \frac{1}{m \, n_{\omega}} \sum\nolimits_{k=1}^{n_{\omega}} \norm{\bbPhi_k(c^*) \, \bba^*_k(c^*) - \bby_k}^2 ,
\end{equation}
as a measure of the quality of reconstruction in $\Omega$.

% ------------------------------------------------------ %
\subsection{Plane Wave Elastography Algorithm} \label{sec:}
\begin{figure}[t!]
  \centering
    \includegraphics[width=0.4\textwidth]{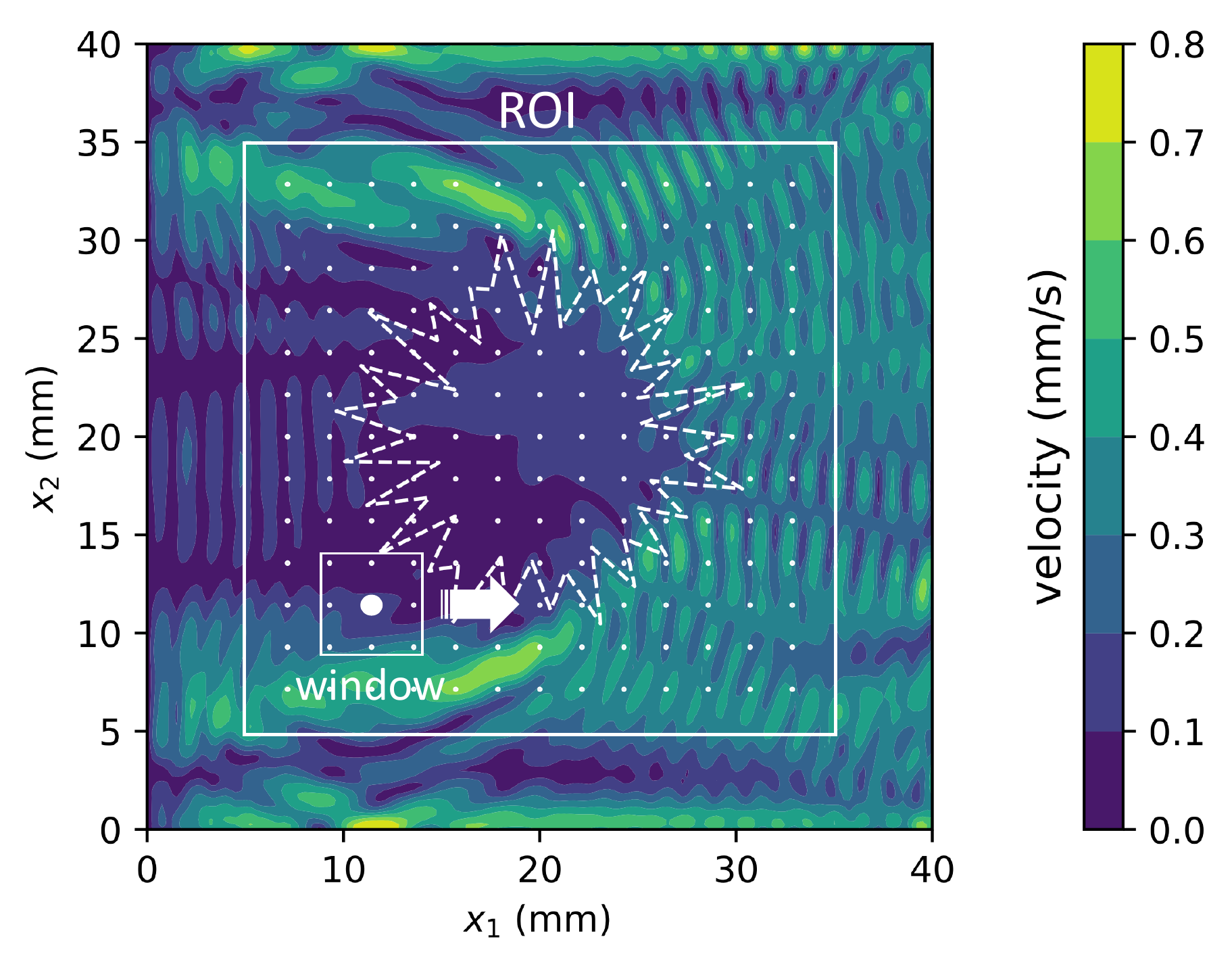}
  \caption{Schematic of the PWE algorithm to estimate the shear modulus field in an inhomogeneous ROI. The white dots depict $n_s$ discretization points within the ROI. The main idea is to estimate the field at each discretization point by a window $\Omega_l$ of size $w \times w$ centered at that point. %The window $\Omega_l$ is assumed homogeneous and \eqref{eq:PWEsol} can be used to estimate the constant wave-speed.
 The contour plot shows the magnitude of Fourier transformed data at $800$\,Hz for the digital phantom of Section \ref{sec:digital-phantom} as an example, where the white dashed lines delineate the boundary of the inclusion. In practice, the PWE Algorithm \ref{alg:PWE} utilizes $n_{\omega}$ dominant frequencies within each window.} \label{fig:PWE}
\end{figure}

Our solution in Section \ref{sec:solution} was for a homogeneous subdomain $\Omega$ with constant shear modulus. To estimate the shear modulus field $\mu_f(\bbx)$ over an inhomogeneous ROI, we discretize the ROI into $n_s$ grid points and use windows of size $w \in \reals_{++}$ with constant shear moduli. Particularly, for a point $\bbx_l \in \Omega_{\text{ROI}}$ we define a window (subdomain) $\Omega_l$ as
\begin{equation} \label{eq:domDecWindow}
\Omega_l = \bbx_l + \left[ -\frac{w}{2}, \frac{w}{2} \right] \times \left[ -\frac{w}{2}, \frac{w}{2} \right] .
\end{equation}
We solve \eqref{eq:PWEsol} to estimate the wave-speed $c_l$ and the corresponding shear modulus $\mu_l$ within $\Omega_l$ and assign the value to point $\bbx_l$, constructing in this way, a discretized vector of estimations $\bbmu \in \reals_{++}^{n_s}$ for the shear modulus field $\mu_f(\bbx)$; see Fig. \ref{fig:PWE}.

The Plane Wave Elastography (PWE) approach is summarized in Algorithm \ref{alg:PWE}.
\begin{algorithm}[t]
\caption{Plane Wave Elastography Algorithm}
\label{alg:PWE}
\begin{algorithmic}[1]
\small

\REQUIRE Discretization of the ROI and window size $w$;

\REQUIRE Number of basis functions $n_b$ and dominant frequencies $n_{\omega}$;

%\REQUIRE Number of dominant frequencies $n_{\omega}$;

\REQUIRE Shear wave measurement signal $\hhatbby_{\text{ROI}}(t)$;

\STATE Compute the Fourier transformed data $\bby_{\text{ROI}}$;

\STATE Select the regularization parameter $\tau$;	\label{line:regParam}

\FOR {$l \in \set{1, \dots, n_s}$}

	\STATE Get measurements $\bby_k \in \mbC^m$ within window $\Omega_l$ for dominant frequencies $k \in \set{1, \dots, n_{\omega}}$;		\label{line:domFreq}
		
	\STATE Compute the wave-speed $c_l$ from \eqref{eq:PWEsol} and the corresponding MSE value from \eqref{eq:MSE};		\label{line:computeC}
	
\ENDFOR
	
\STATE Compute the shear modulus vector $\bbmu \in \reals_{++}^{n_s}$ from \eqref{eq:shearM};	\label{line:muVec}

\STATE Return $\bbmu$ and the corresponding MSE vector \eqref{eq:MSE};

\end{algorithmic}
\end{algorithm}
The algorithm starts by requiring the $n_s$ discretization points and window size $w$, as well as the number of plane wave basis functions $n_b$ and dominant frequencies $n_{\omega}$.
In line \ref{line:regParam}, given the (discrete) Fourier transformed displacement signal $\bby_{\text{ROI}}$, it selects the regularization parameter $\tau$; see Section \ref{sec:paramSelec} for details. Then, the algorithm loops over the $n_s$ discrete points within the ROI. In line \ref{line:domFreq}, given the measurements within window $\Omega_l$ for $l \in \set{1, \dots, n_s}$, it extracts the $n_{\omega}$ dominant frequencies contributing the highest amount of energy to the Fourier spectrum, and the corresponding measurements $\bby_k \in \mbC^m$ for $k \in \set{1, \dots, n_{\omega}}$. Then, in line \ref{line:computeC}, it computes the constant wave-speed $c_l$ for window $\Omega_l$ from \eqref{eq:PWEsol} and the corresponding MSE from \eqref{eq:MSE}.
In line \ref{line:muVec}, $\bbmu$ collects the estimated shear moduli corresponding to all discretization points $\bbx_l$ within the ROI. Given $\bbmu$, we can approximate the shear modulus field $\mu_f(\bbx)$ at any point $\bbx$ via interpolation.

% ------------------------------------------------------ %
\subsection{Parameter Selection} \label{sec:paramSelec}
%
% paragraph theme: basis number
Next, we discuss the important parameters that affect the performance of the PWE Algorithm \ref{alg:PWE}. As discussed in Section \ref{sec:intro}, an important advantage of the PWE approach is that it does not rely on the directionality of the propagation, due to the plane wave representation in \eqref{eq:basisExpansion}. In the absence of any prior knowledge on the directionality of the propagation, we choose the directions $\bbd_j$ to uniformly sample $[0, 2\pi]$, i.e.,
\begin{equation} \label{eq:dirDisc}
\bbd_j = \left[ \cos\left(2\pi \frac{j}{n_b} \right), \sin\left(2\pi \frac{j}{n_b} \right) \right] .
\end{equation}
As we demonstrate in Section \ref{sec:exp}, in practice often $n_b\le12$ directions are sufficient to resolve the propagating waves.

% paragraph theme: frequency parameters and window size
An important parameter for the PWE method is the number of dominant frequencies $n_{\omega}$. In principle, increasing $n_{\omega}$ adds more information and leads to more accurate reconstructions. However, one should take caution not to include very high frequencies with unfavorable signal to noise ratios, i.e., $\omega_{\max} = 2\pi f_{\max}$ must be upper-bounded, where $f_{\max}$ is the corresponding frequency in Hertz.
Another important parameter is the lower-bound $\omega_{\min} = 2\pi f_{\min}$ on dominant frequencies, which determines the longest wavelength $\lambda_{\max}$ in the data. More specifically, $\lambda_{\max} = \hhatc_{\max}/f_{\min}$ where $\hhatc_{\max}$ is the unknown maximum wave-speed in the ROI. As a general rule of thumb, for data with a high signal-to-noise ratio (SNR), the window size must be larger than $\lambda_{\max}/5$ to ensure that waves can be resolved with a window of size $w$. Note that $w$ also determines the number of measurements $m$ used to estimate the wave-speed in each subdomain $\Omega_l$. Using at least $m>n_b$ measurements is required to ensure that the design matrix \eqref{eq:designMat} is well-conditioned. When the SNR is low, larger values of $w$ should be used to ensure that the measurements contain adequate information about the unknown wave speed.
In practice, the range of dominant frequencies $[f_{\min}, f_{\max}]$ and the spatial resolution of measurements are fixed for a given shear wave data but both $\hhatc_{\max}$ and SNR are unknown. As a result, for best reconstructions we might need to adjust $w$.
The quality of reconstructions by the PWE Algorithm \ref{alg:PWE} are often better for data with higher SNR and $f_{\min}$ for which smaller values of $w$ can be used.

%For a given window size $w$, there exists a value of $f_{\min}$ below which a wave cannot be resolved with a window of size $w$. The performance of the PWE Algorithm \ref{alg:PWE} is often better for data with higher dominant frequencies, i.e., higher $f_{\min}$.

% paragraph theme: window size
%A closely related parameter is the window size $w$. Given data with a fixed set of dominant frequencies, use of overly small values of $w$ compared to the longest wavelength, results in poor reconstruction. On the other hand, use of excessively large values of $w$ results in over-smoothing and poor contrast. %In practice there is a range of values for $w$ that result in a reasonable reconstruction.
%Moreover, $w$ determines the number of measurements $m$ used to estimate the wave-speed in each subdomain $\Omega_l$. Using $m>n_b$ is often required to ensure that the design matrix \eqref{eq:designMat} is well-conditioned. %This imposes an additional lower-bound on $w$ that is related to the spatial resolution of measurements.
%\red{In practice, the range of dominant frequencies $[f_{\min}, f_{\max}]$ and the spatial resolution of measurements are determined given shear wave data and $w$ is selected accordingly.}

% paragraph theme: regularization parameter
When data is noisy, a major parameter that affects the reconstruction is the regularization parameter $\tau$. As discussed in Section \ref{sec:elastProb}, proper selection of $\tau$ allows us to simultaneously perform filtering and reconstruction. In this paper, we utilize the L-curve approach to select $\tau$. This involves plotting the regularization term $\sum\nolimits_{k=1}^{n_{\omega}} \norm{\bba^*_k}^2$ in \eqref{eq:PWEsol} versus the sum-of-squares value (a constant multiple of MSE \eqref{eq:MSE}) as a function of $\tau$ and selecting the regularization parameter corresponding to the point of maximum curvature in the L-curve;\footnote{Due to the extra minimization with respect to wave-speed in \eqref{eq:elastProb}, the curve generated in this way is not exactly the L-curve but as we show in Section \ref{sec:exp}, this version can still be used to select appropriate regularization parameters.}see \cite{DIP2010H}.
In Section \ref{sec:paramStudy}, we present parameter study results to further clarify the discussion of this section.

% ------------------------------------------------------ %
\subsection{Prior Knowledge of ROI Geometry} \label{sec:prior}
In practice, the prior knowledge of the location and shape of inclusions (inhomogeneities) within the ROI might be available, e.g., from B-mode images. In that case, we can estimate the shear modulus field with a considerably fewer solves than what is needed in Algorithm \ref{alg:PWE}, which requires one solution per discretization point. Particularly, consider a decomposition of the ROI into $n_s$ non-overlapping subdomains such that
$ \Omega_{\text{ROI}} = \bigcup\nolimits_{l=1}^{n_s} \Omega_l , $
%
%\begin{equation} \label{eq:decomposition}
%\Omega_{\text{ROI}} = \bigcup\nolimits_{l=1}^{n_s} \Omega_l ,
%\end{equation}
%
where within $\Omega_l$ the shear modulus is constant and equal to $\mu_l$.
Then, solving \eqref{eq:PWEsol} with measurements belonging to $\Omega_l$, we obtain an estimate of the wave-speed $c_l$ and the corresponding shear modulus $\mu_l$ from \eqref{eq:shearM} and we can estimate the shear modulus field as 
\begin{equation}
\mu_f(\bbx) = \sum\nolimits_{l=1}^{n_s} \mu_l \, \mathbf{1}_{\Omega_l}(\bbx) ,
\end{equation}
where the indicator function $\mathbf{1}_{\Omega_l}(\bbx) = 1 \ \text{if} \ \bbx \in \Omega_l$ and is zero otherwise.
%$\mathbf{1}_{\Omega_1}(\bbx)$ is defined by
%%
%\begin{equation}
%\mathbf{1}_{\Omega_1}(\bbx) =
%\left\{
%\begin{array}{cl}
%1  & \text{if} \ \bbx \in \Omega_l     \\
%0  & \text{o.w.}     
%\end{array}
%\right. .
%\end{equation}

% ------------------------------------------------------------------------------------------------------------------------------ %
\section{Experiments} \label{sec:exp}
In this section, we present phantom studies to demonstrate the performance of the PWE Algorithm \ref{alg:PWE}. Particularly, we first study a digital phantom with a complex inclusion mimicking a malignant tumor to induce reflections and refractions, demonstrating the ability of PWE to resolve complicated wave patterns without prior filtering.
Then, we consider two categories of phantom experiments. The first category involves four simultaneous ARF push beams applied using a curved-array ultrasound transducer at different angles and validates in practice, the ability of the PWE approach to resolve waves traveling at unknown arbitrary directions. We also consider the more common case of two parallel push beams generated by a linear transducer. For these phantom experiments, we calculate the shear wave data from the in-phase-quadrature data using an autocorrelation algorithm \cite{RTBFIAT1985KNKO}.
Finally, we study the effect of various parameters on the performance of PWE to further illustrate the discussion of Section \ref{sec:paramSelec}. %Finally, we evaluate the significance of the MSE feedback in evaluating the reconstruction performance.

% paragraph theme: FSC method
In each case, in addition to nominal values, we report reconstructions by the fast shear compounding (FSC) method, proposed by \citeasnoun{FSCSWSC2014SMZU}, to validate the PWE reconstructions and demonstrate that PWE performs at least as well as the state-of-the-art\footnote{FSC method is commercially used on General Electric LOGIQ E9 SWE system \cite{TSWECU2015SMBL}.}, even though it does not require prior filtering or post-processing.
Note that the FSC method strongly relies on directional filtering to ensure the directionality of the waves and radial filtering in the spatial frequency domain, to enhance the SNR; see \ref{app:directionalFilter} for details. In the case of a multi-push excitation, the FSC method reconstructs shear modulus fields individually for each filtered direction and then uses compounding to combine the reconstructions.
In the following, we fine-tune the parameters of this method for best possible reconstructions. We particularly report the window size defined similar to \eqref{eq:domDecWindow}, and the patch size which is the distance between pairs of points used for cross-correlation to determine the time-of-flight; see \cite{FSCSWSC2014SMZU} for details.

To measure the reconstruction performance, we report the average shear moduli $\mu_b = \avg(\bbmu_{\Omega_b}) \and \mu_i = \avg(\bbmu_{\Omega_i})$ over the background $\Omega_b$ and inclusion $\Omega_i$ along with the standard deviations $\std(\bbmu_{\Omega_b}) \and \std(\bbmu_{\Omega_i})$, where $\bbmu_{\Omega_b} \and \bbmu_{\Omega_i}$ denote the estimated shear modulus vector $\bbmu$ confined to subdomains $\Omega_b \and \Omega_i$, respectively.
We also report the contrast-to-noise ratio (CNR), defined as
\begin{equation} \label{eq:CNR}
\text{CNR} = 20 \log_{10} \frac{ \abs{ \avg(\bbmu_{\Omega_b}) - \avg(\bbmu_{\Omega_i}) } }{ \sqrt{ \std^2(\bbmu_{\Omega_b}) + \std^2(\bbmu_{\Omega_i}) } }.
\end{equation}
%
%To measure the reconstruction performance, we use two different metrics. First, we report the normalized $\ell_2$-error compared to the ground truth given by
%%
%\begin{equation} \label{eq:err}
%\text{err}_{\Omega_l} = \frac{\norm{ \bbmu_{\Omega_l} - \bbmu^{\text{true}}_{\Omega_l} }}{ \norm{ \bbmu^{\text{true}}_{\Omega_l} } } ,
%\end{equation}
%%
%where $\bbmu_{\Omega_l} \and \bbmu^{\text{true}}_{\Omega_l}$ are the estimated and exact shear modulus vectors at discrete points within subdomain $\Omega_l$. Particularly, since in each study there exists only one inclusion, we report three error values: total error err$_t$ for $\Omega_{\text{ROI}}$, error err$_b$ at the background for $\Omega_b$, and error err$_i$ at the inclusion for $\Omega_i$.
%%
%We also report the contrast-to-noise ratio (CNR), defined as
%%
%\begin{equation} \label{eq:CNR}
%\text{CNR} = 20 \log_{10} \frac{ \abs{ \avg(\bbmu_{\Omega_b}) - \avg(\bbmu_{\Omega_i}) } }{ \std(\bbmu_{\Omega_b}) },
%\end{equation}
%%
%where $\avg(\cdot) \and \std(\cdot)$ denote the mean and standard deviation functions and $\bbmu_b \and \bbmu_i$ denote the estimated shear modulus vector $\bbmu$ confined to subdomains $\Omega_b \and \Omega_i$, respectively.

Throughout this section, for PWE reconstructions we use $n_{\omega} = 10$ dominant frequencies and $n_b = 12$ basis functions and set the maximum frequency to $f_{\max} = 1500$\,Hz; see Section \ref{sec:paramStudy} for the reasoning behind this selection. We also set the wave-speed bounds in \eqref{eq:PWEsol} to $c_{\min} = 1$\,m/s and $c_{\max} = 10$\,m/s, which is a reasonable range for soft tissue.
Moreover, to conform to the ultrasound coordinate system convention, where the transducer is located on top, in the following plots we use axial or depth axis $z$ and lateral axis $x$ with the origin located on the top-left corner. %instead of the standard $x_1-x_2$ coordinate system used in the previous sections.

% --------------------------------------------------------- %
\subsection{Single-push Digital Phantom} \label{sec:digital-phantom}
\begin{figure*}
\centering
\setlength{\tabcolsep}{2pt}
\begin{tabular}{ccccc}
					&	(a) FSC		&	(b) PWE					&	(c) MSE	&	(d) cross-section	\\
\rotatebox{90}{\hspace{3mm}noiseless data}			&
\includegraphics[height=25mm]{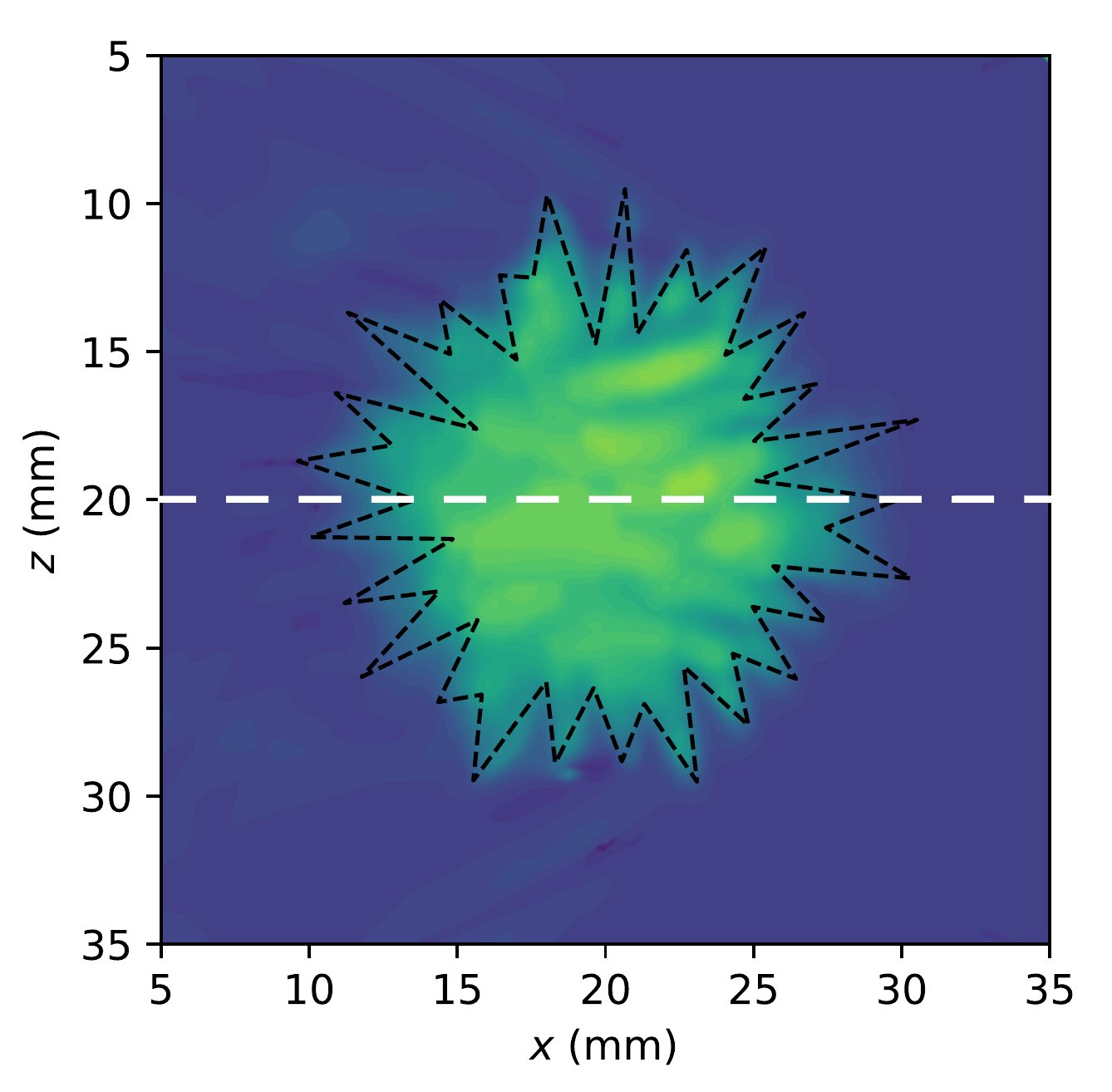}			& 
\includegraphics[height=25mm]{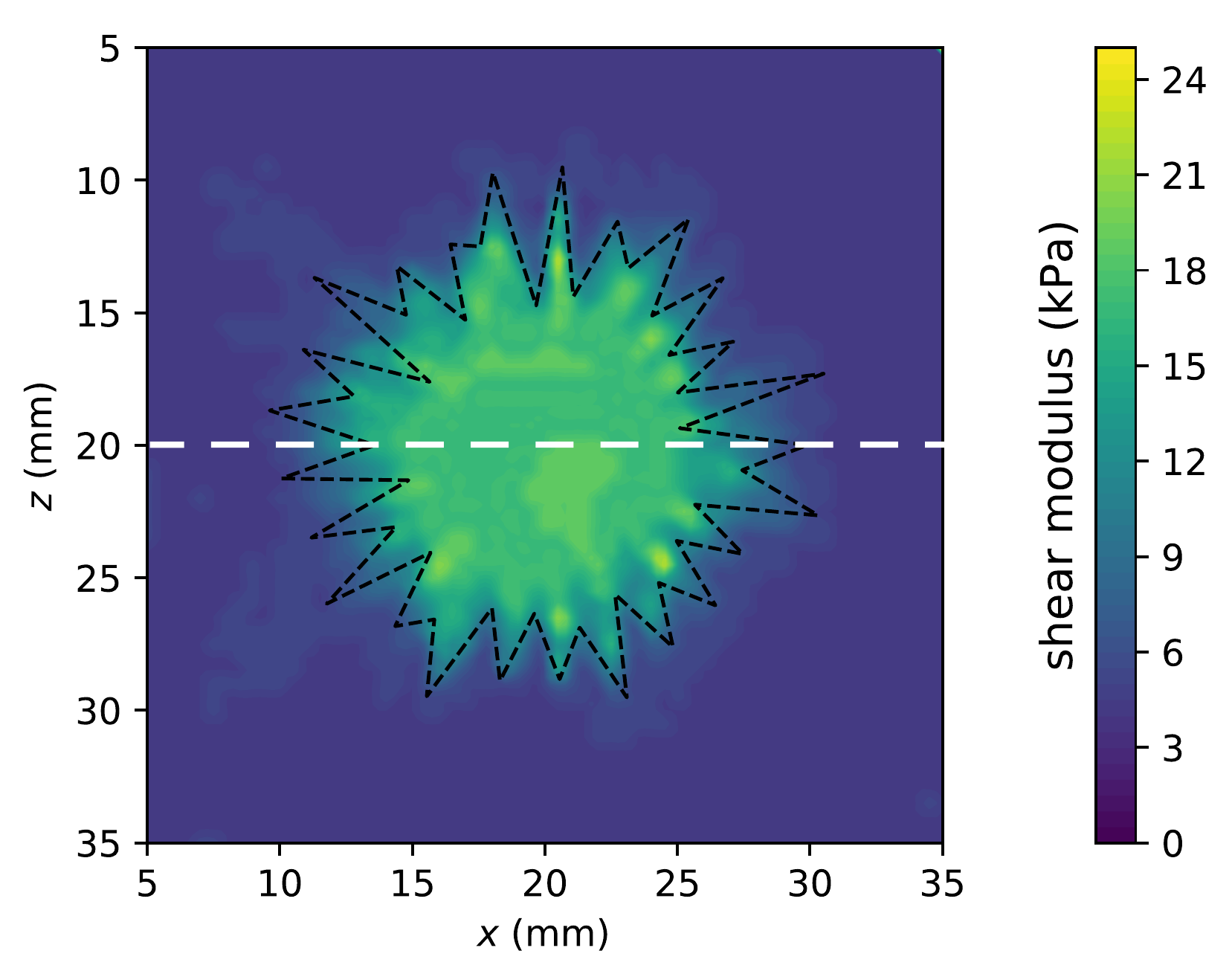}			& 
\includegraphics[height=25mm]{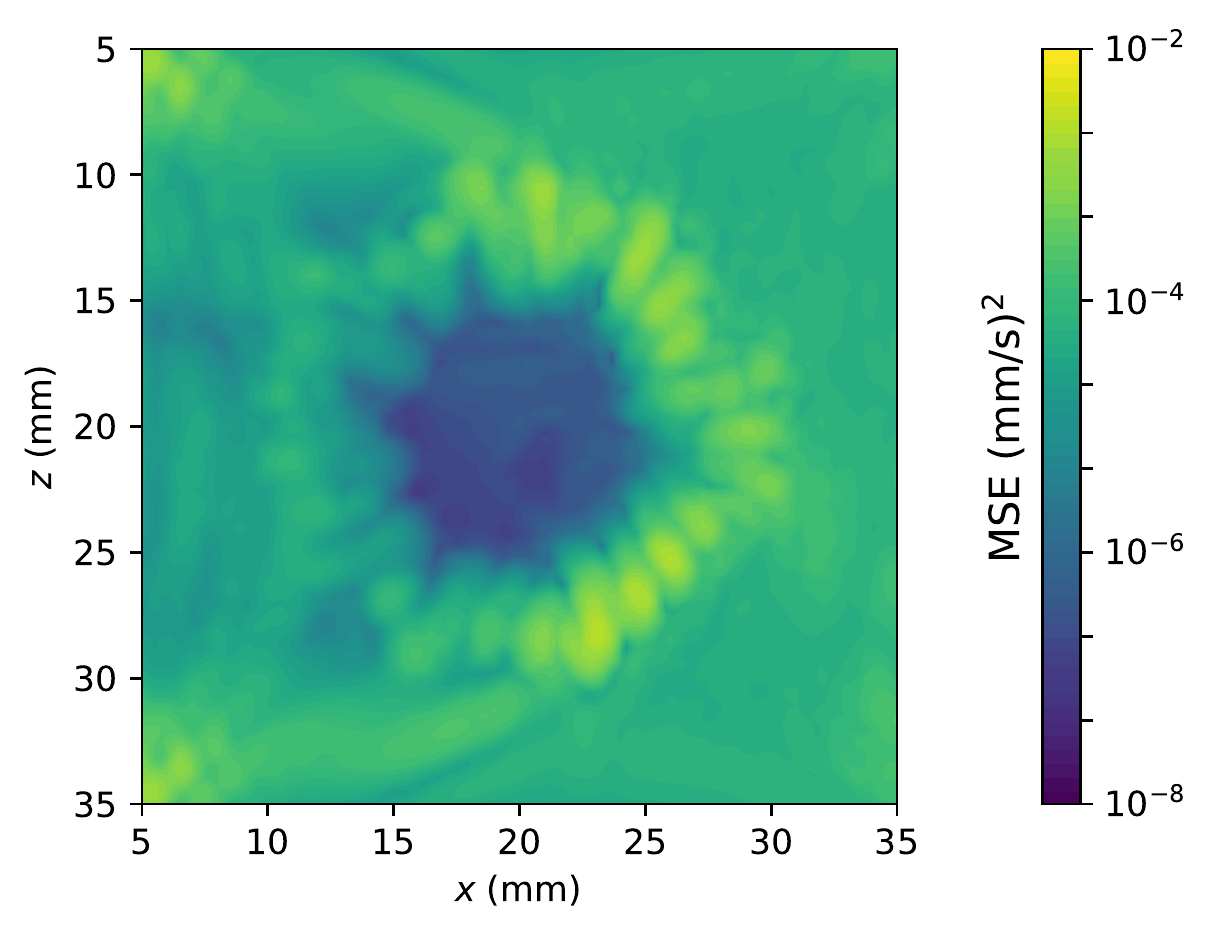}			& 
\includegraphics[height=25mm]{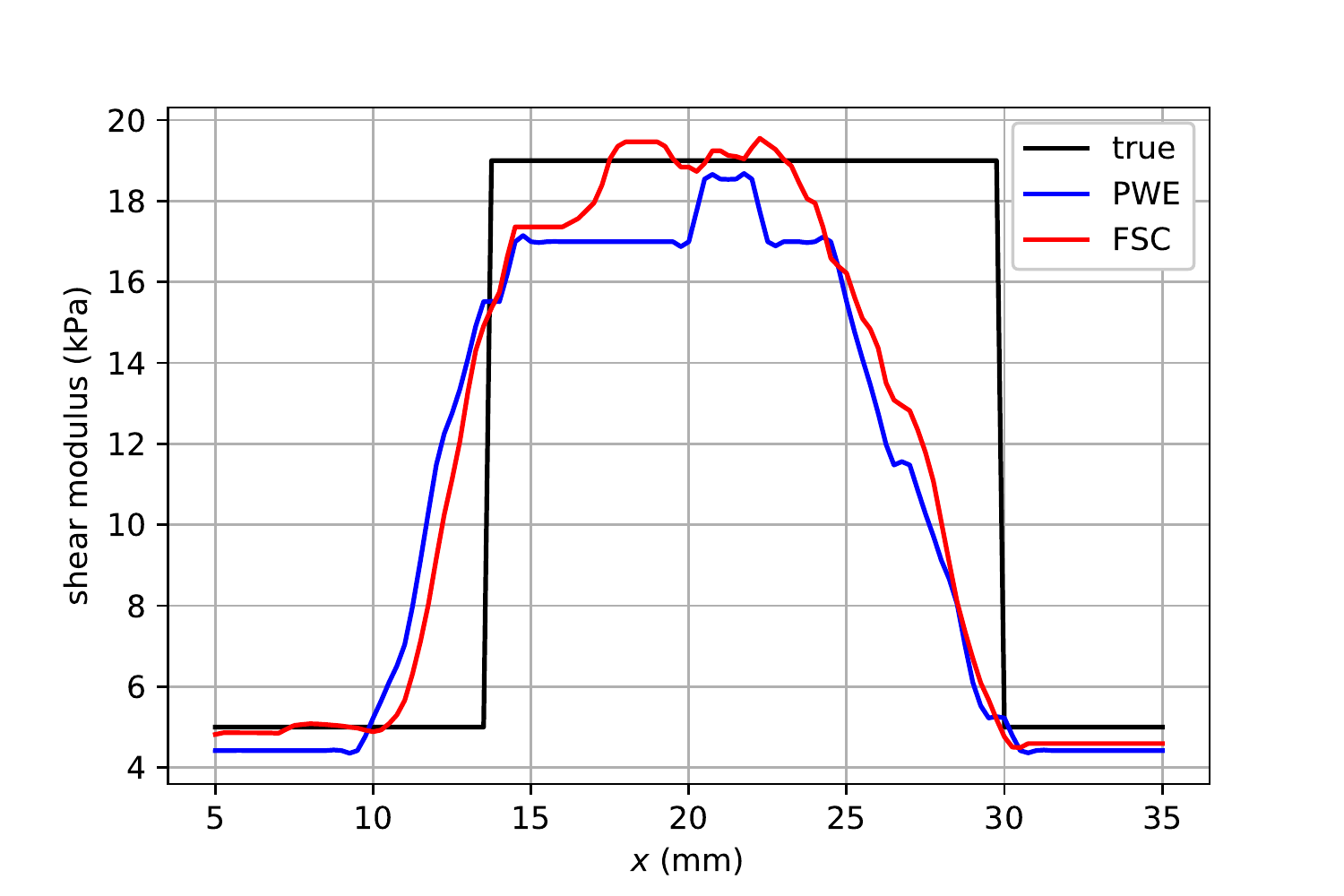} 	\\
\rotatebox{90}{\hspace{6mm}noisy data}						&
\includegraphics[height=25mm]{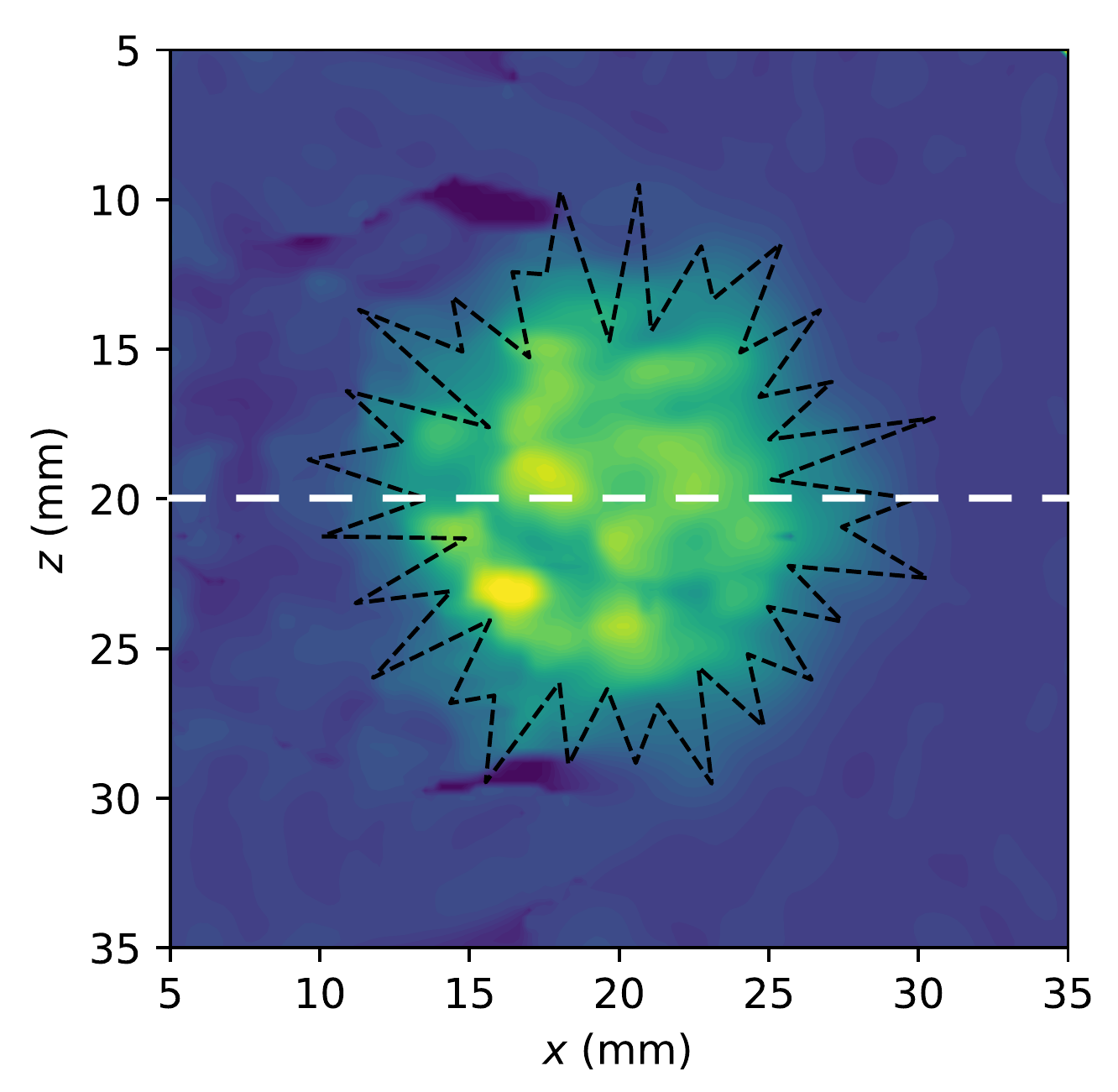}			& 
\includegraphics[height=25mm]{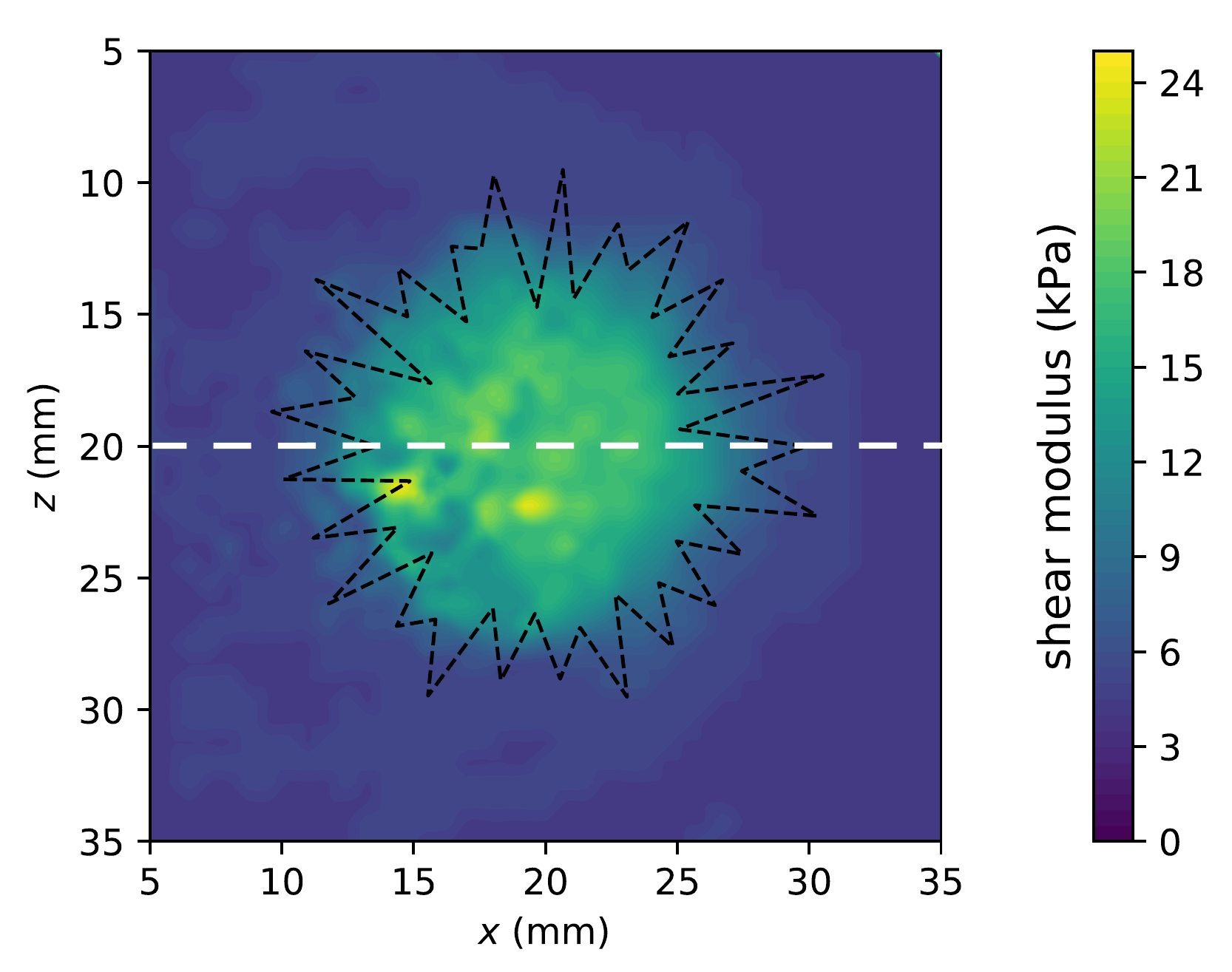}			& 
\includegraphics[height=25mm]{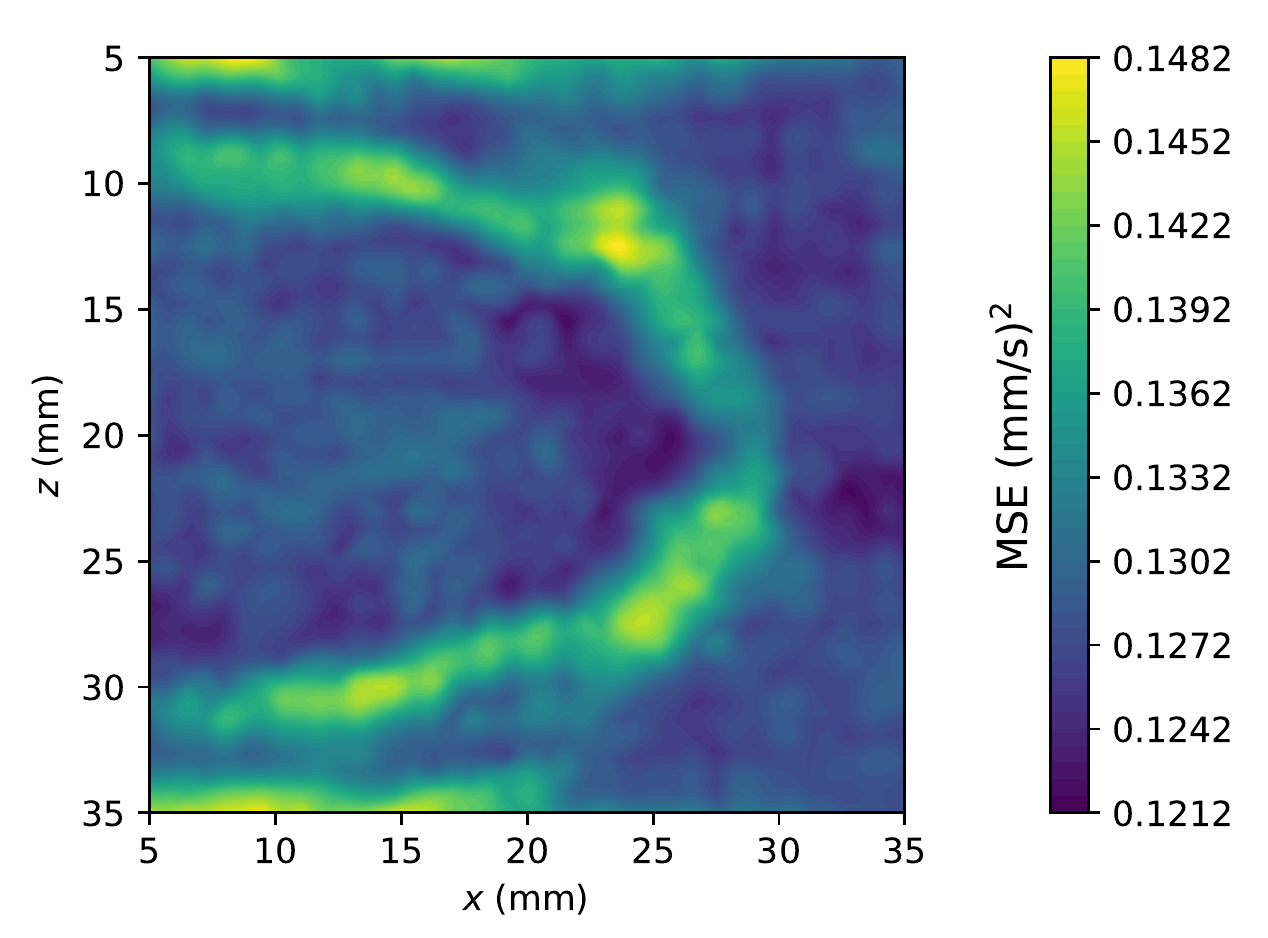}		& 
\includegraphics[height=25mm]{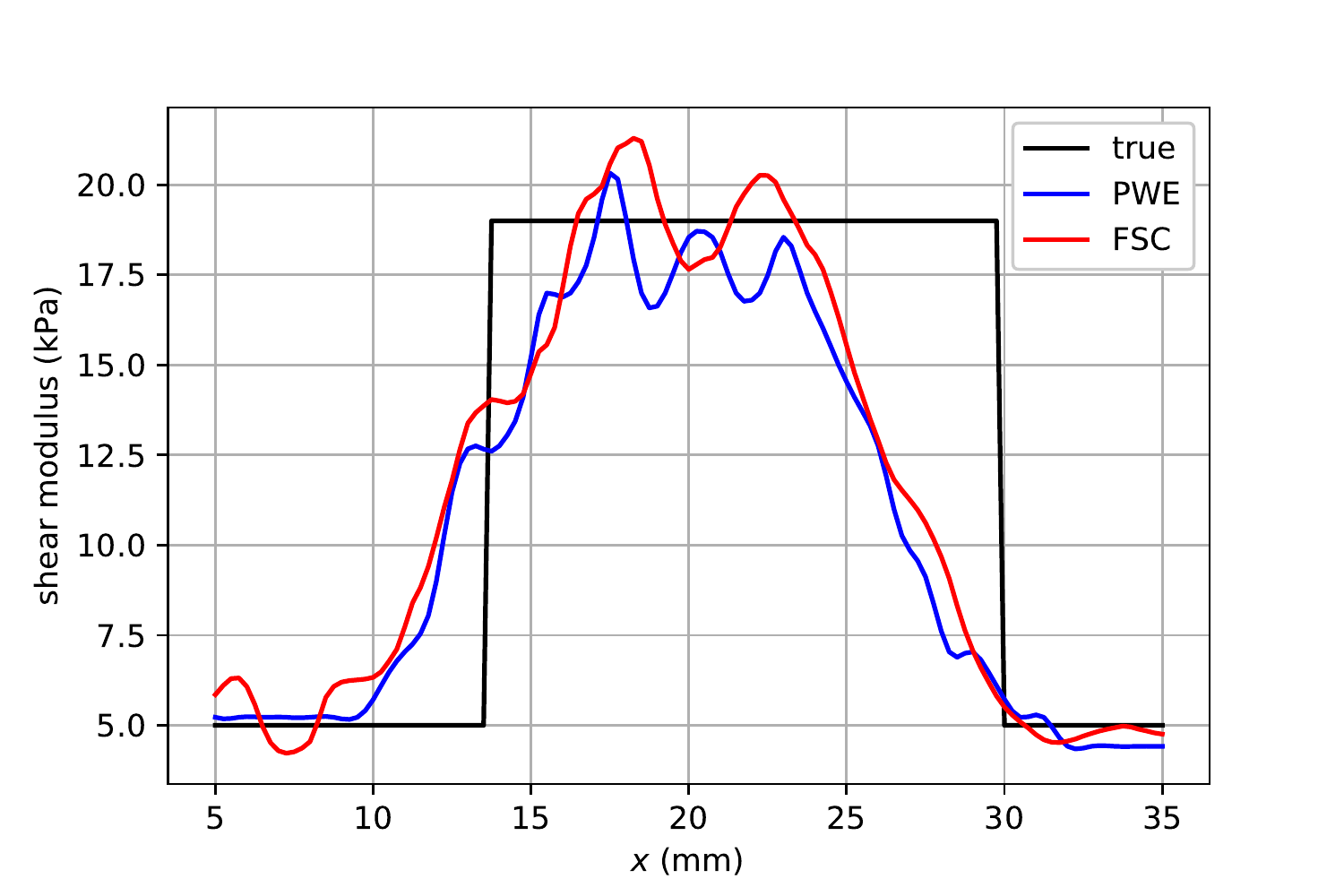} 	
\end{tabular}
\caption{PWE and FSC reconstructions for the digital phantom. The first row corresponds to noiseless data while the second row corresponds to data contaminated with additive Gaussian noise. Column (a) includes the reconstructions obtained using the FSC method whereas column (b) includes the PWE reconstructions. Column (c) depicts the MSE fields \eqref{eq:MSE} for the PWE reconstructions. In column (d), we compare the ground-truth shear modulus with the estimated values along the horizontal line passing through the center of the ROI, depicted by the white dashed lines in the shear modulus contour plots.} \label{fig:Malignant}
\end{figure*}

In this section, we study the performance of PWE for a $40$\,mm$\times40$\,mm digital phantom with an inclusion mimicking a malignant tumor, see e.g. \cite[Fig. 1]{CPSMMLSB2016LWXL}, with a background shear modulus of $5$\,kPa and inclusion shear modulus of $19$\,kPa; see Fig. \ref{fig:Malignant} for the shape of the inclusion.
To simulate the shear wave propagation, we discretize the domain with a spatial interval of $\Delta x = 250$\,\textmu m and temporal step-size of $\Delta t = 100$\,\textmu s and solve the 2D incompressible elastodynamic PDE \cite{AECETE2019AB} in \textsc{FEniCS} \cite{alnaes2015fenics} using a mixed finite element method, subject to an unfocused ARF impulse modeled by a sinusoidal traction with frequency of $1000$\,Hz along the right side of the domain. The duration of the impulse was $10 \, \Delta t$ and the duration of the simulation was $20$\,ms. In the following, we study the performance of PWE without and with noise over a $30$\,mm$\times30$\,mm ROI.

The plots in the first row of Fig. \ref{fig:Malignant} show the reconstructions for noiseless data.
For PWE results, we set the minimum frequency to $f_{\min} = 500$\,Hz after inspecting the Fourier transfer spectrum, and 
 the window size to $w = 1.50$\,mm, accordingly. Because the data is noiseless, a wide range of values are appropriate for the regularization parameter; we set $\tau = 10^{-4}$ from the L-curve analysis.
From the PWE reconstruction, the average background and inclusion shear moduli are $\mu_b = 4.89 \pm 1.27$\,kPa and $\mu_i = 14.76 \pm 3.36$\,kPa, respectively. The contrast-to-noise ratio is CNR $= 8.78$\,dB.
Referring to the MSE field in column (c) of Fig. \ref{fig:Malignant}, observe that the regions at the boundary of inclusion and background have the highest MSE values \eqref{eq:MSE} since at those regions, the assumption of homogeneity breaks down (the windows cover parts of both the inclusion and background). Moreover, generally the MSE values are smaller inside the inclusion indicating better agreement with the wave equation \eqref{eq:scalarWE}.
For the FSC method \cite{FSCSWSC2014SMZU}, we get $\mu_b = 5.15 \pm 1.25$\,kPa and $\mu_i = 15.12 \pm 3.28$\,kPa and CNR $= 9.08$\,dB, where we use a window size of $1.00$\,mm and patch size of $0.75$\,mm.
Referring to cross-section plots in column (d) of Fig. \ref{fig:Malignant}, observe that the inclusion edges are smoothed out due to windowing, resulting in slightly skewed estimations compared to the ground truth.

To investigate the effect of noise, we contaminated the simulated data with additive Gaussian noise decreasing the signal-to-noise ratio to SNR $=-9.05$\,dB.\footnote{Although Gaussian noise is not an accurate model of the ultrasound noise, it has been used in the literature regardless; see e.g., \cite{RPVDEVM2018KQC}. In the next sections, we study phantom experiments that inherently contain realistic ultrasound noise.}
The plots in the second row of Fig. \ref{fig:Malignant} show the reconstructions for this case. In the presence of noise, more measurements are required for the same minimum frequency $f_{\min} = 500$\,Hz, to ensure reasonable reconstructions. Using $w = 5.50$\,mm and $\tau = 10^{-2}$ from the L-curve analysis, we get $\mu_b = 5.09 \pm 1.20$\,kPa and $\mu_i = 13.35 \pm 3.71$\,kPa, and CNR $= 6.52$\,dB from the PWE method.
Referring to the MSE feedback in column (c), we can see that the MSE values are orders of magnitude higher in this case indicating considerable disagreement with the wave equation \eqref{eq:scalarWE}. Because the noise is uniformly added across all spatial locations, the MSE values are very close throughout the ROI.
Finally using the FSC method, we get $\mu_b = 5.42 \pm 1.52$\,kPa and $\mu_i = 14.87 \pm 3.97$\,kPa and CNR $= 6.94$\,dB, where we use a window size of 2.00\,mm and patch size of 1.75\,mm. Note that prior filtering to improve the SNR is essential to obtaining reasonable reconstructions using the FSC method in the presence of noise.

% --------------------------------------------------------- %
\subsection{Multi-push CIRS Phantom} \label{sec:multi-push}
\begin{figure}[t!]
  \centering
    \includegraphics[width=0.5\textwidth]{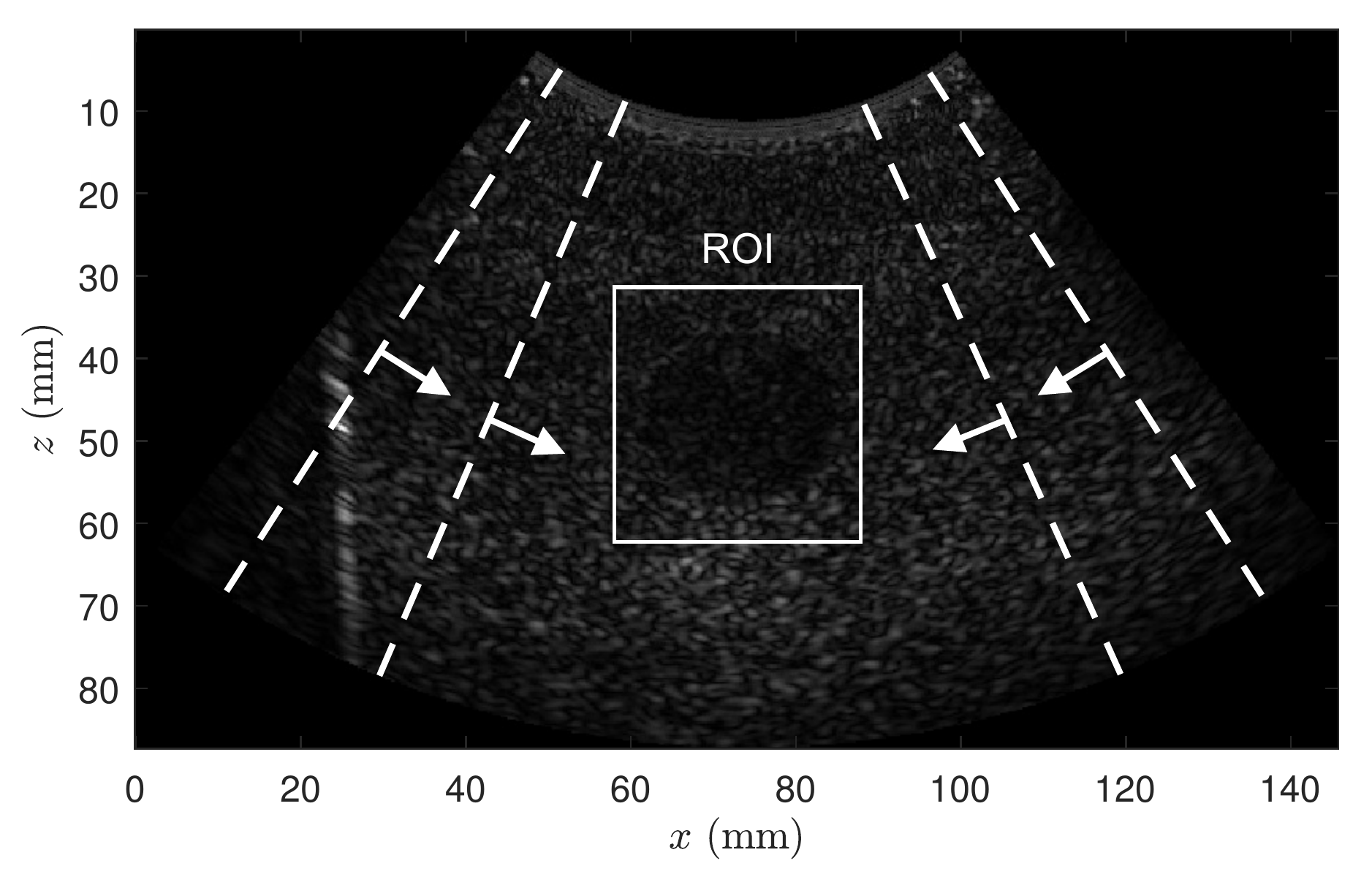}
  \caption{The geometry of the Type IV CIRS phantom and the location of the $20$\,mm spherical inclusion for multi-push data. The dashed lines delineate the approximate orientation of the four push beams while the arrows show the directions of propagation and the approximate focal depths ($40$\,mm) along the push beam axes. Finally, the square box depicts the $30$\,mm$\times30$\,mm ROI.} \label{fig:multiPush-Bmode}
\end{figure}

Next, we consider a phantom with nominal background shear modulus of $8.33$\,kPa and a single spherical inclusion with shear modulus of $26.66$\,kPa and diameter of $20$\,mm (Model 049, CIRS, Inc., Norfolk, VA), excited by four simultaneous ARF push beams transmitted by a C5-2 curved-array transducer (Verasonics, Inc., Kirkland, WA) and measured with a research scanner (V1, Verasonics, Inc., Kirkland, WA). Each push beam used $16$ elements, and the beams were moved to the edge of the transducer such that two beams were formed on the extreme left and right sides of the array and $64$ elements in the center were inactive.
We study two ARF push configurations with push frequency of $4.09$\,MHz: (i) one push with duration of $400$\,\textmu s; (ii) four repeated pushes of duration $200$\,\textmu s separated by $800$\,\textmu s of waiting, generating a repeated push of $1000$\,Hz. This repeated push is intended to excite a wider frequency range \cite{HPEMD2008UG}.
A movie of the vertical shear wave velocity component generated under these two push configurations can be found in \cite{meJ6_video1}.
Fig. \ref{fig:multiPush-Bmode} depicts the geometry of the phantom and the $30$\,mm$\times30$\,mm ROI along with the approximate location of the inclusion and the focused push beams (focal depth is $40$\,mm).
Note that the prior knowledge of the push beam angles and equivalently the propagation directions, is not required for PWE.
The spatial spacing of the shear wave data was $240$\,\textmu m while the temporal interval was $360$\,\textmu s. The duration of the signal was $20$\,ms.

\begin{figure*}
\centering
\setlength{\tabcolsep}{2pt}
\begin{tabular}{ccccc}
					&	(a) FSC		&	(b) PWE				&	(c) MSE		&	(d) cross-section	\\
\rotatebox{90}{\hspace{1mm}push config. (i)}				&
\includegraphics[height=25mm]{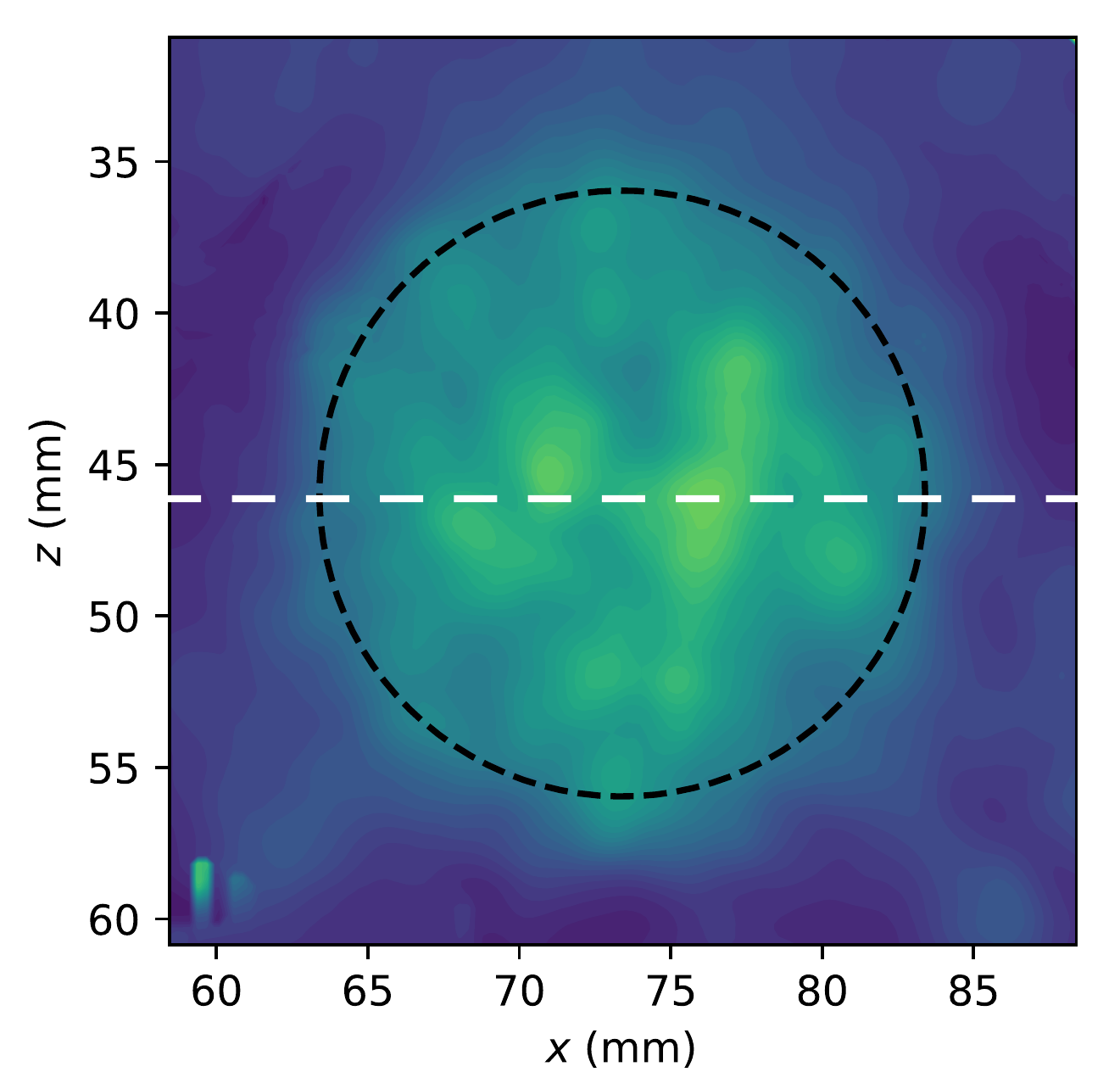}			& 
\includegraphics[height=25mm]{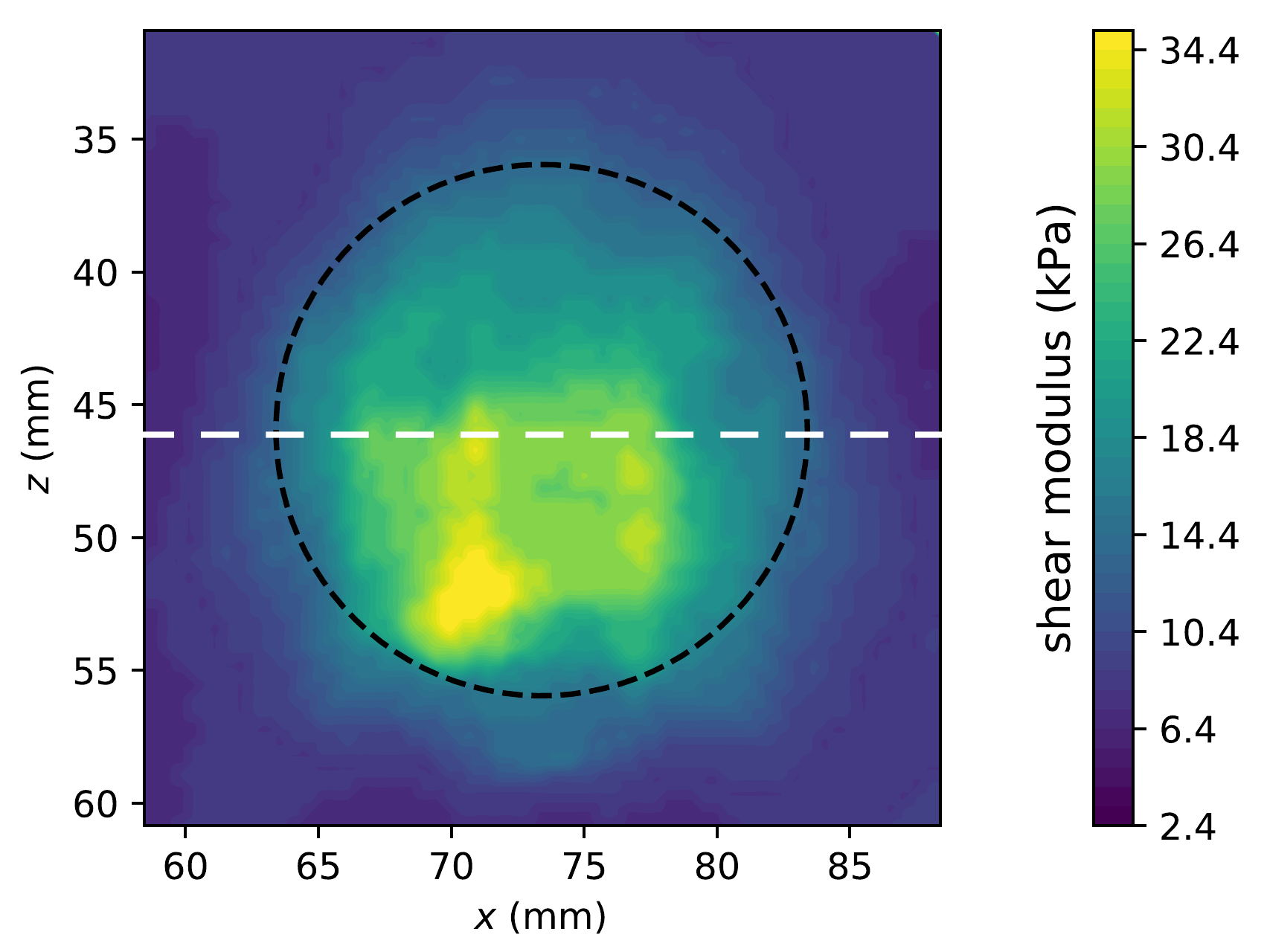}			& 
\includegraphics[height=25mm]{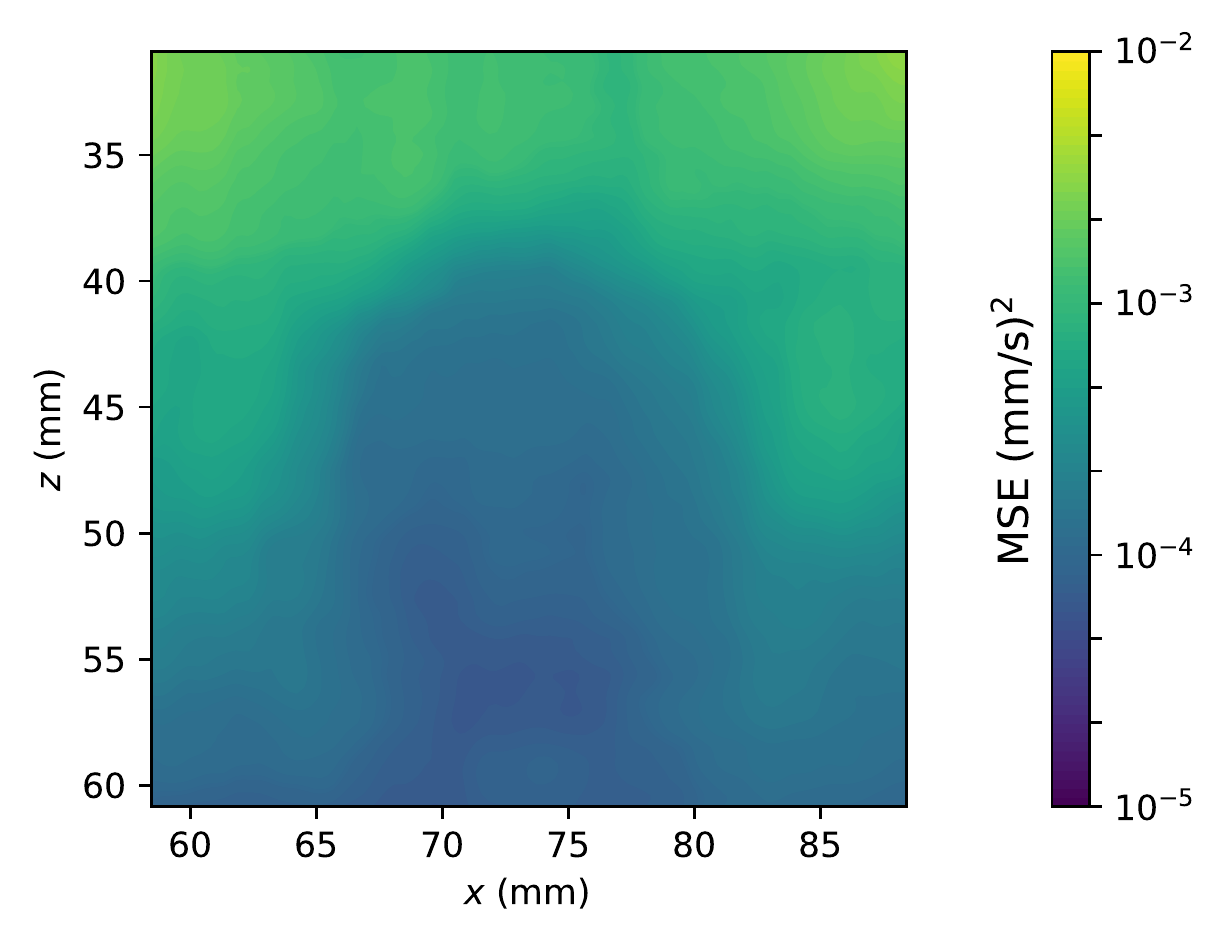}			& 
\includegraphics[height=25mm]{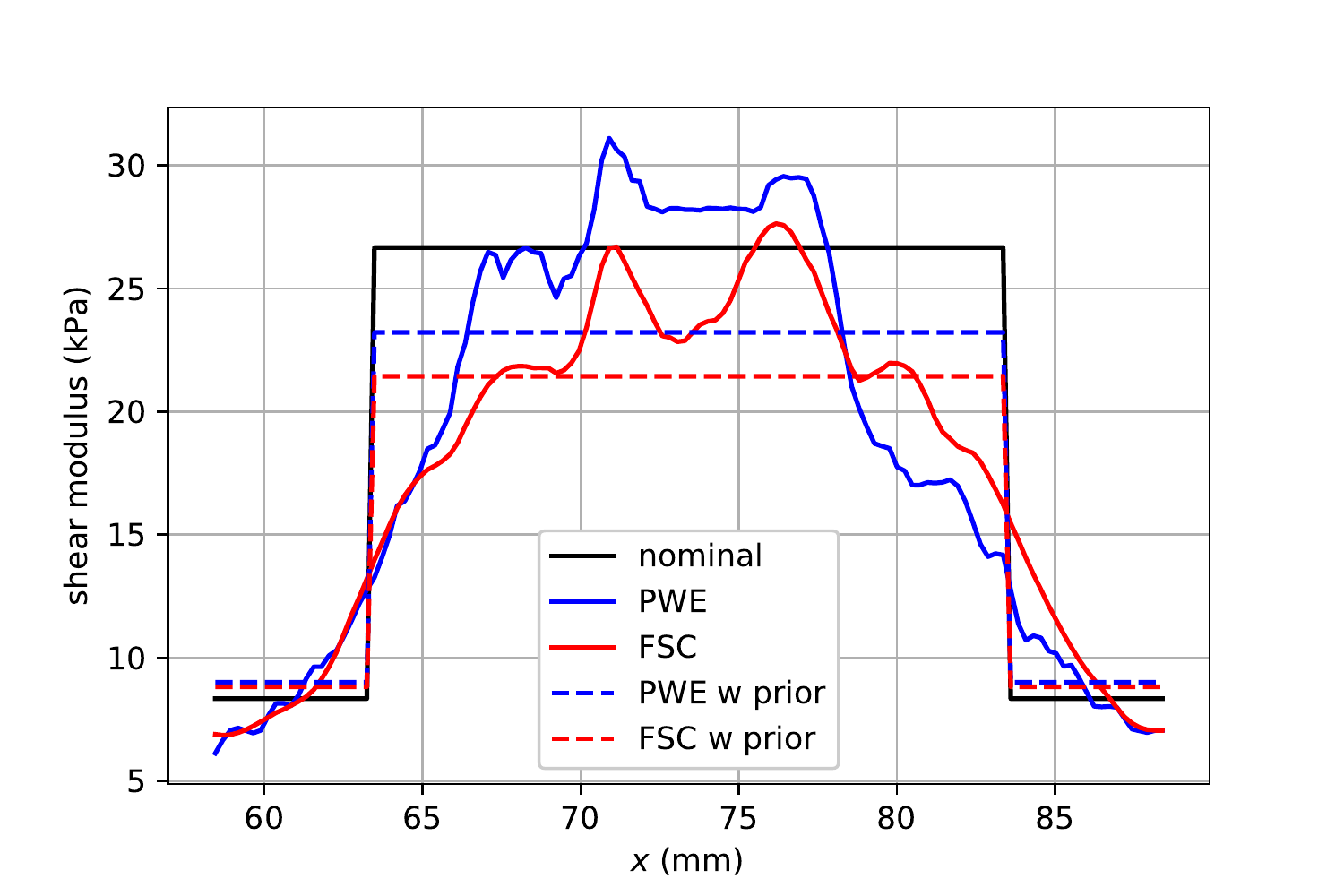} 	\\
\rotatebox{90}{\hspace{0.5mm}push config. (ii)}						&
\includegraphics[height=25mm]{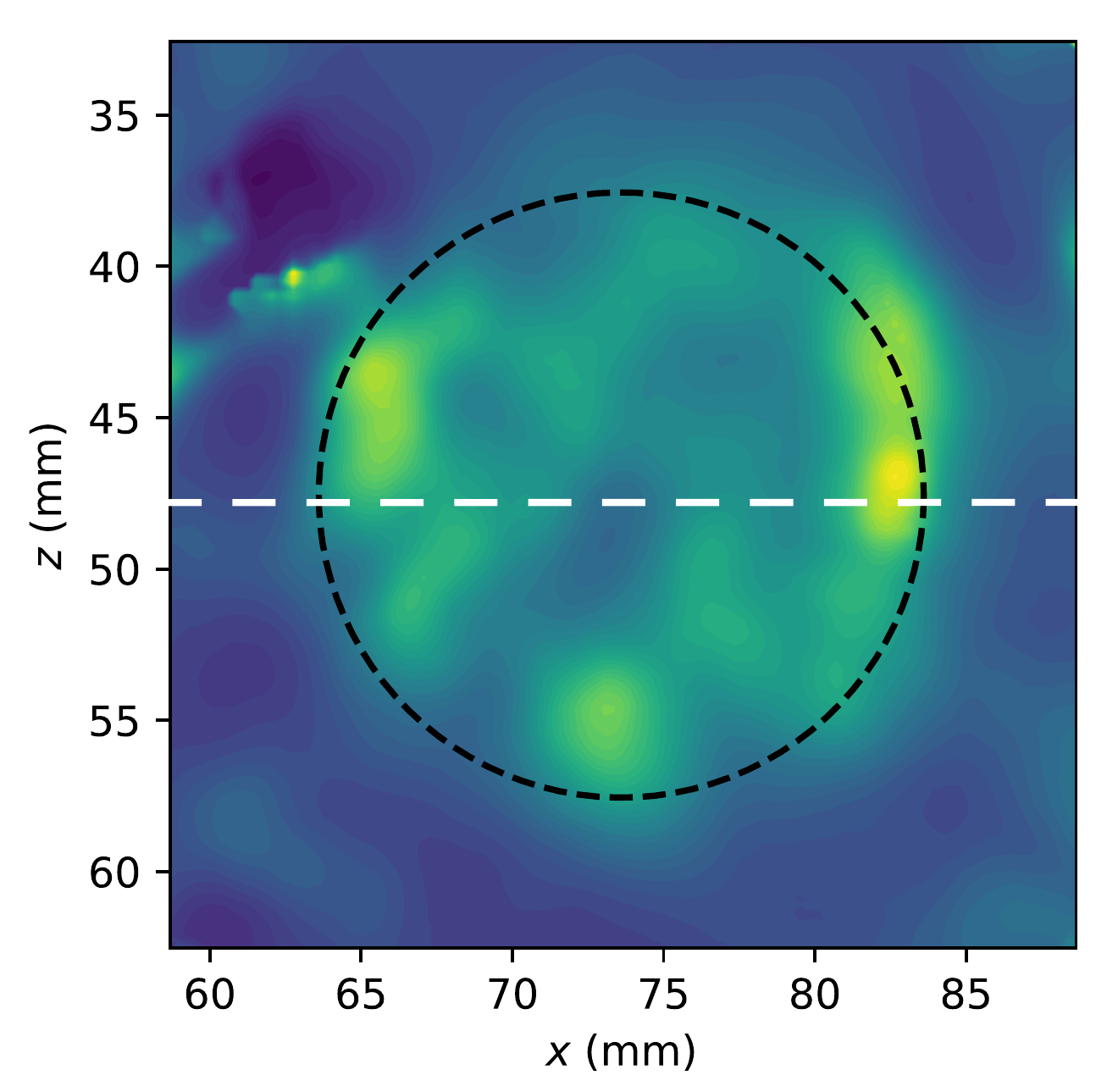}			& 
\includegraphics[height=25mm]{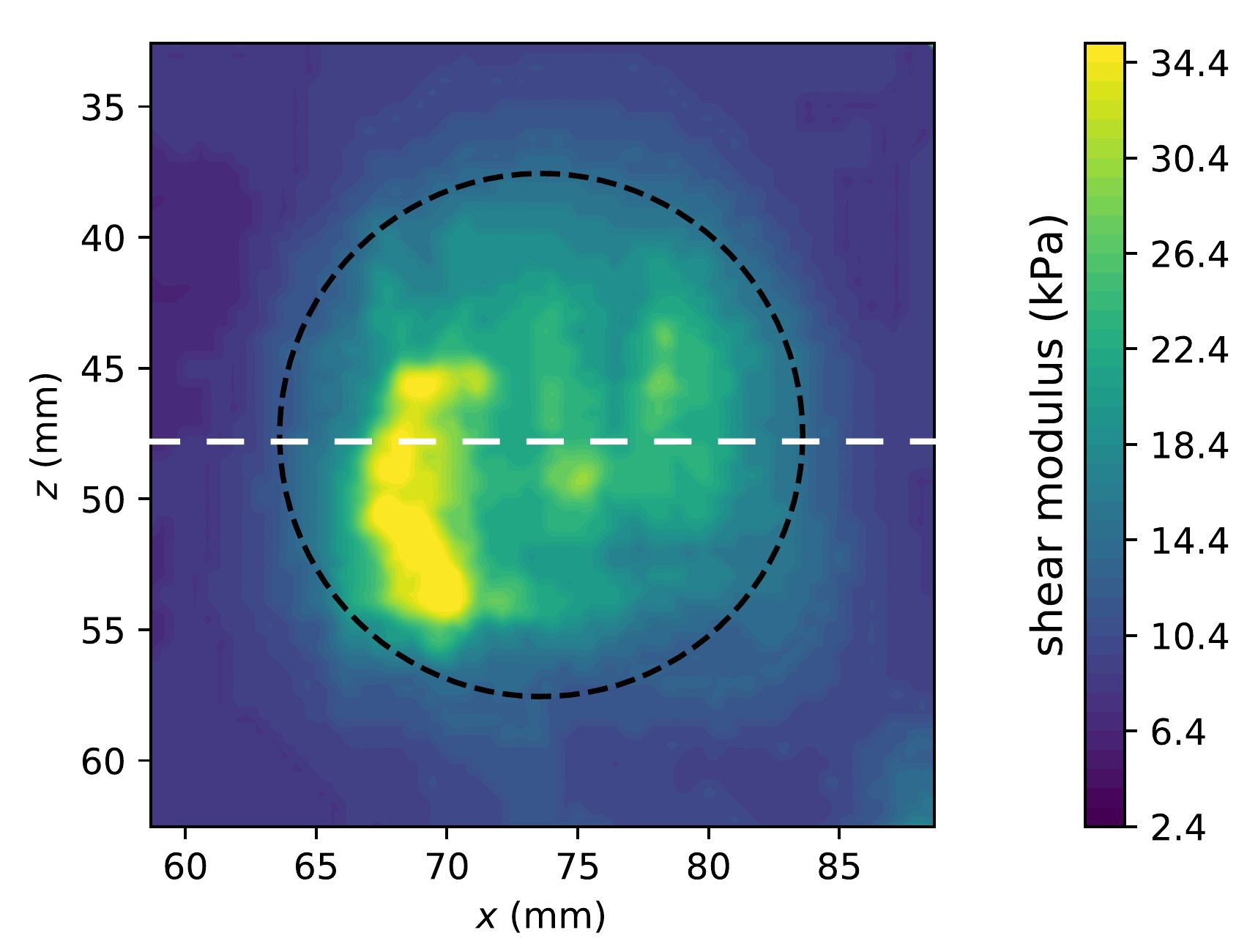}			& 
\includegraphics[height=25mm]{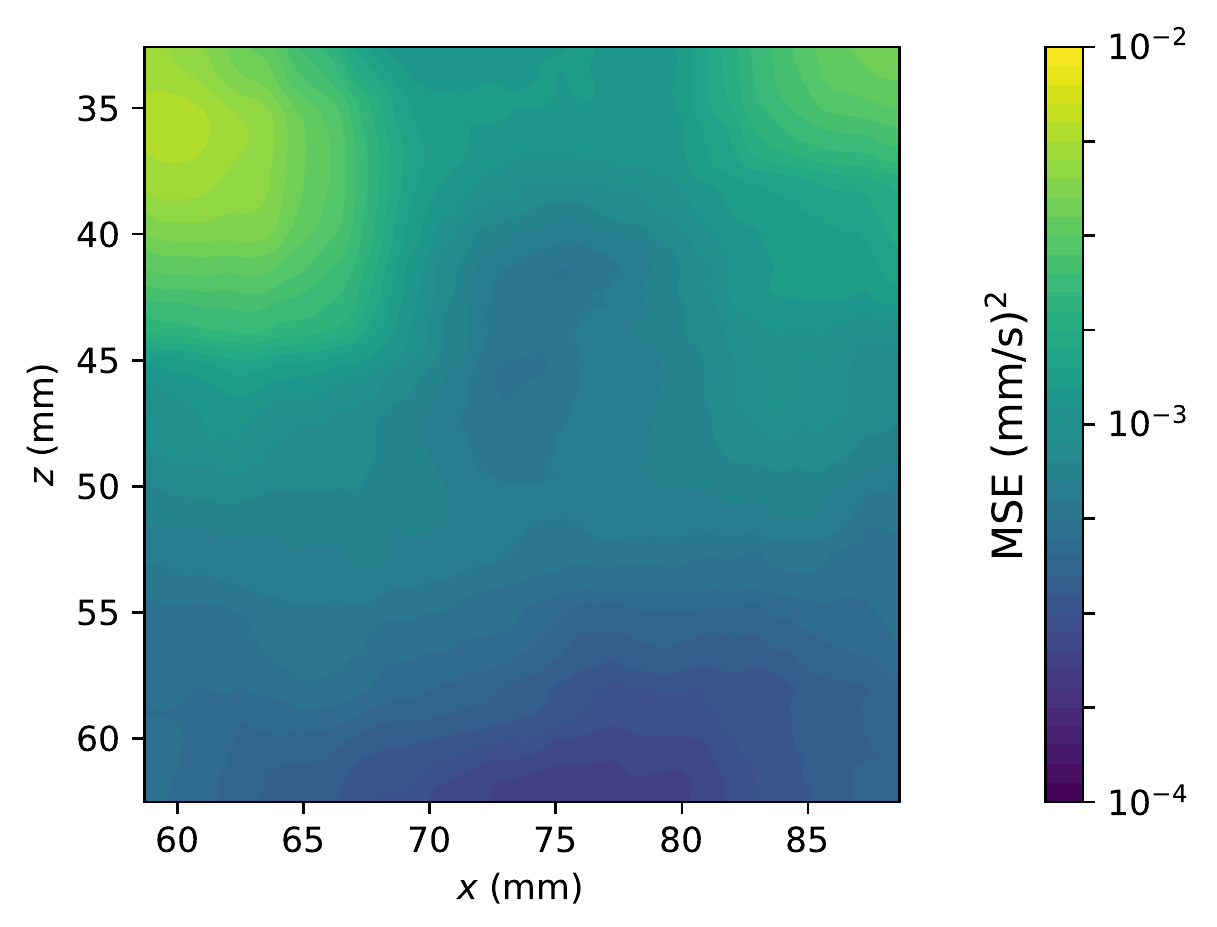}			& 
\includegraphics[height=25mm]{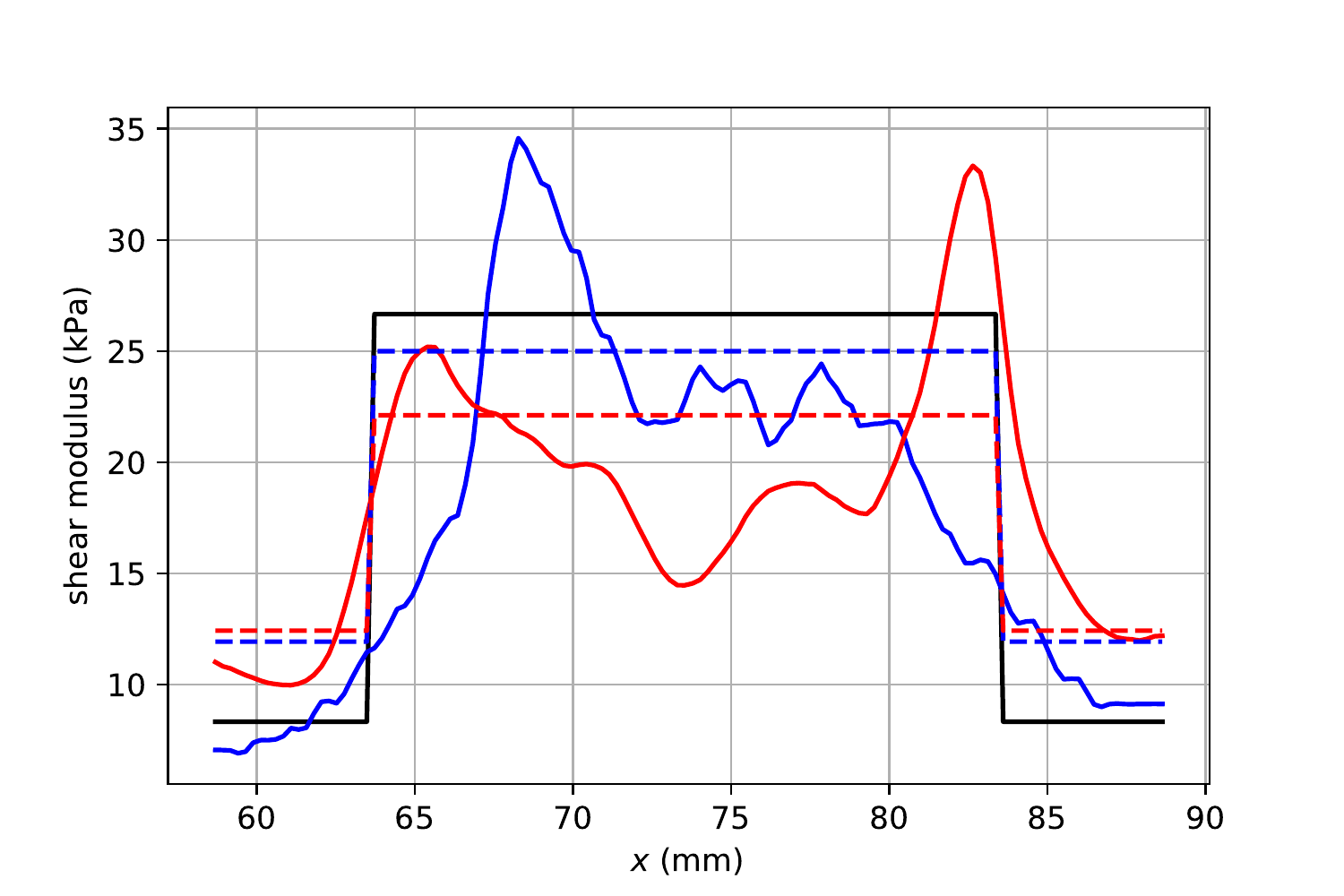} 	\\
\end{tabular}
\caption{PWE and FSC reconstructions of the shear modulus field for the multi-push data where the results in each row correspond to one ARF configuration. Column (a) includes the FSC reconstructions whereas column (b) includes the PWE reconstructions. Column (c) plots the MSE feedback \eqref{eq:MSE} corresponding to the PWE reconstructions. In column (d) we compare the estimated shear modulus fields without and with the prior knowledge of inclusion geometry, to the nominal values along the horizontal line passing through the center of the ROI, depicted by the white dashed lines in the shear modulus contour plots.} \label{fig:multiPushResults}
\end{figure*}

\begin{figure}[t!]
  \centering
    \includegraphics[width=0.5\textwidth]{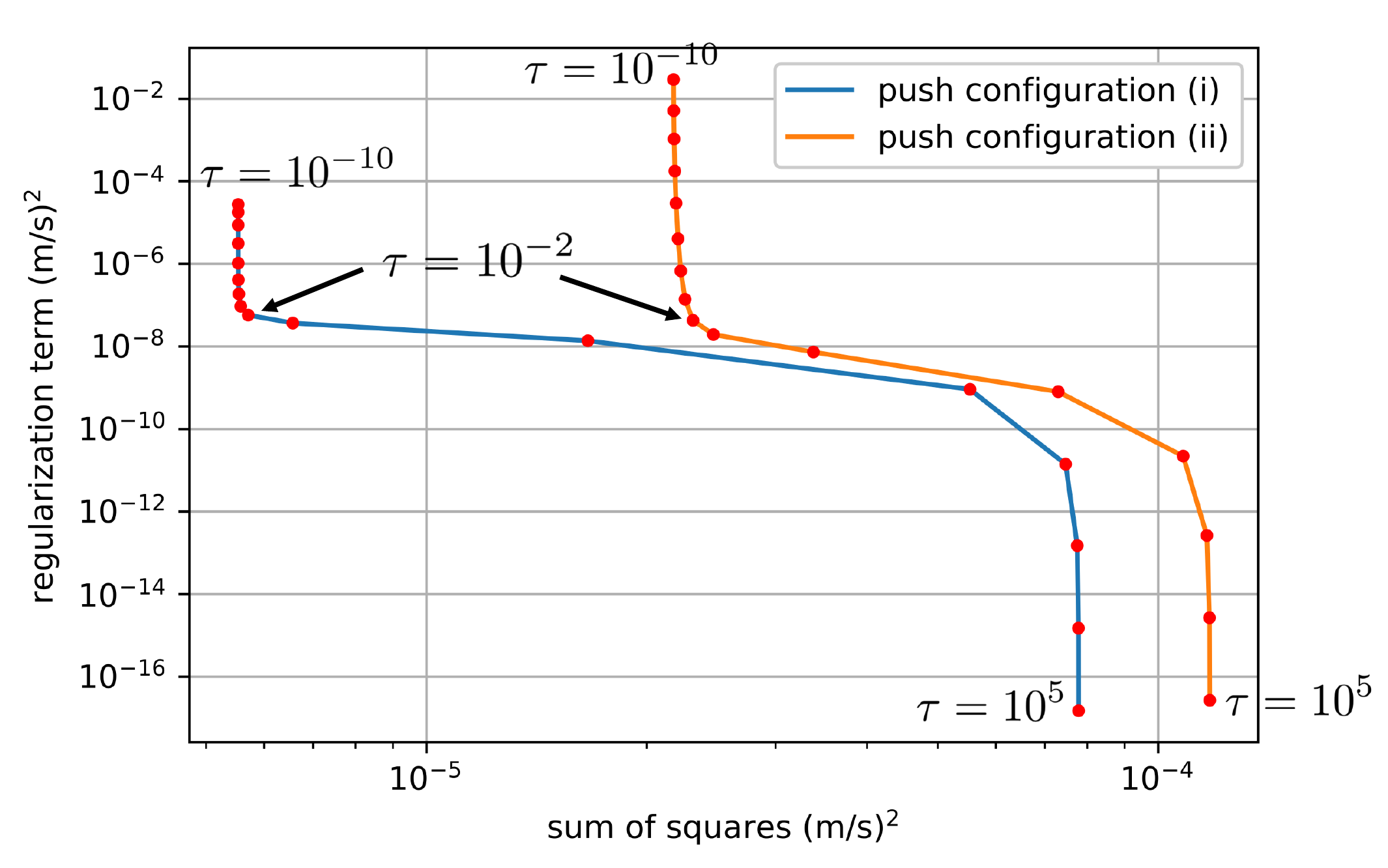}
  \caption{L-curves for the multi-push data with the two ARF configurations. The ideal value of the regularization parameter is the point of maximum inflection corresponding to $\tau = 10^{-2}$ for both cases; see Section \ref{sec:paramSelec} for more details.} \label{fig:multiPush_tau}
\end{figure}

The plots in the first row of Fig. \ref{fig:multiPushResults} depict the reconstructions for the push configuration (i).
For the PWE results in column (b), we use $f_{\min} = 300$\,Hz and $w = 7.67\, \text{mm} \approx 32 \times 240$\,\textmu m and set $n_b = 12$ as before. Note that $n_b = 12$ basis functions are sufficient to reconstruct the field even though the PWE Algorithm \ref{alg:PWE} is unaware of the propagation directions. 
We use a regularization parameter of $\tau = 10^{-2}$, obtained from the L-curve in Fig. \ref{fig:multiPush_tau}; see Section \ref{sec:paramSelec} for details.
The average shear moduli are $\mu_b = 9.33 \pm 2.16$\,kPa and $\mu_i = 21.69 \pm 5.79$\,kPa and the contrast-to-noise ratio \eqref{eq:CNR} is CNR $= 6.02$\,dB.
%
%The contour plot in column (c) shows the reconstructed shear modulus field given the prior knowledge of inclusion geometry from the B-mode image in Fig. \ref{fig:multiPush-Bmode}; see Section \ref{sec:prior} for details.
Given the prior knowledge of inclusion geometry from the B-mode image in Fig. \ref{fig:multiPush-Bmode}, we get more accurate estimates $\mu_b = 9.00$\,kPa and $\mu_i = 23.22$\,kPa, where we use settings similar to the previous reconstruction and set $\tau = 10^{-2}$ from the L-curve analysis; see Section \ref{sec:prior} for details.

The contour plot in column (a) of Fig. \ref{fig:multiPushResults} shows the corresponding FSC reconstruction for push configuration (i). To obtain the FSC estimate, using directional filters we decompose the shear wave into four directional waves traveling along angles $-32.1^o, -26.6^o, 206.6^o, \and 212.1^o$ and rely on a radial filter to enhance the SNR. We use window and patch sizes of $3.83$\,mm and $3.59$\,mm, respectively. After constructing the shear modulus estimates for the individual waves, we combine them using compounding; see \cite{FSCSWSC2014SMZU} for details.
The average shear moduli in this case are $\mu_b = 10.01 \pm 2.50$\,kPa and $\mu_i = 19.70 \pm 1.89$\,kPa and the contrast-to-noise ratio \eqref{eq:CNR} is CNR $= 8.08$\,dB.
We also performed a reconstruction given the prior knowledge of the geometry, where we average the values from windows that completely fall within the background or inclusion. Using a similar window size, we get the improved estimates $\mu_b = 8.82$\,kPa and $\mu_i = 21.43$\,kPa.
%
%The plot in column (d) of Fig. \ref{fig:multiPushResults} compares the PWE and FSC reconstructions with the nominal values along a horizontal line through the center of the ROI.

% paragraph theme: PWE - repeated push
%
The plots in the second row of Fig. \ref{fig:multiPushResults} depict the reconstructions for push configuration (ii). This ARF configuration results in a wider frequency range at the expense of lower SNR which requires larger window sizes for acceptable reconstructions. For PWE, we use $f_{\min} = 100$\,Hz and $w = 10.54$\,mm and $\tau = 10^{-2}$ from Fig. \ref{fig:multiPush_tau}, resulting in $\mu_b = 9.92 \pm 1.95$\,kPa and $\mu_i = 20.59 \pm 5.43$\,kPa and CNR $= 5.33$\,dB. Relying on the prior knowledge of inclusion geometry and with $\tau = 10^{-3}$ obtained from the L-curve analysis, we get $\mu_b = 11.93$\,kPa and $\mu_i = 25.00$\,kPa. It can be seen that using the prior knowledge improves the estimation inside the inclusion but deteriorates it for the background. This is due to the fact that parts of the background have higher noise levels and solving the SWE problem \eqref{eq:elastProb} only once, lumps all measurements into a single estimate. In practice it might be beneficial to decompose the background into multiple subdomains.

Plots in column (c) of Fig. \ref{fig:multiPushResults} show the MSE \eqref{eq:MSE} feedback from the PWE method for the two ARF configurations.
%Fig. \ref{fig:multiPush_mse} shows the MSE \eqref{eq:MSE} feedback from PWE for the two ARF configurations.
%
%\begin{figure}
%	\centering
%	\begin{subfigure}[b]{0.235\textwidth}
%		\includegraphics[width=\textwidth]{Multi-push/OnePush/PWE_mse.eps}
%	\caption{} \label{fig:multiPush_mse1}
%	\end{subfigure}
%	\begin{subfigure}[b]{0.235\textwidth}
%		\includegraphics[width=\textwidth]{Multi-push/RepeatedPush/PWE_mse.eps}
%	\caption{} \label{fig:multiPush_mse2}
%	\end{subfigure}
%\caption{The MSE feedback for the multi-push data given in Fig. \ref{fig:multiPush_mse1} and \ref{fig:multiPush_mse2} for push configurations (i) and (ii), respectively. Observe the higher MSE values for configuration (ii) indicating more discrepancy between the data and the wave equation \eqref{eq:scalarWE} due to noise.}	\label{fig:multiPush_mse}
%\end{figure}
%
Note the higher variations in background MSE values for both cases.
%, supporting the fact that the ultrasound noise is non-identically distributed; compare with column (c) of Fig. \ref{fig:Malignant} where noise is identically distributed resulting in symmetric contours in the background.
Note also that the values for configuration (ii) are an order of magnitude higher indicating further inconsistency with the physics of the wave propagation. This lower SNR is also evident from Fig. \ref{fig:multiPush_tau} where the L-curve for configuration (ii) is shifted to the right (higher sum of squares).

% paragraph theme: FSC - repeated push
%
Finally, column (a) in the second row of Fig. \ref{fig:multiPushResults} shows the FSC reconstruction where we perform similar filtering procedures to decompose the wave and improve the SNR. With window and patch sizes of $4.31$\,mm and $4.07$\,mm, we get $\mu_b = 12.39 \pm 3.48$\,kPa and $\mu_i = 19.93 \pm 3.62$\,kPa and CNR $= 3.54$\,dB.
Relying on the prior knowledge of the geometry and with window and patch sizes of $2.87$\,mm and $2.64$\,mm, we get $\mu_b = 12.43$\,kPa and $\mu_i = 22.11$\,kPa.
It seems that in this specific case, the performance of the FSC method degrades more with noise than the PWE method; similar behavior was observed for other data with similar push configuration.
Table \ref{table:multiPushResults} summarizes the estimated shear modulus values for the reconstructions of this section. Note that PWE reconstructions are generally more accurate.
\begin{table*}
\centering
\setlength{\tabcolsep}{4pt}
\renewcommand{\arraystretch}{1.3}
\caption{Average shear moduli $\mu_b \and \mu_i$ for the background and inclusion and the CNR \eqref{eq:CNR}, for the PWE and FSC methods, corresponding to the multi-push data and Fig. \ref{fig:multiPushResults}. The nominal values for the shear moduli of background and inclusion are $8.33$\,kPa and $26.66$\,kPa, respectively.} \label{table:multiPushResults}
\footnotesize
\begin{tabular}{|c||c|c|c||c|c|c|}
\hline
\multirow{2}{*}{method}	&	\multicolumn{3}{c||}{push configuration (i)}	&	\multicolumn{3}{c|}{push configuration (ii)}	\\ \cline{2-7}
	&	$\mu_b$ (kPa)	&	$\mu_i$ (kPa)	&	CNR (dB)		
	&	$\mu_b$ (kPa)	&	$\mu_i$ (kPa)	&	CNR (dB)	\\ \hline

PWE		&	$9.33\pm2.16$			&	$21.69\pm5.79$	&	$6.02$
		&	$9.92\pm1.95$			&	$20.59\pm5.43$	&	$5.33$		\\ \hline
FSC		&	$10.01\pm2.50$		&	$19.70\pm1.89$	&	$8.08$
		&	$12.39\pm3.48$		&	$19.93\pm3.62$	&	$3.54$	\\ \hline
PWE with prior	&	$9.00$			&	$23.22$			&	-
			&	$11.93$			&	$25.00$			&	-	\\ \hline
FSC	with prior	&	$8.82$			&	$21.43$			&	-
			&	$12.43$			&	$22.11$			&	-	\\ \hline
\end{tabular}
\end{table*}

% --------------------------------------------------------- %
\subsection{Parallel Double-push CIRS Phantoms} \label{sec:stateArt_comp}
\begin{figure}[t!]
  \centering
    \includegraphics[width=0.4\textwidth]{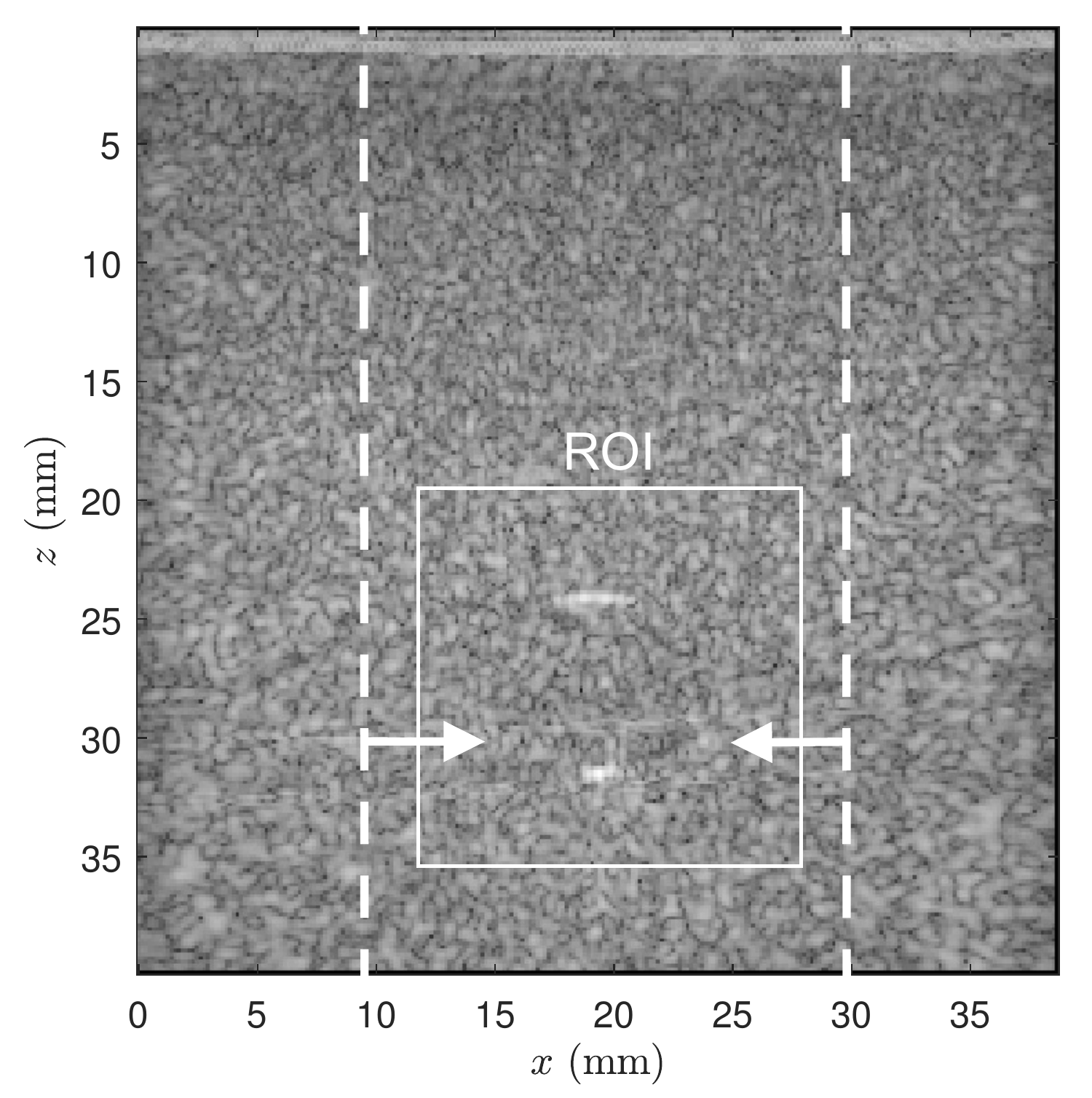}
  \caption{The geometry of the Type IV CIRS phantom and the location of the 6.49\,mm cylindrical inclusion for the parallel double-push case. The dashed lines delineate the approximate locations of the two push beams while the arrows show the directions of propagation and the approximate focal depths (30\,mm) along the push beam axes. The rectangular box depicts the 16\,mm$\times$16\,mm ROI.} \label{fig:Bmode-double-push}
\end{figure}

In this section, we consider phantoms with two parallel ARF pushes applied using a Verasonics V1 system with a L7-4 transducer (Philips Healthcare, Andover, MA) on the sides of the phantom at $30$\,mm focal depth.
Specifically, we consider a homogeneous phantom with nominal background shear modulus of $8.33$\,kPa (Model 039, CIRS, Inc., Norfolk, VA), a soft Type I cylindrical inclusion with diameter $10.40$\,mm and nominal shear modulus of $2.66$\,kPa, and three stiff Type IV cylindrical inclusions with diameters $10.40$\,mm, $6.49$\,mm, and $4.05$\,mm and nominal shear modulus of $26.66$\,kPa (Model 049A, CIRS, Inc., Norfolk, VA). The push duration was $400$\,\textmu s and the push frequency was $4.09$\,MHz. The push beams were generated by $32$ active elements located at the edges of L7-4 probe.
Fig. \ref{fig:Bmode-double-push} shows the B-mode image for the phantom with inclusion size of $6.49$\,mm along with the position of the push beams and the $16$\,mm$\times16$\,mm ROI.
The spatial spacing of the shear wave data was $154$\,\textmu m while the temporal intervals were $240$\,\textmu s and $80$\,\textmu s for the homogeneous phantom and with inclusions, respectively. The duration of the signal was $10$\,ms in all cases.

\begin{figure*}
\centering
\setlength{\tabcolsep}{2pt}
\begin{tabular}{ccccc}
					&	(a) FSC		&	(b) PWE			&	(c) MSE	& (d) cross-section	\\
\rotatebox{90}{\hspace{4mm}homogeneous}					&
\includegraphics[height=25mm]{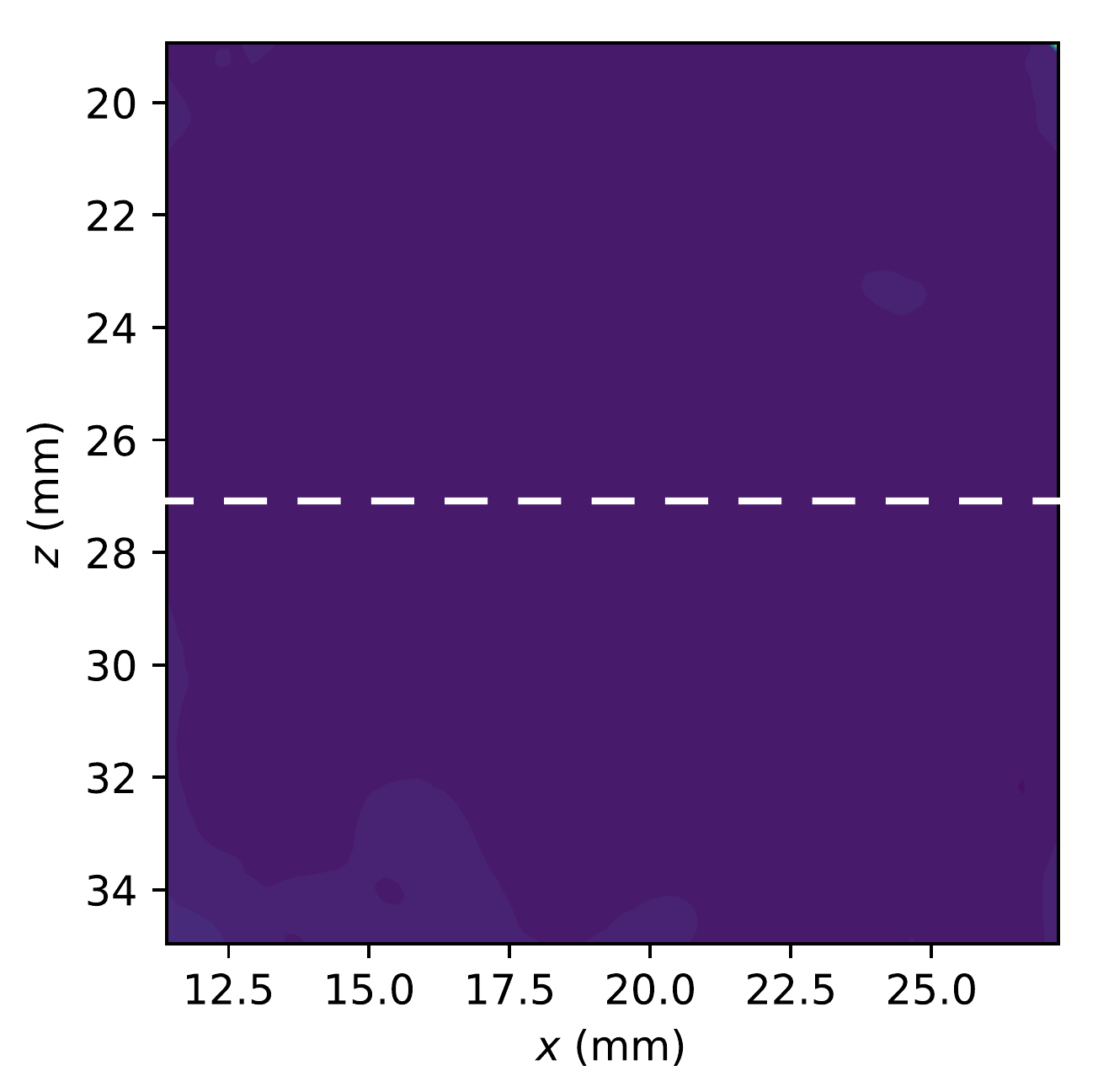}			& 
\includegraphics[height=25mm]{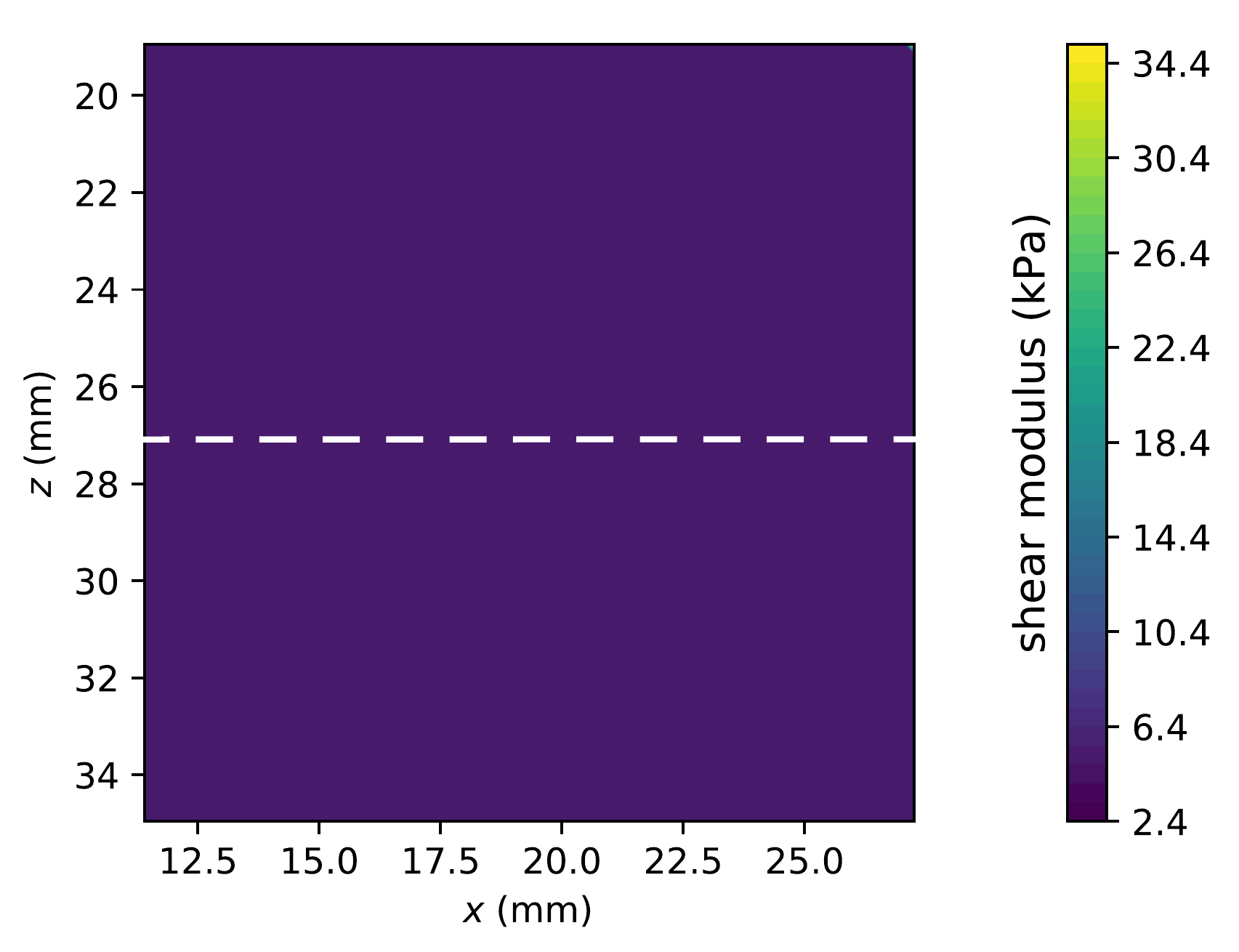}			&
\includegraphics[height=25mm]{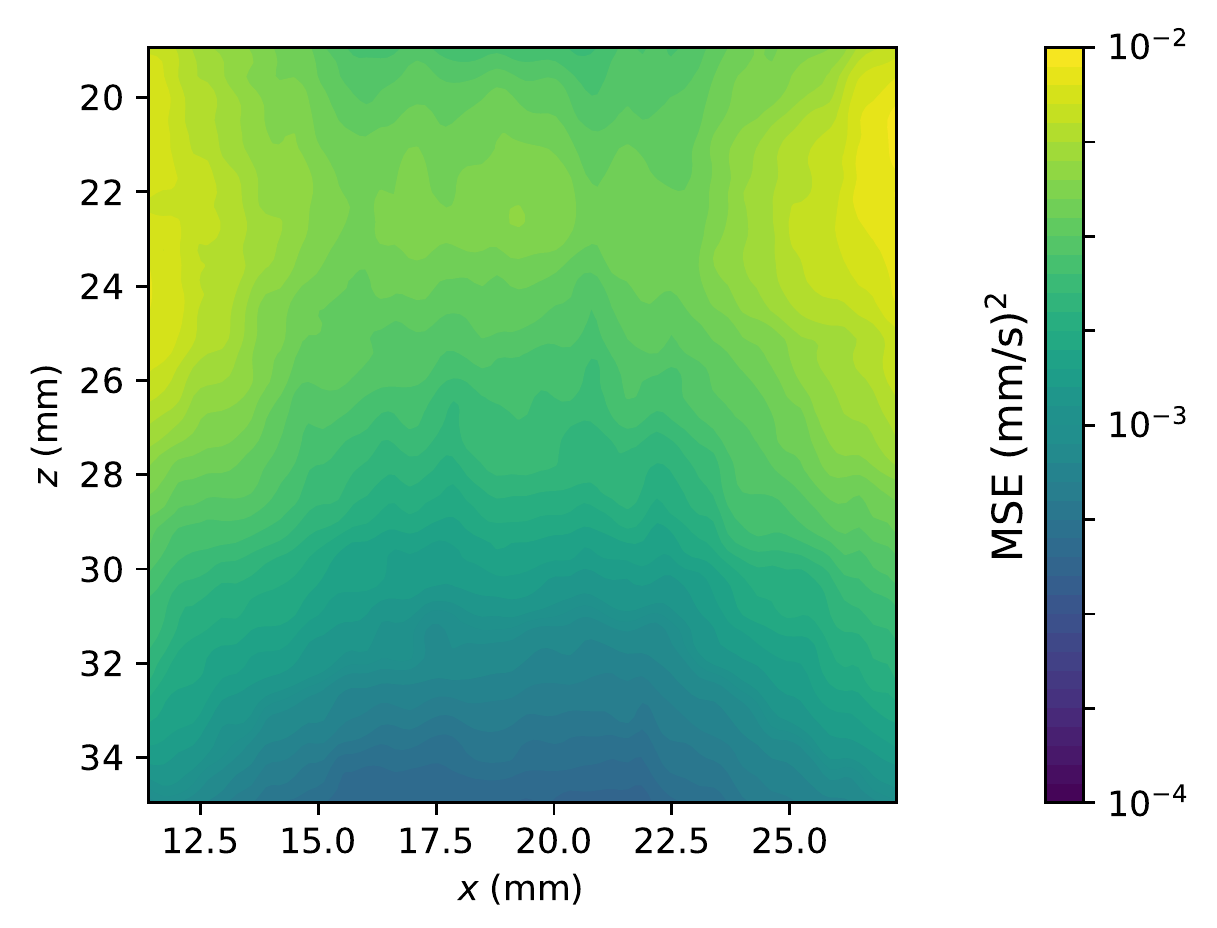}			&
\includegraphics[height=25mm]{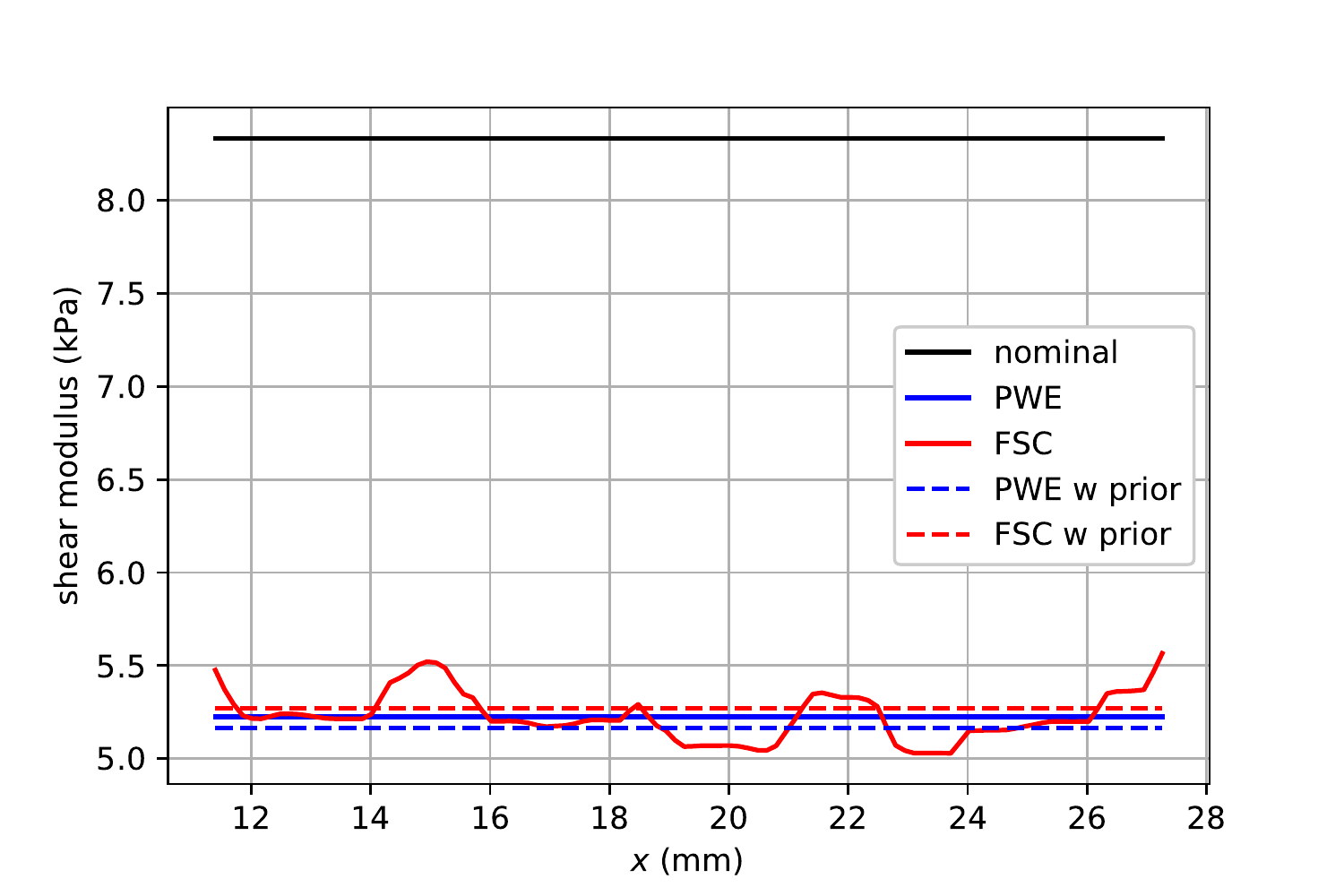} 	\\
\rotatebox{90}{\hspace{4mm}soft inclusion}						&
\includegraphics[height=25mm]{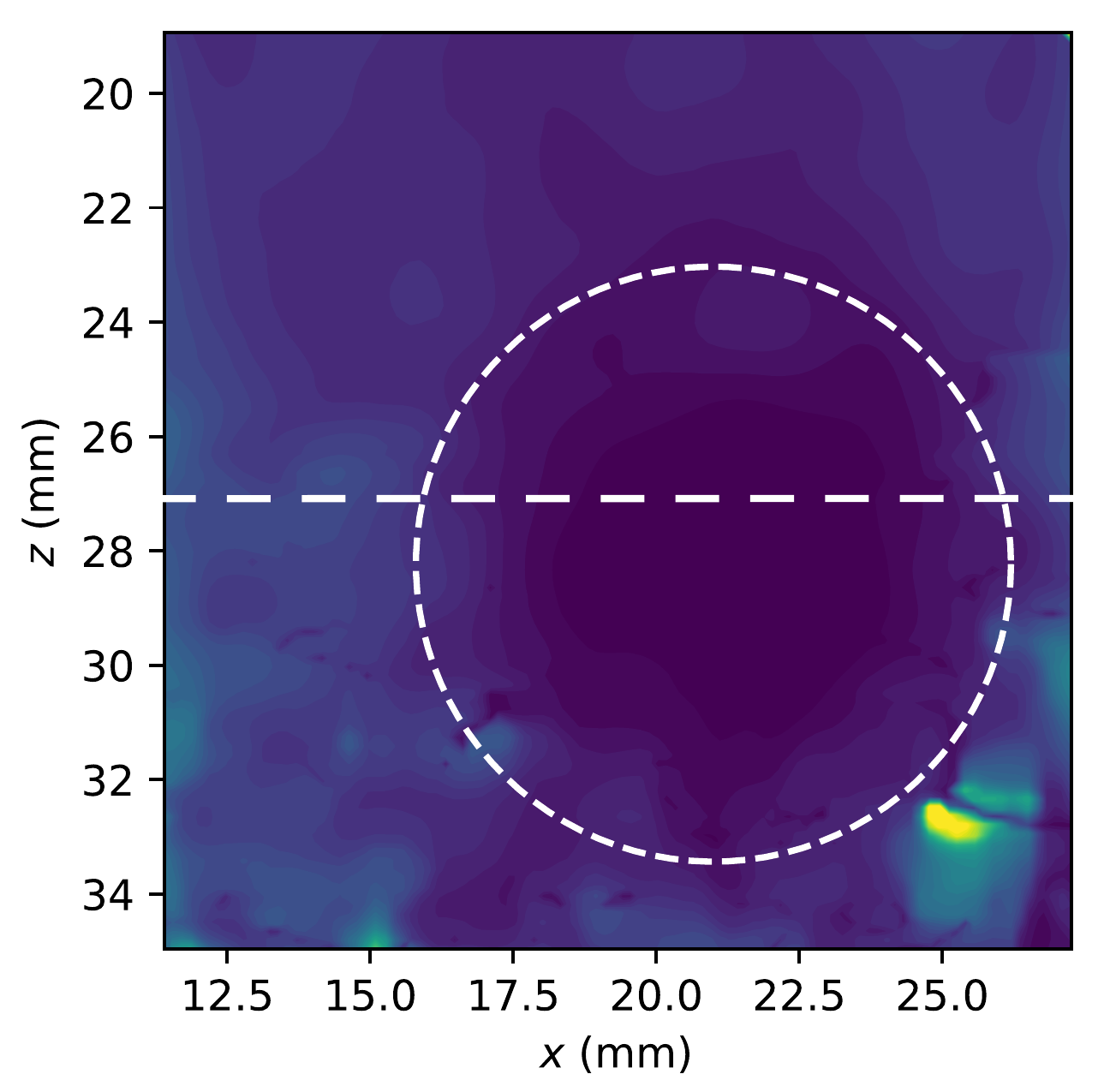}				& 
\includegraphics[height=25mm]{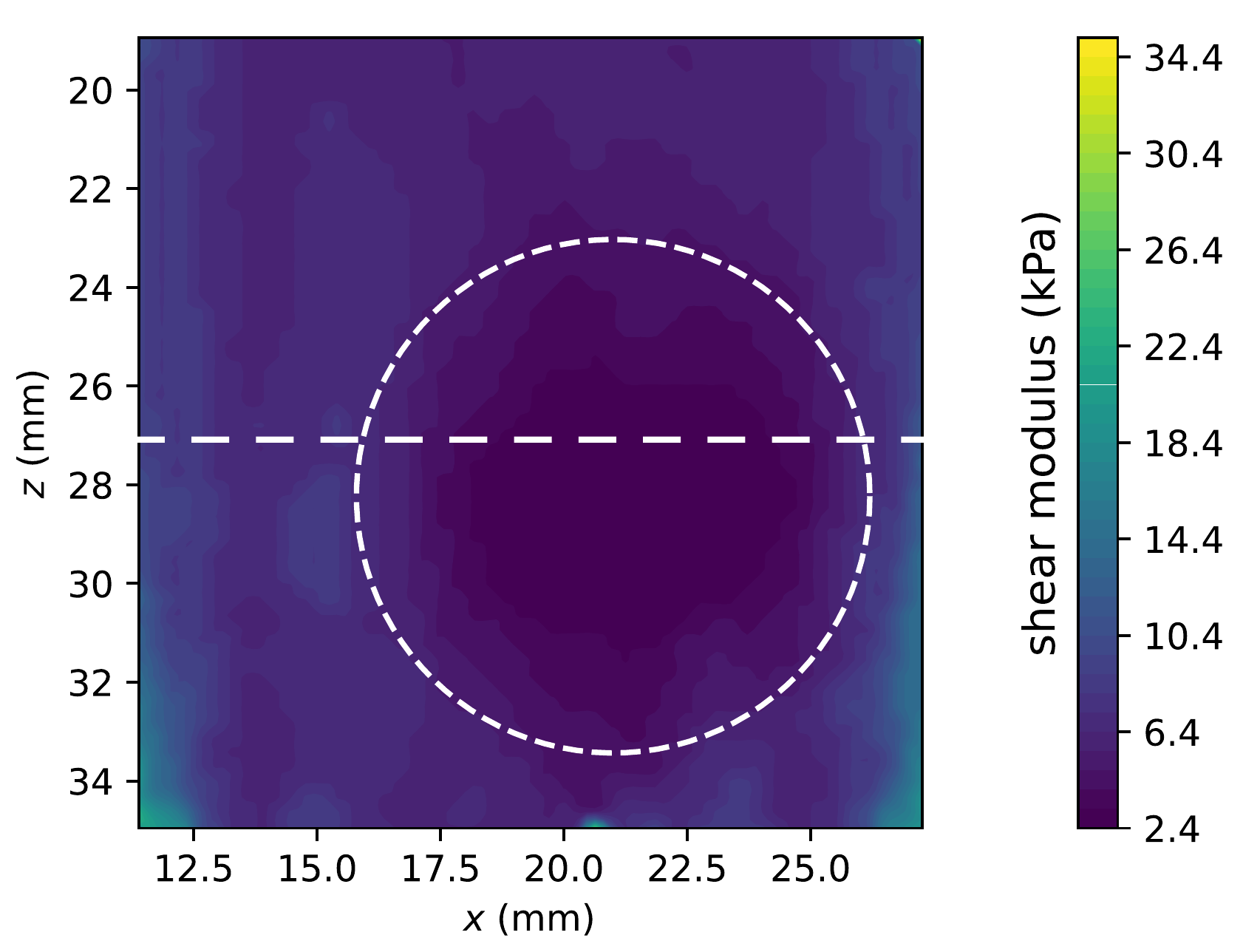}				&
\includegraphics[height=25mm]{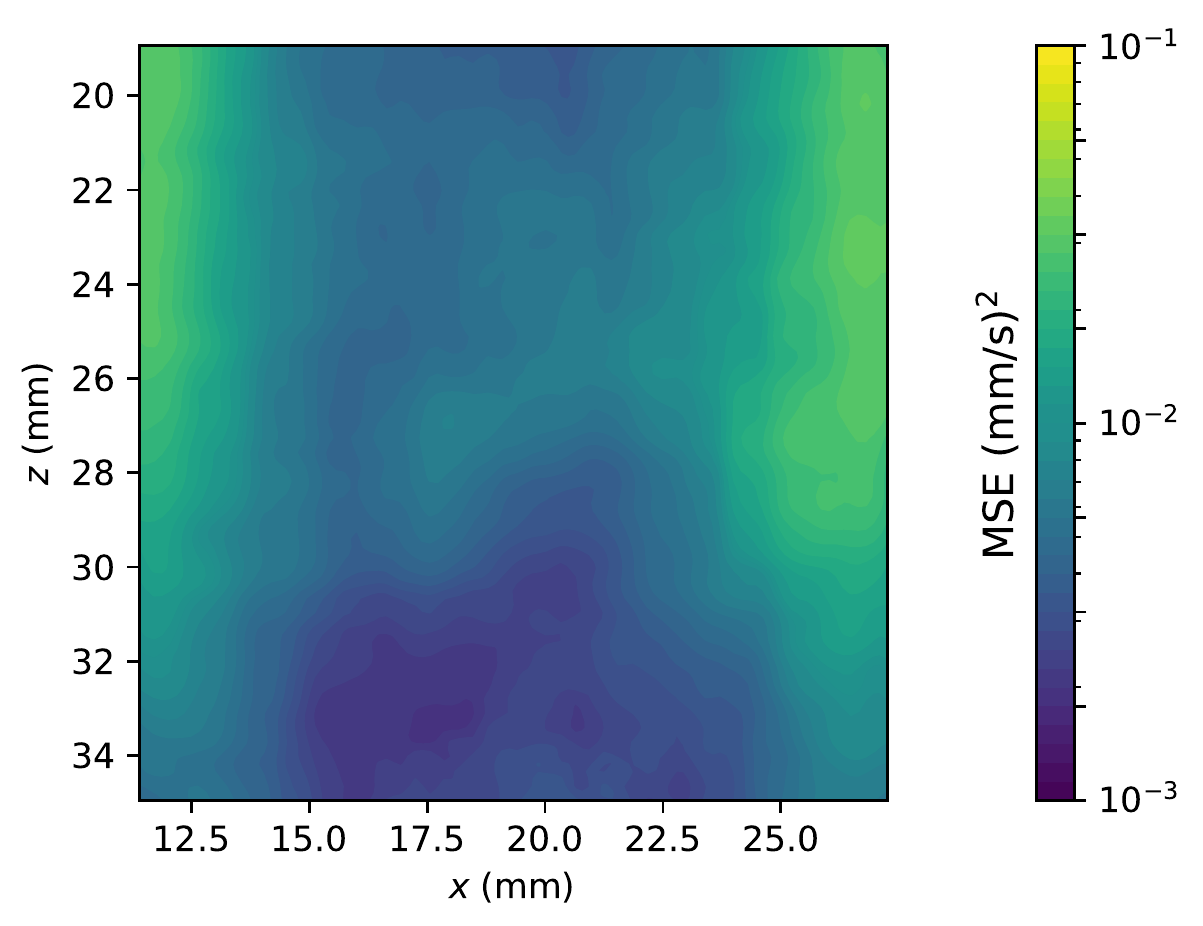}				&
\includegraphics[height=25mm]{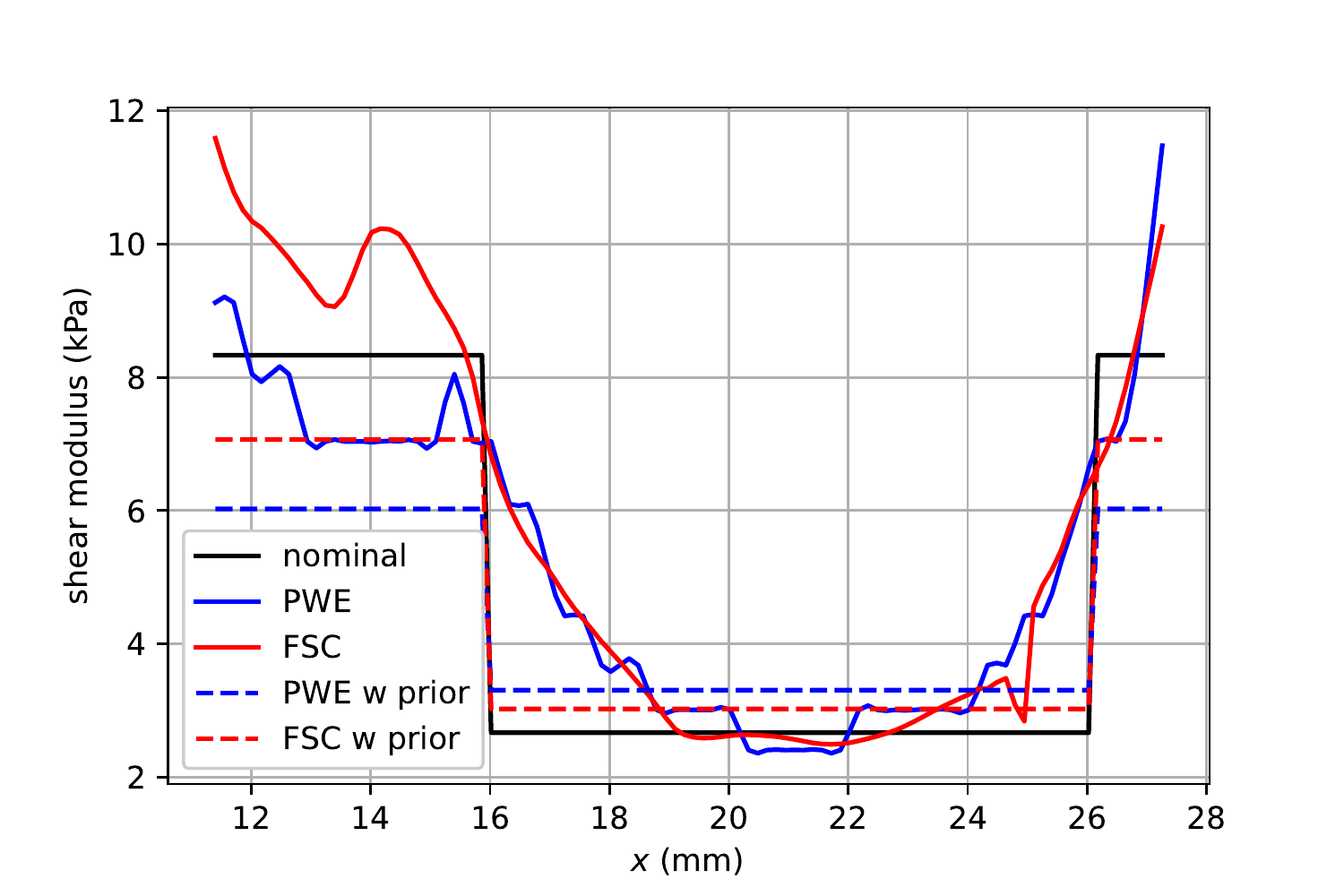} 			\\
\rotatebox{90}{\hspace{7mm}10.40\,mm}							&
\includegraphics[height=25mm]{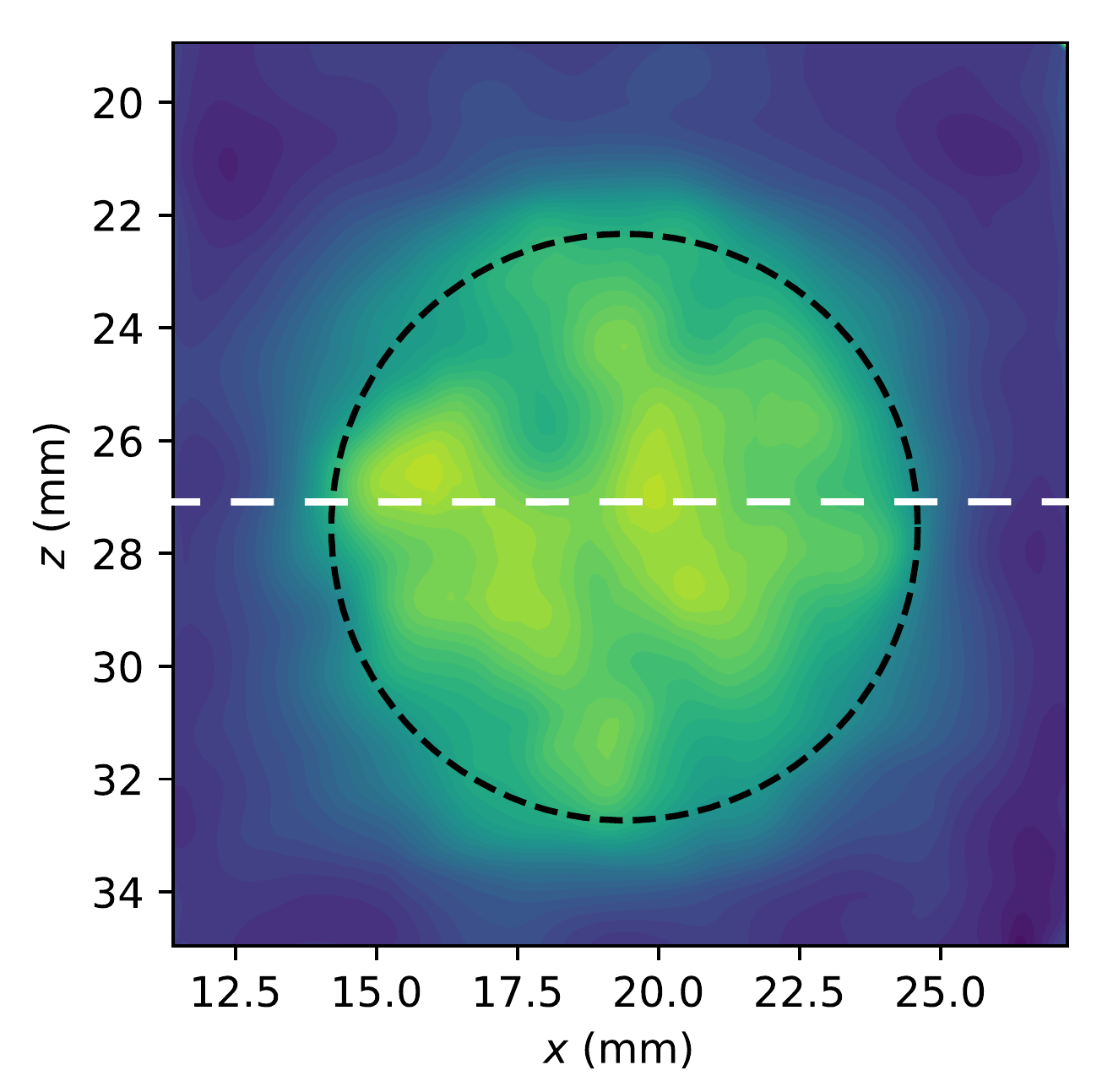}				& 
\includegraphics[height=25mm]{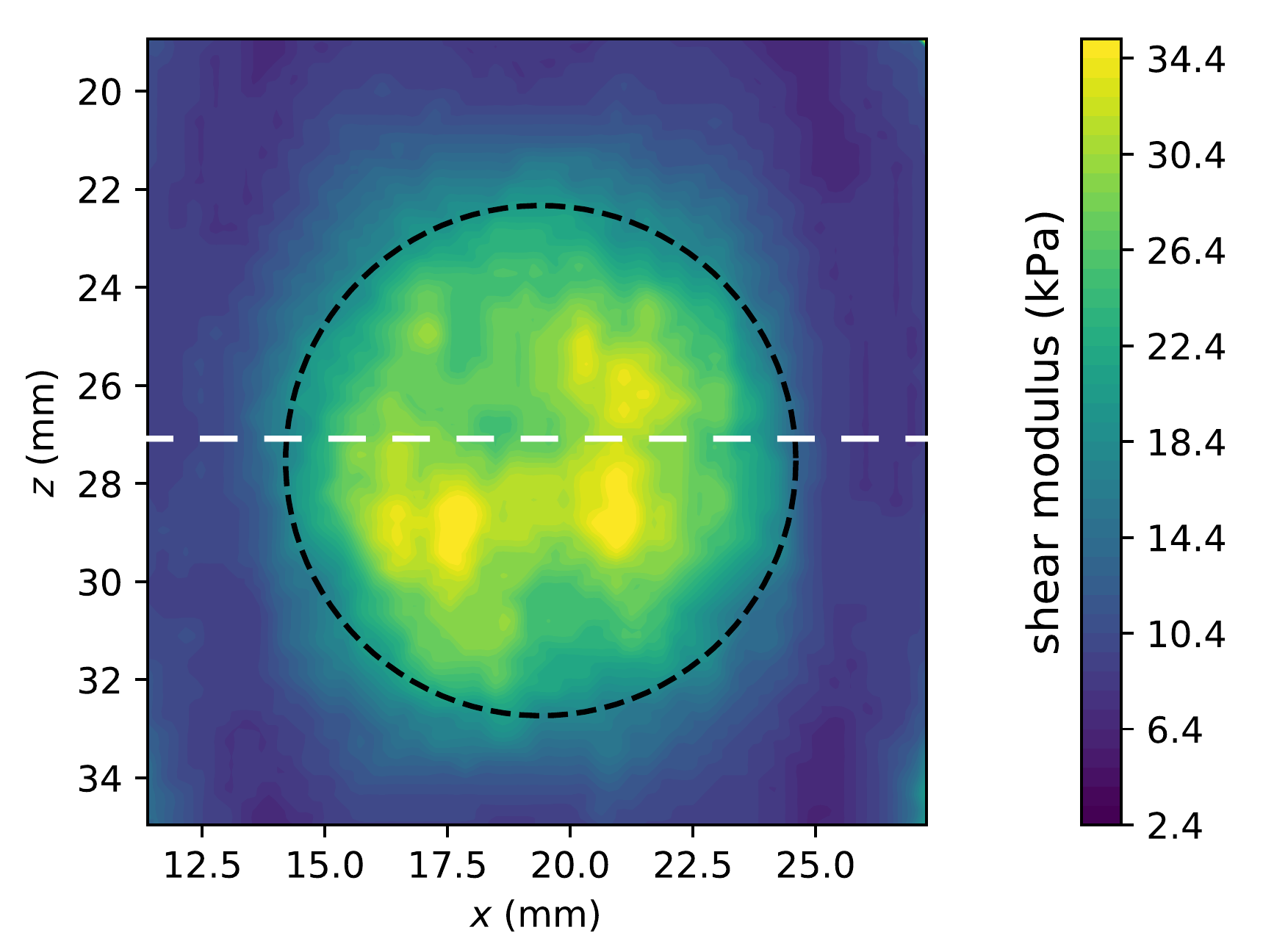}				&
\includegraphics[height=25mm]{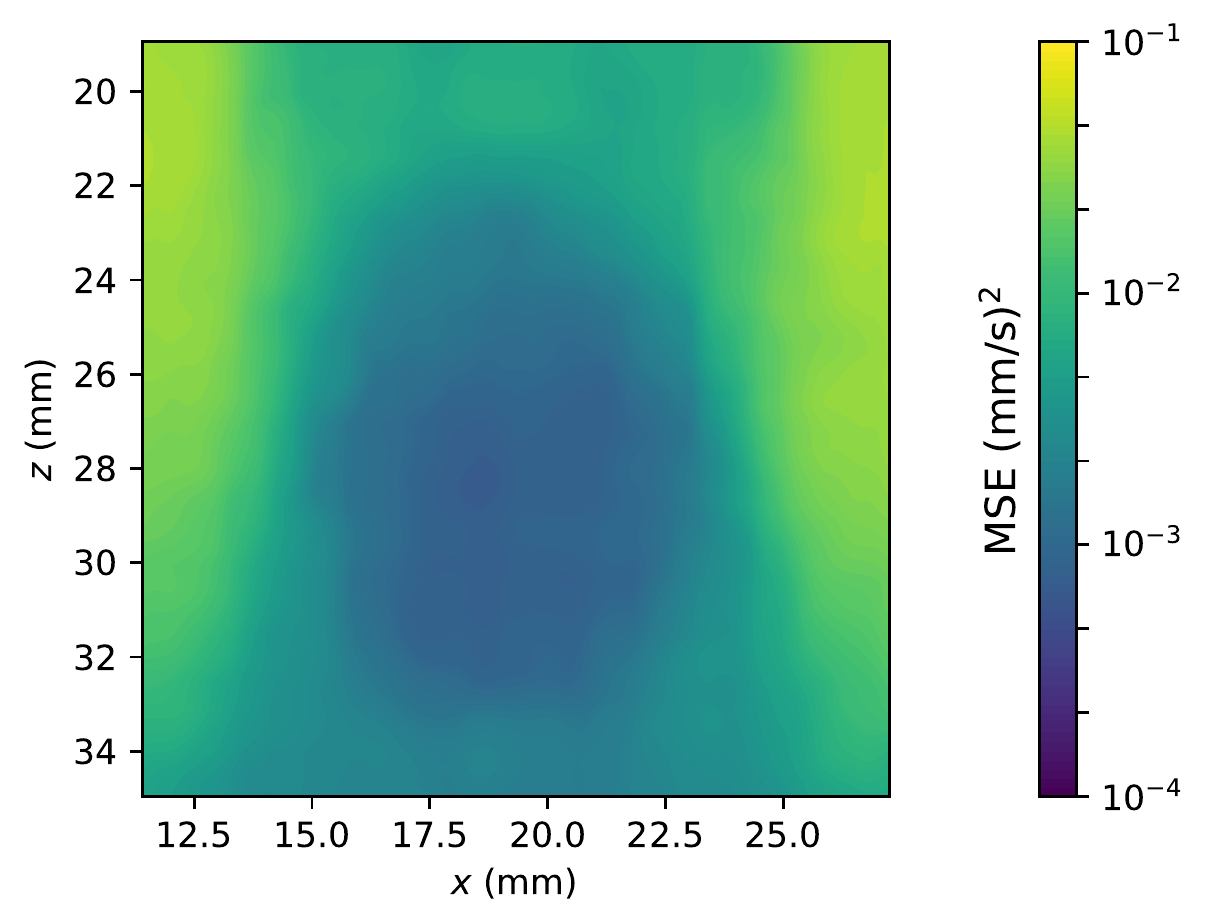}				&
\includegraphics[height=25mm]{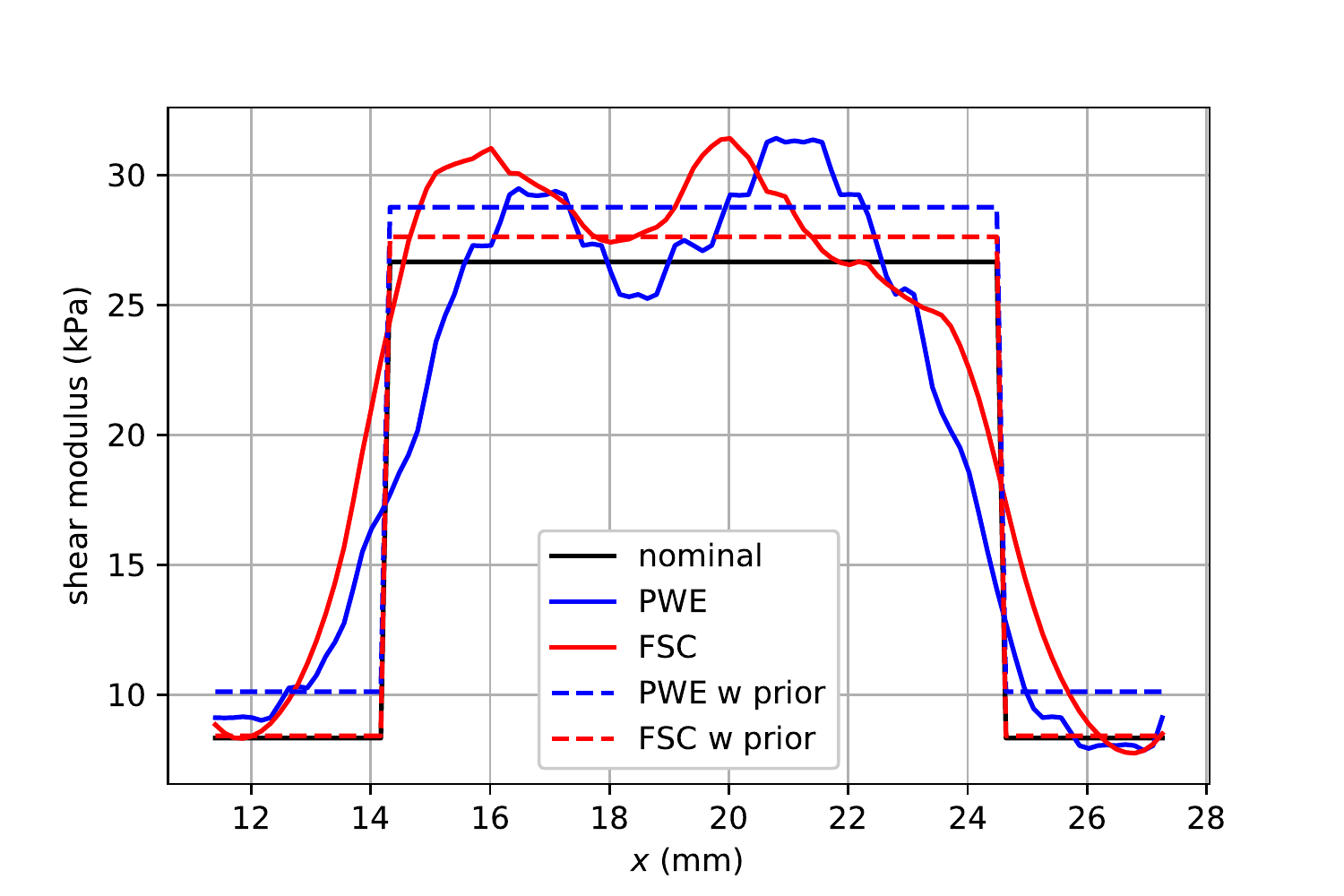} 		\\
\rotatebox{90}{\hspace{7mm}6.49\,mm}							&
\includegraphics[height=25mm]{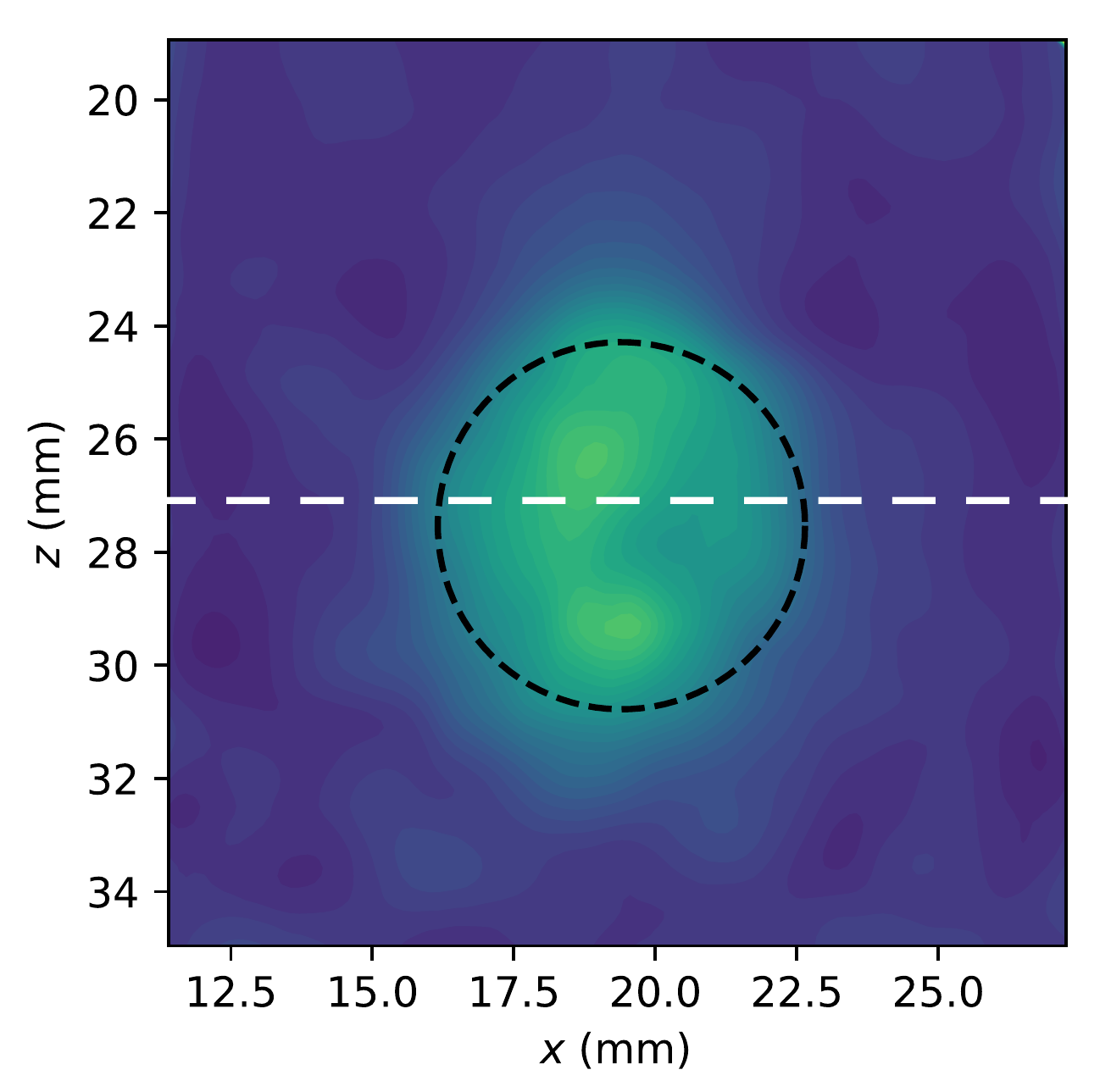}				& 
\includegraphics[height=25mm]{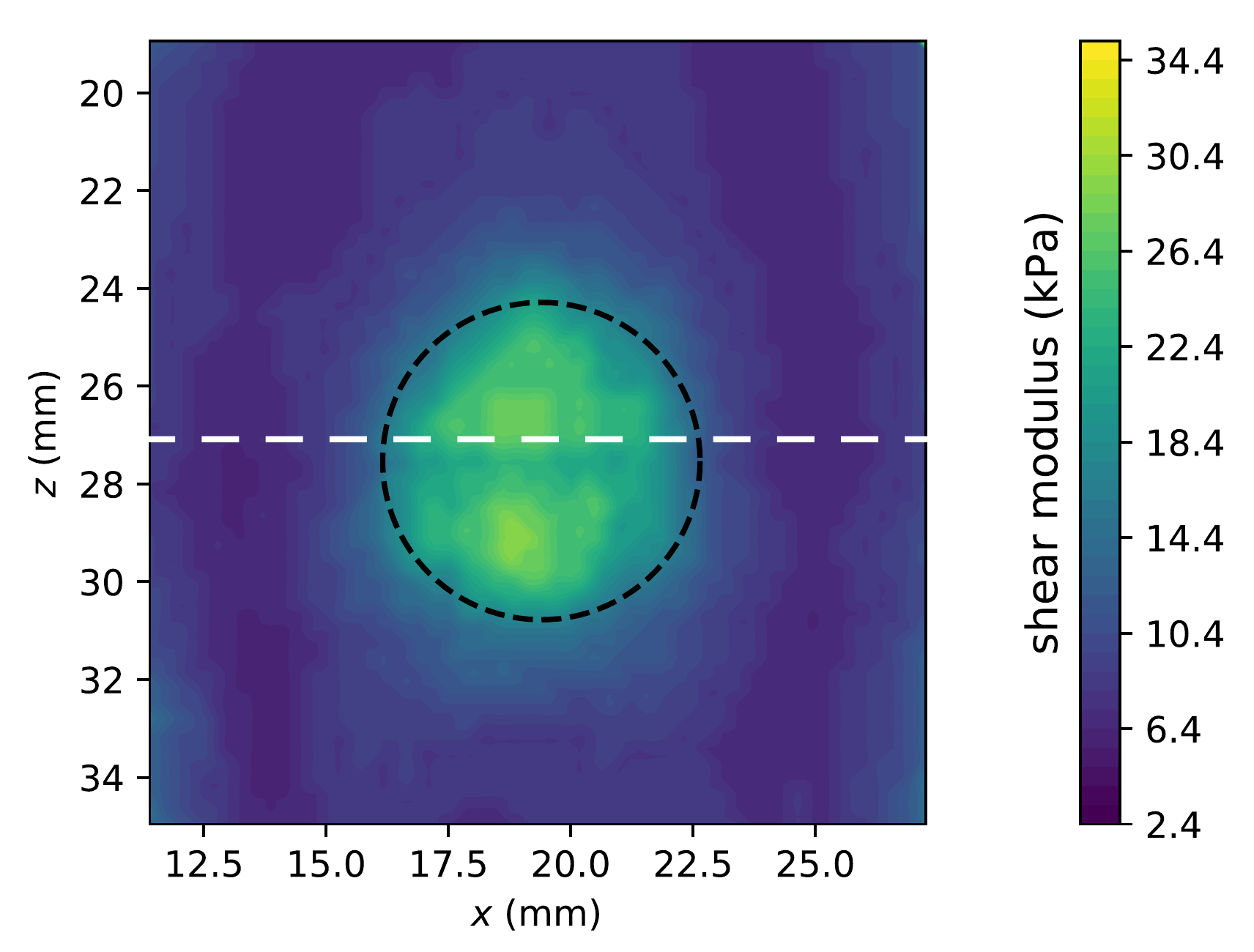}				&
\includegraphics[height=25mm]{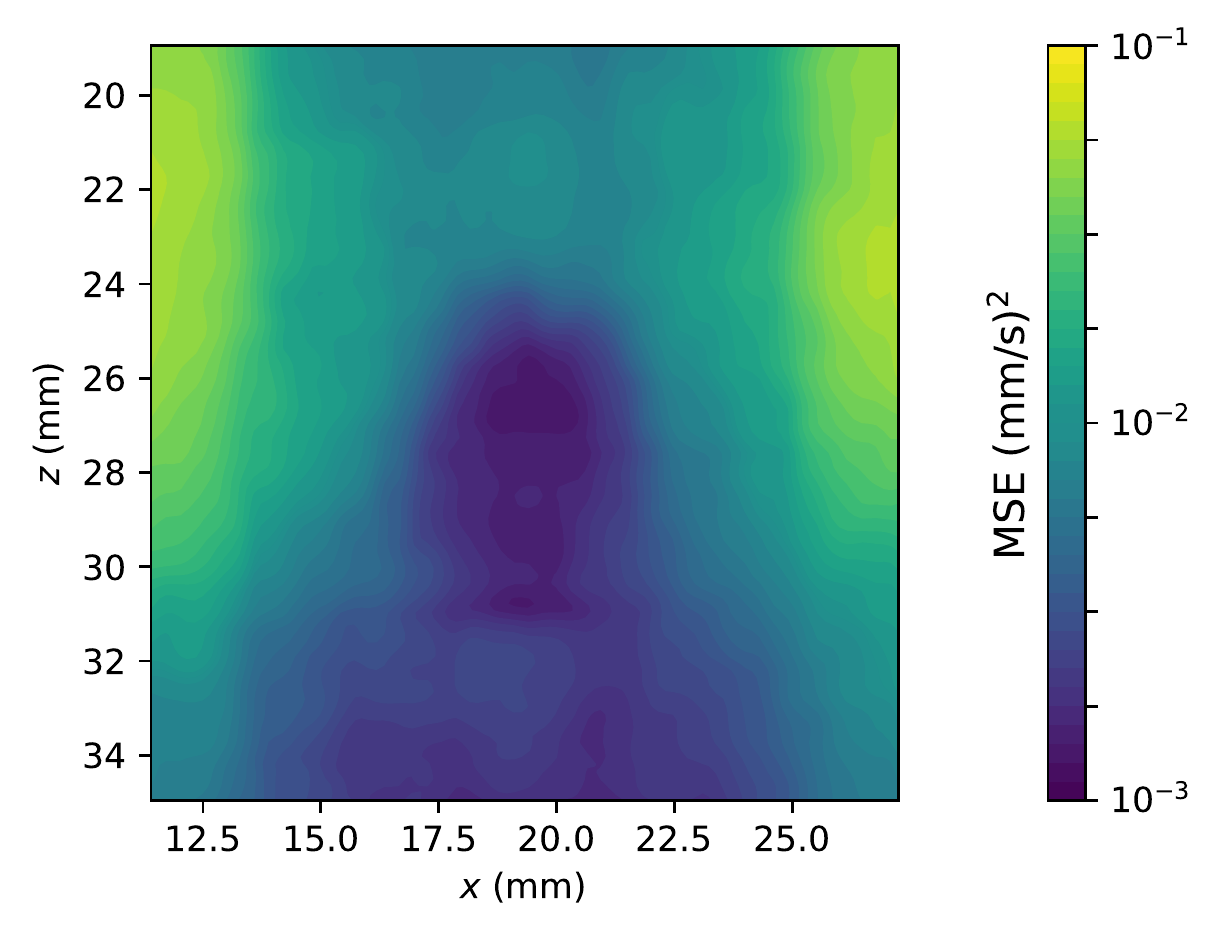}				&
\includegraphics[height=25mm]{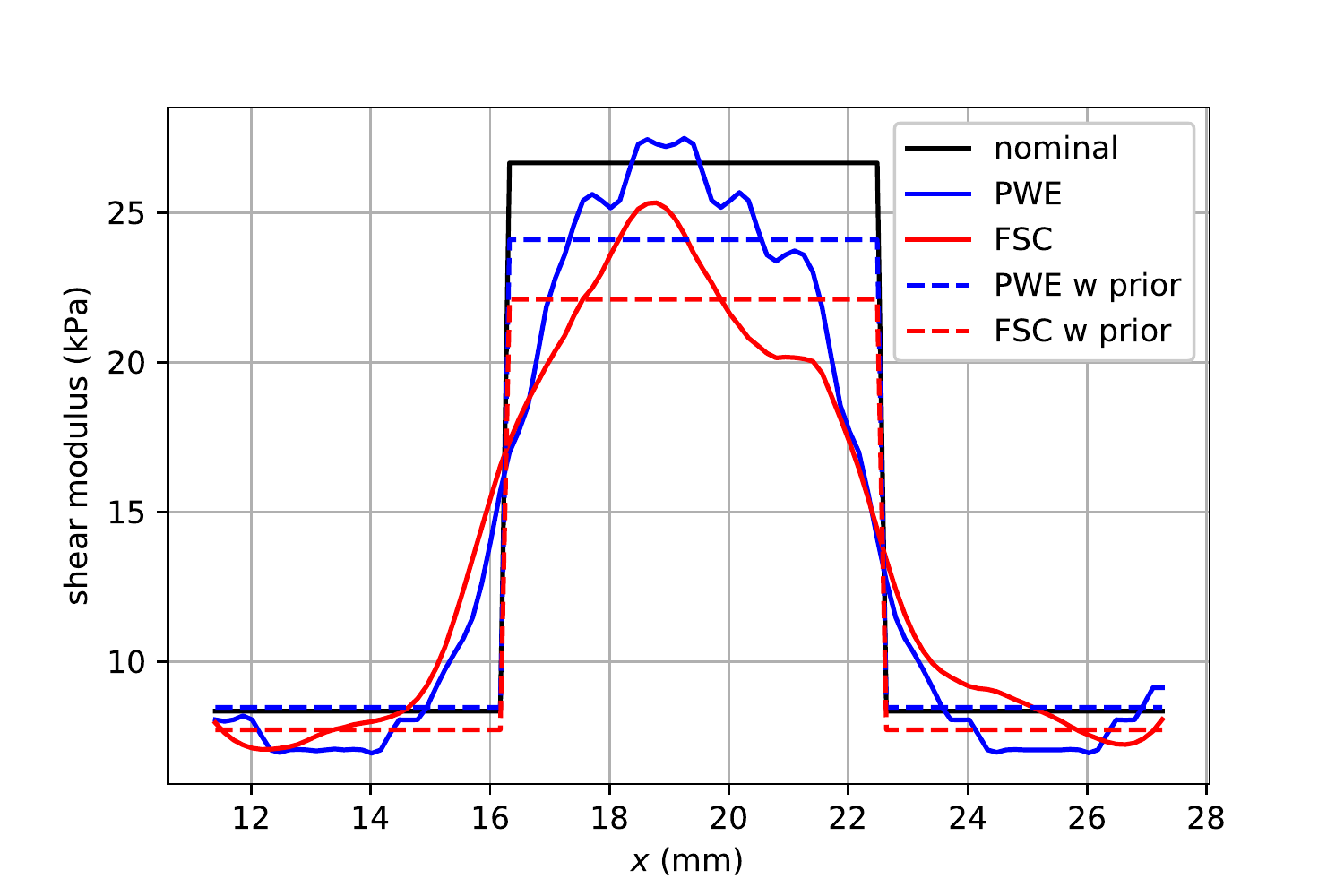} 		\\
\rotatebox{90}{\hspace{7mm}4.05\,mm}							&
\includegraphics[height=25mm]{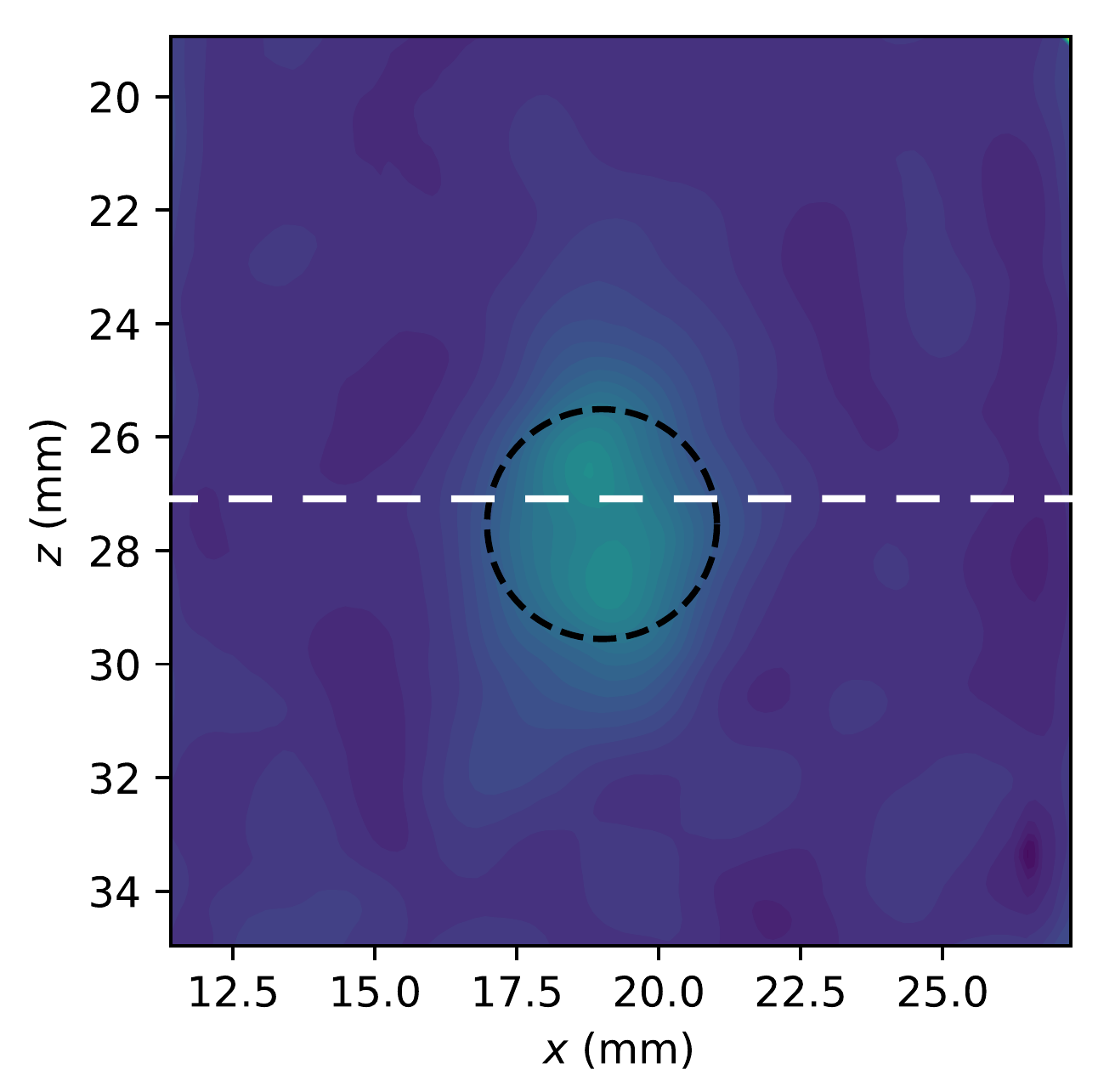}				& 
\includegraphics[height=25mm]{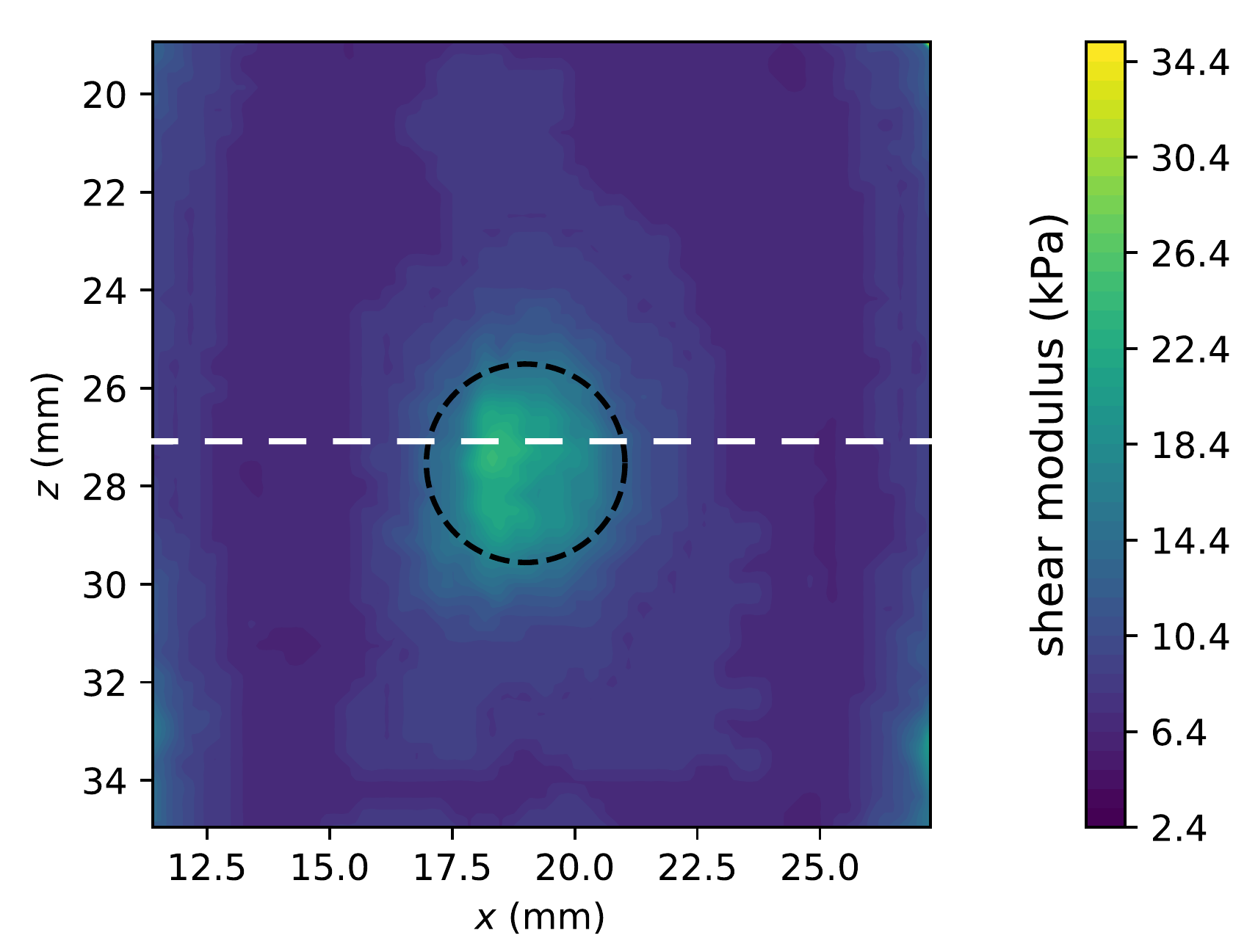}				&
\includegraphics[height=25mm]{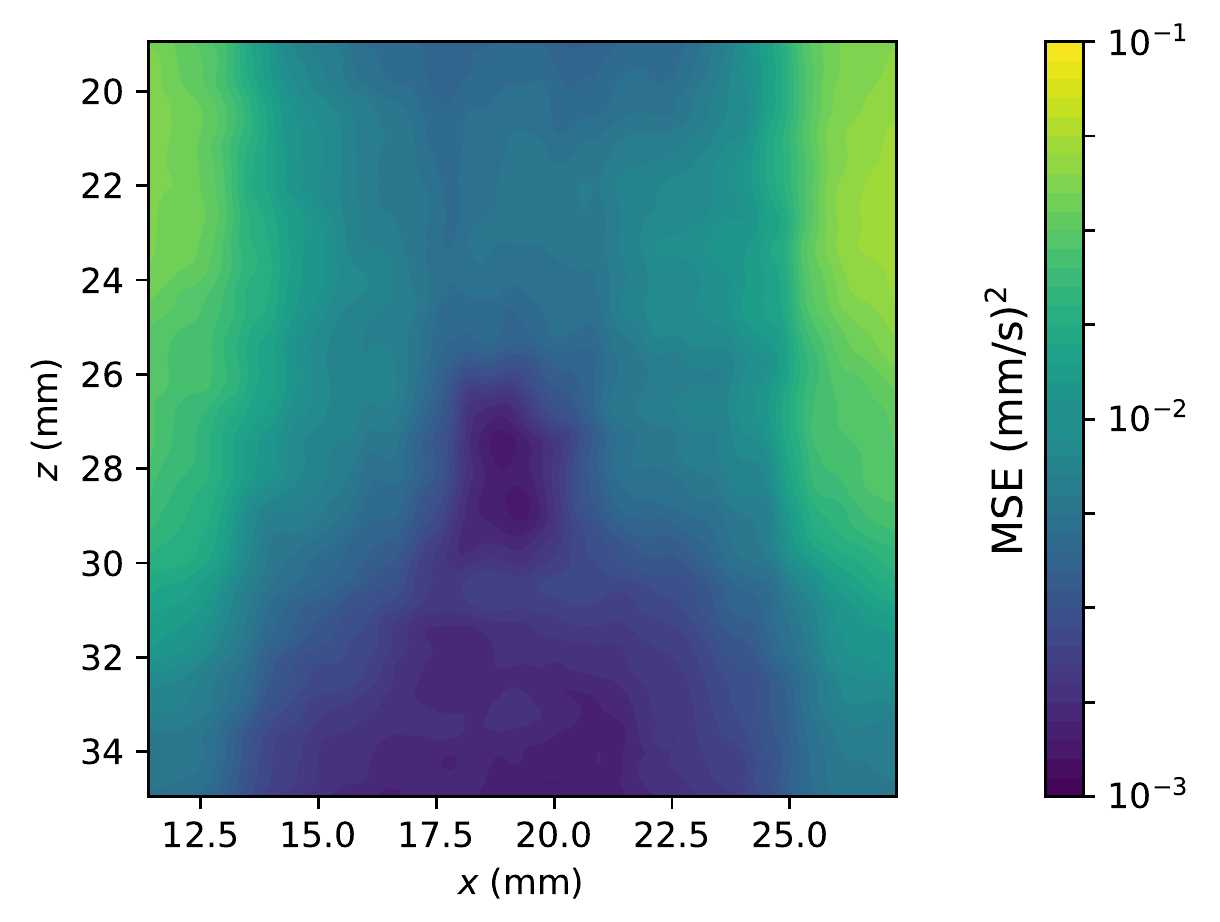}				&
\includegraphics[height=25mm]{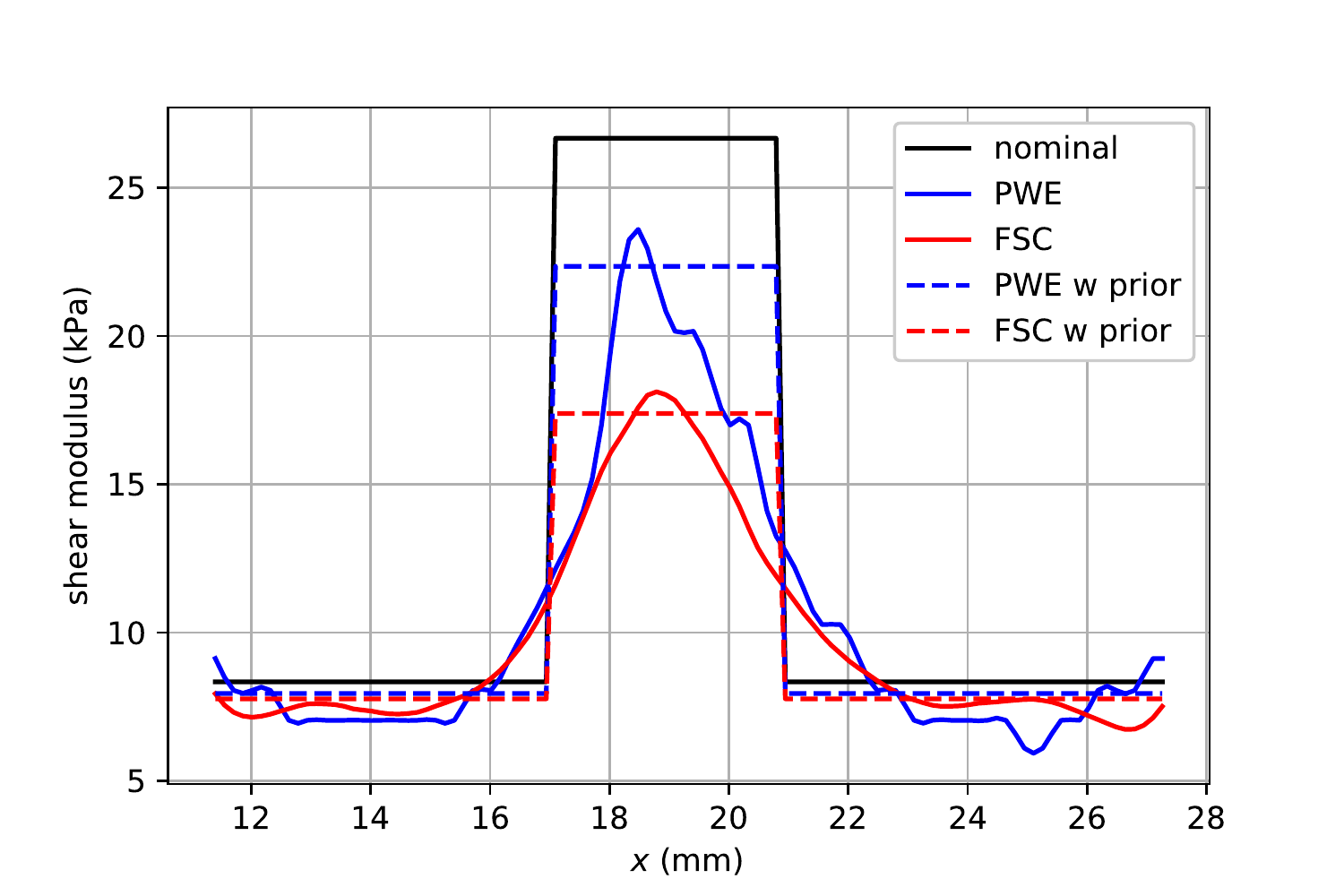} 		\\
%
%\rotatebox{90}{\hspace{1mm}inclusion size: 2.53\,mm}				&
%\includegraphics[width=0.285\textwidth]{TypeIV/2p53/SWE.pdf}			& 
%\includegraphics[width=0.285\textwidth]{TypeIV/2p53/PWE.pdf}			&
%\includegraphics[width=0.33\textwidth]{TypeIV/2p53/crossections.eps} 	
%
\end{tabular}
\caption{PWE and FSC reconstructions for the double-push phantom data with a soft Type I inclusion and three stiff Type IV inclusions. Each row includes the reconstructions for one inclusion. Columns (a) and (b) include the FSC and PWE reconstructions, respectively. Column (c) plots the MSE feedback \eqref{eq:MSE} corresponding to the PWE reconstructions. In column (d) we compare the estimated shear modulus fields without and with the prior knowledge of inclusion geometry, to the nominal values along the horizontal line passing through the center of the ROI, depicted by the white dashed lines in the shear modulus contour plots. % For all PWE reconstructions, we set $f_{\min}$ = 500\,Hz and window size to $w$ = 3.70\,mm. The values of the regularization parameter $\tau$ for each case are obtained from the L-curve analysis and are given in Table \ref{table:parallel-tau}. For the FSC reconstructions, we use window and patch sizes of 1.85\,mm and 1.69\,mm, respectively.
} \label{fig:TypeIV}
\end{figure*}

Fig. \ref{fig:TypeIV} shows the reconstructions for both PWE and FSC methods over a $16$\,mm$\times16$\,mm ROI without and with the prior knowledge of inclusion geometry, while Table \ref{table:TypeIV} reports the corresponding average shear moduli and CNR values \eqref{eq:CNR}.
\begin{table*}
\renewcommand{\arraystretch}{1.3}
\setlength{\tabcolsep}{3pt}
\centering
\footnotesize
\caption{Average shear moduli $\mu_b \and \mu_i$ for the background and inclusion and the CNR \eqref{eq:CNR}, for the PWE and FSC methods, corresponding to the double-push data and Fig. \ref{fig:TypeIV}. The nominal values for the shear moduli of background and soft and stiffer inclusions are $8.33$\,kPa, $2.66$\,kPa, and $26.66$\,kPa.% For all PWE reconstructions, we set $f_{\min}$ = 500\,Hz and window size to $w$ = 3.70\,mm. The values of the regularization parameter $\tau$ are given in Table \ref{table:parallel-tau}.
} \label{table:TypeIV}
\begin{tabular}{|c||c||c|c|c||c|c|c|}
\hline
\multirow{3}{*}{method}	&	homogeneous		&	\multicolumn{3}{c||}{soft inclusion ($10.40$\,mm)}		&	\multicolumn{3}{c|}{stiff inclusion ($10.40$\,mm)}		\\ \cline{2-8}
	&	$\mu_b$
	&	$\mu_b$	&	$\mu_i$	&	CNR		
	&	$\mu_b$	&	$\mu_i$	&	CNR		\\
	&	(kPa)
	&	(kPa)	&	(kPa)	&	(dB)	
	&	(kPa)	&	(kPa)	&	(dB)		\\ \hline
PWE		&	$5.22\pm0.00$
		&	$7.17\pm1.94$		&	$3.93\pm1.11$	&	$3.20$
		&	$10.69\pm2.88$		&	$25.89\pm4.53$	&	$9.04$		\\ \hline
FSC		&	$5.32\pm0.20$
		&	$7.86\pm2.73$		&	$4.11\pm1.29$	&	$1.88$
		&	$11.24\pm3.81$		&	$25.62\pm2.97$	&	$9.47$	\\ \hline
PWE with prior	&	$5.17$
			&	$6.03$			&	$3.31$			&	-
			&	$10.12$			&	$28.77$			&	-	\\ \hline
FSC	with prior	&	$5.27$
			&	$7.07$			&	$3.02$			&	-
			&	$8.42$			&	$27.63$			&	-	\\ \hline \hline
\multirow{3}{*}{method}	&	-		&	\multicolumn{3}{c||}{stiff inclusion ($6.49$\,mm)}	&	\multicolumn{3}{c|}{stiff inclusion ($4.05$\,mm)}	\\ \cline{2-8}
	&	-
	&	$\mu_b$	&	$\mu_i$	&	CNR	
	&	$\mu_b$	&	$\mu_i$	&	CNR	\\
	&	-
	&	(kPa)	&	(kPa)	&	(dB)	
	&	(kPa)	&	(kPa)	&	(dB)	\\ \hline
PWE		&	-
		&	$8.72\pm2.07$			&	$21.92\pm3.58$	&	$10.07$
		&	$8.10\pm1.57$			&	$17.71\pm2.90$	&	$9.30$		\\ \hline
FSC		&	-
		&	$8.95\pm2.23$			&	$20.81\pm2.72$	&	$10.56$
		&	$8.01\pm1.16$			&	$15.46\pm1.80$	&	$10.84$	\\ \hline
PWE with prior	&	-
			&	$8.46$			&	$24.10$			&	-
			&	$7.94$			&	$22.35$			&	-	\\ \hline
FSC	with prior	&	-
			&	$7.71$			&	$22.11$			&	-
			&	$7.76$			&	$17.39$			&	-	\\ \hline
\end{tabular}
\end{table*}
For the PWE reconstructions, by inspecting the Fourier spectrums, for all cases except for the soft inclusion, we set the minimum frequency to $f_{\min} = 500$\,Hz, for the soft inclusion we set $f_{\min} = 100$\,Hz, and the window size to $w = 3.70$\,mm in all cases. In each case, the regularization parameter $\tau$ is selected according to an L-curve similar to Fig. \ref{fig:multiPush_tau}; see Table \ref{table:parallel-tau} for numerical values.
\begin{table} [t]
\renewcommand{\arraystretch}{1.3}
\caption{Regularization parameter $\tau$, selected by the L-curve analysis, for the PWE reconstructions in Fig. \ref{fig:TypeIV} corresponding to the double-push data.}
\centering
\footnotesize
\begin{tabular}{|c||c|c|}
\hline
inclusion type					&	without prior	& 	with prior			\\ \hline\hline
homogeneous					& 	$10^{-3}$		& 	$10^{-2}$		 	\\ \hline
soft inclusion ($10.40$\,mm)		& 	$10^{-2}$		& 	$10^{-3}$		 	\\ \hline
stiff inclusion ($10.40$\,mm)		& 	$10^{-2}$		& 	$10^{-1}$		 	\\ \hline
stiff inclusion ($6.49$\,mm	)		& 	$10^{-3}$		&	$10^{-2}$		 	\\ \hline
stiff inclusion ($4.05$\,mm)		& 	$10^{-6}$		& 	$10^{-6}$		 	\\ \hline
%2.53mm		& 	$10^{-5}$				& 	$10^{-4}$		 	\\ \hline
\end{tabular}
\label{table:parallel-tau}
\end{table}
For the FSC method, we use window and patch sizes of $1.85$\,mm and $1.69$\,mm, respectively. 
Note that as in the previous cases, the PWE method is unaware of the directions of propagation. Nevertheless, the PWE reconstructions without prior directional filtering are competitive with the FSC method and often more accurate. For the homogeneous case in the first row of Fig. \ref{fig:TypeIV}, the estimates are not in agreement with the nominal values due possibly to the change in mechanical properties of the phantom over time\footnote{The homogeneous phantom was manufactured before April 12, 2014 and is over six years old.} \cite{TWEMFAIP2011BSHS} but the PWE and FSC estimates are in agreement with each other. From Table \ref{table:TypeIV}, observe that the FSC reconstructions have slightly better CNR values due to compounding, which naturally increases the CNR because of averaging and reduced variance.
Also, the reconstructions with prior knowledge of the geometry often lead to a better contrast between the background and inclusion.

\subsection{Parameter Study} \label{sec:paramStudy}
\begin{figure}[t!]
	\centering
%	\begin{subfigure}[b]{0.43\textwidth}
%		\includegraphics[width=\textwidth]{Simulations/Reflection-refraction/basisNumParamStudy.eps}
%	\caption{} \label{fig:paramBasisNum_digital}
%	\end{subfigure}
	\begin{subfigure}[b]{0.43\textwidth}
		\includegraphics[width=\textwidth]{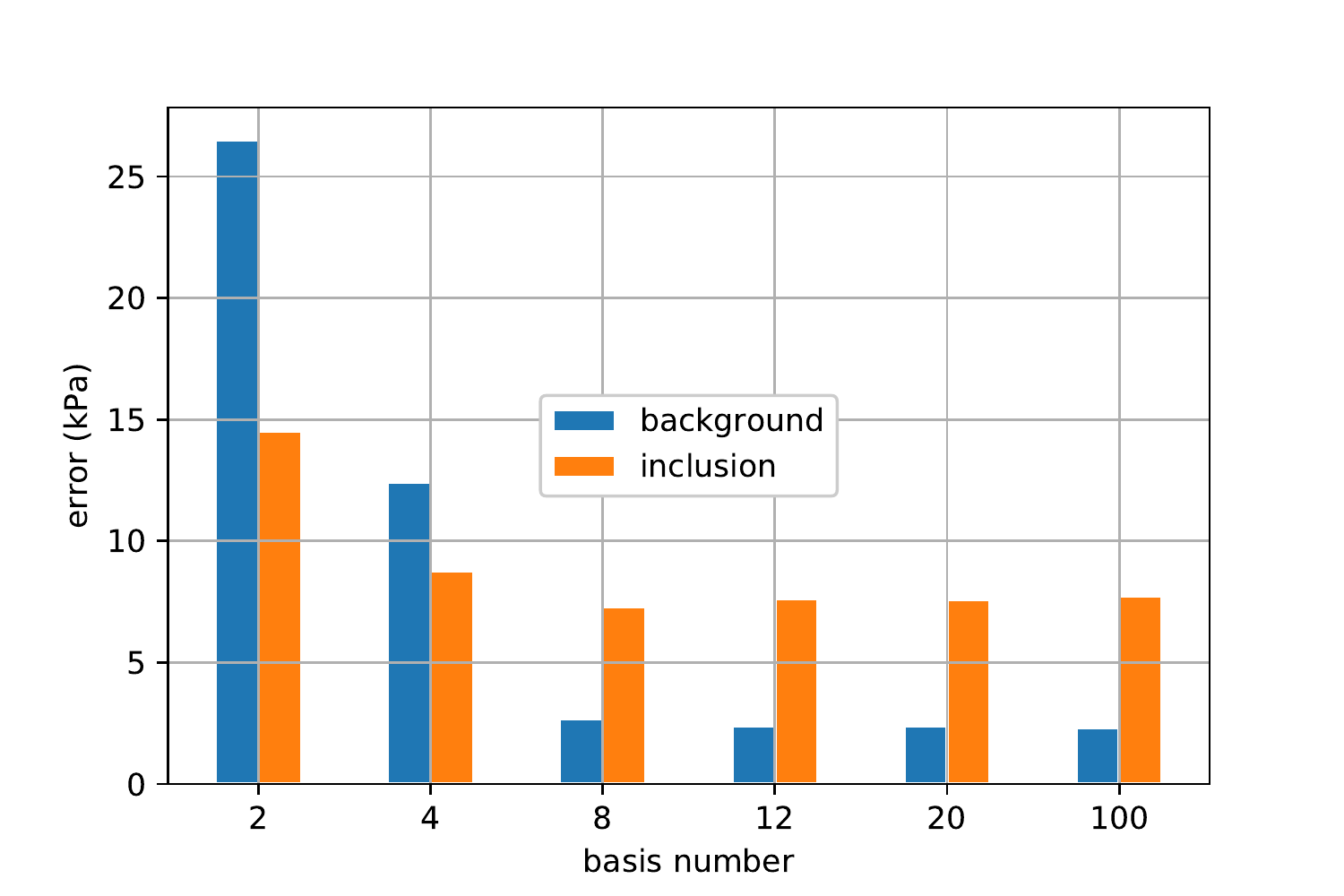}
	\caption{} \label{fig:paramBasisNum_multi}
	\end{subfigure}
	\quad
	\begin{subfigure}[b]{0.43\textwidth}
		\includegraphics[width=\textwidth]{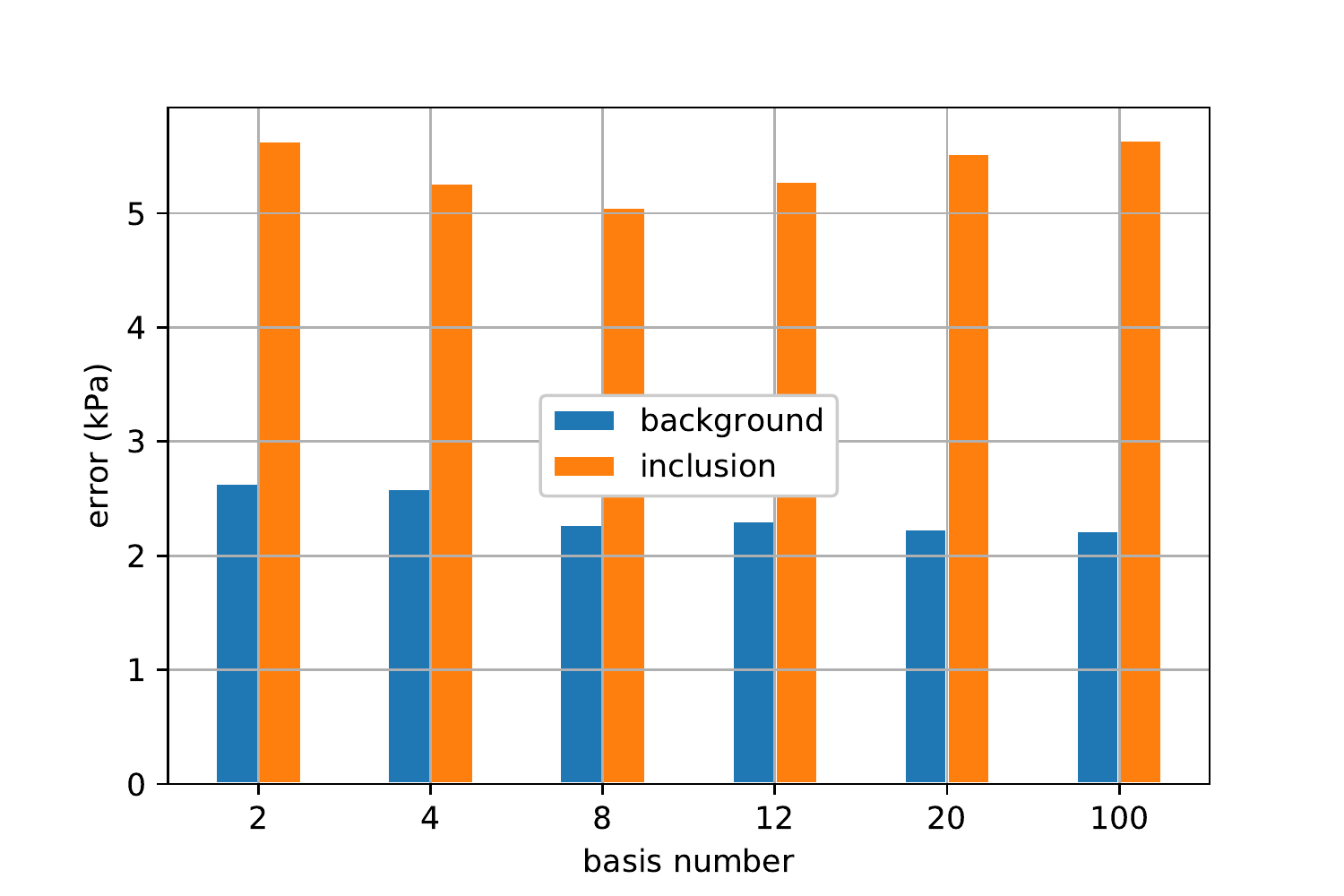}
	\caption{} \label{fig:paramBasisNum_double}
	\end{subfigure}
	\caption{Absolute errors of the average background and inclusion shear moduli as a function of the basis number $n_b$. (a) Fig. \ref{fig:paramBasisNum_multi} corresponds to the multi-push data of Section \ref{sec:multi-push} with push configuration (i). (b) Fig. \ref{fig:paramBasisNum_double} corresponds to the double-push data of Section \ref{sec:stateArt_comp} for inclusion diameter of $6.49$\,mm.} \label{fig:paramBasisNum}
\end{figure}

In this section, we study the effect of important parameters, discussed in Section \ref{sec:paramSelec}, on PWE reconstructions. In each case, we optimally select all other parameters including the regularization parameter.
First, we consider the effect of basis number $n_b$ on PWE reconstructions.
%This parameter is important since in the absence of prior knowledge of propagation directions, the PWE method searches within the $n_b$ directions specified by \eqref{eq:dirDisc}, to find dominant directions that best describe the shear wave data; see Appendix \ref{app:directionalFilter} for more details.
In Fig. \ref{fig:paramBasisNum}, we plot the absolute errors of $\mu_b \and \mu_i$ compared to the nominal values for the multi-push data with push configuration (i), and the double-push data with $6.49$\,mm inclusion size, discussed in Sections \ref{sec:multi-push} and \ref{sec:stateArt_comp}, respectively. %where we also overlay the standard deviations $\std(\bbmu_{\Omega_b}) \and \std(\bbmu_{\Omega_i})$ on the bars.
%
%Referring to Fig. \ref{fig:paramBasisNum_digital}, observe that in the case of digital phantom, due to reflection and refraction caused by the complicated inclusion shape (cf. Fig. \ref{fig:complicatedShape}), $n_b$ = 2 cannot capture the wave data; this should be the reason for less accurate reconstructions of the FSC method in Section \ref{sec:digital-phantom}. With $n_b$ = 4 the inclusion shear modulus seems to be adequately reconstructed but the background still has a high error value; this is due to the complicated shape of the shear wave within the triangular part of the background; see Section \ref{sec:digital-phantom} for more details. Beyond $n_b$ = 8 the reconstructions become stable.
%
From Fig. \ref{fig:paramBasisNum_multi} corresponding to the multi-push data, it can be seen that $n_b = 2, 4$ are insufficient to resolve the shear wave particularly in the background, but the reconstructions seem to plateau beyond $n_b = 8$.
Unlike the multi-push data, it can be seen from Fig. \ref{fig:paramBasisNum_double} that for the double-push data, only two basis functions are adequate since these two bases happen to align with the directions of propagation.
In both cases, because of appropriate regularization, the solutions stay stable as we keep increasing the number of basis functions.
%
%In the absence of proper regularization, the stability of the elastography problem \eqref{eq:elastProb} degrades and also the solution starts to overfit the noise. This might be the reason for slight increase in error values for larger $n_b$. Regardless, the solution stays stable as we keep increasing $n_b$.
%
Note that throughout the results, we used $n_b = 12$ bases to reconstruct the shear modulus fields although the true propagation directions might not align with such a sparse discretization. This indicates that PWE can reconstruct the shear modulus field even if the dominant propagation directions are not fully recovered. In \ref{app:homogeneous}, we further elaborate on this point.

\begin{figure}[t!]
	\centering
	\begin{subfigure}[b]{0.41\textwidth}
		\includegraphics[width=\textwidth]{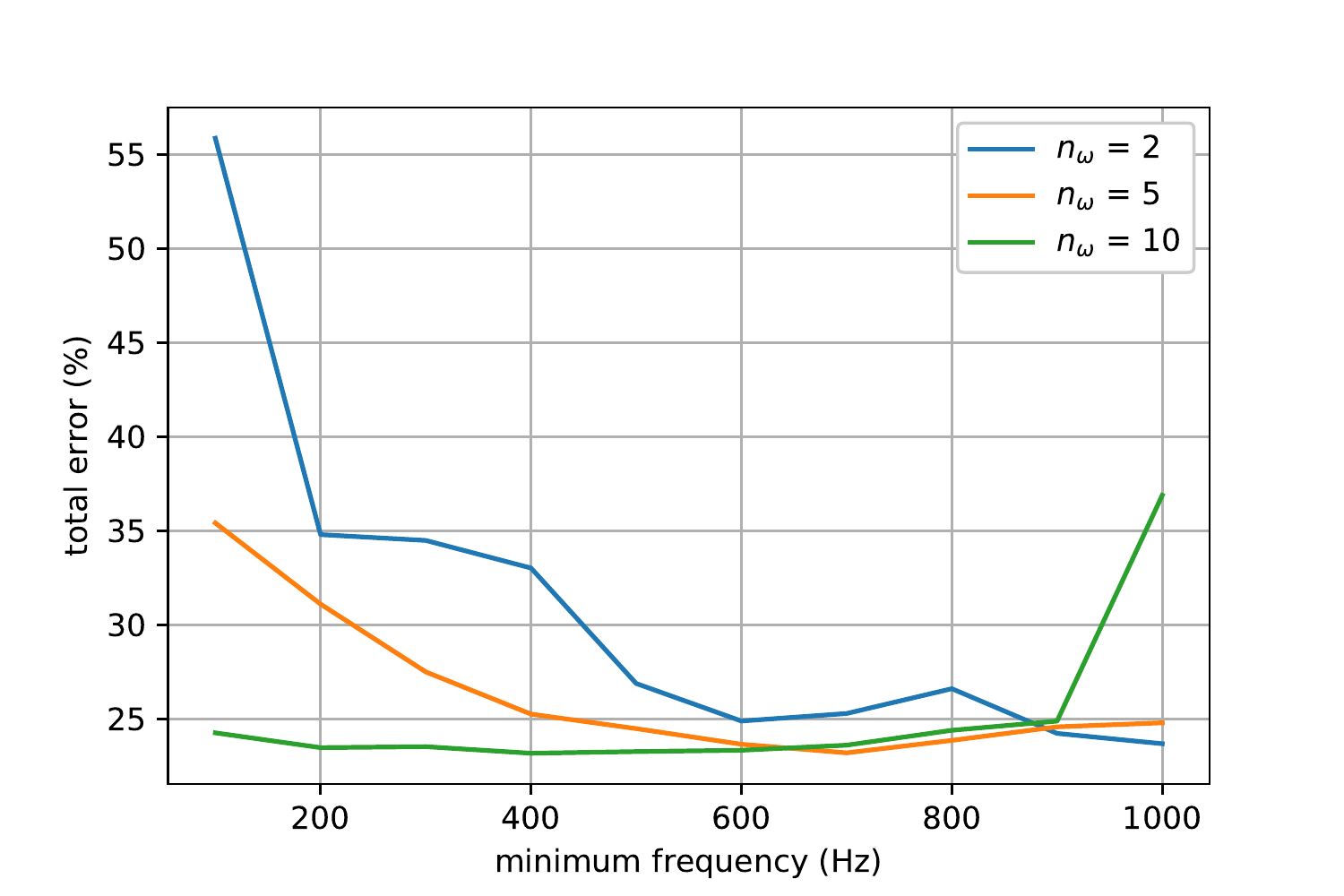}
	\caption{} \label{fig:praramFreq_l2err}
	\end{subfigure}
	\quad
	\begin{subfigure}[b]{0.42\textwidth}
		\includegraphics[width=\textwidth]{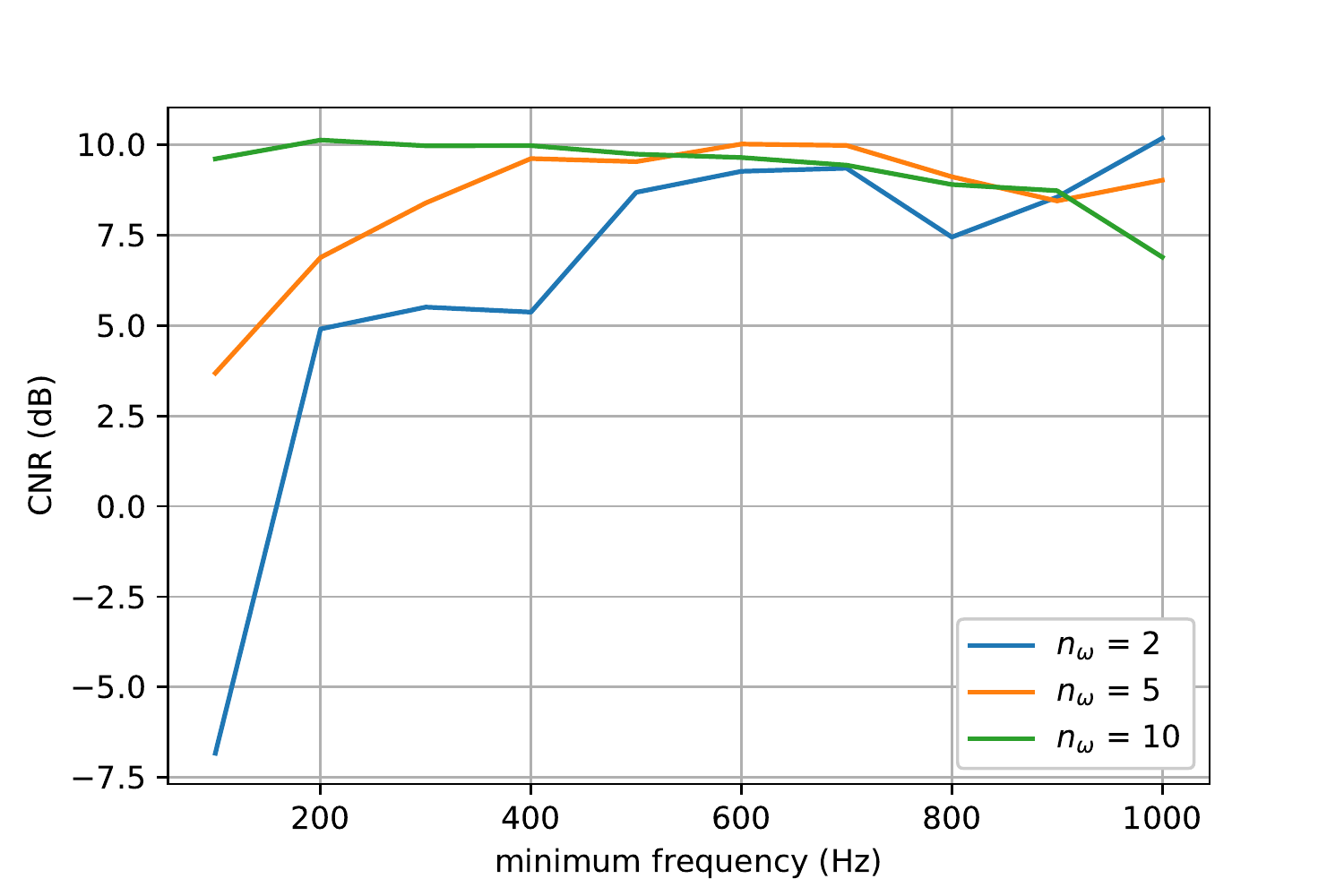}
	\caption{} \label{fig:praramFreq_CNR}
	\end{subfigure}
	\caption{(a) Fig. \ref{fig:praramFreq_l2err} plots the normalized $\ell_2$-error over the ROI as a function of the frequency lower-bound $f_{\text{min}}$ for three different number of dominant frequencies $n_{\omega}$. (b) Fig. \ref{fig:praramFreq_CNR} shows the similar plot for the CNR \eqref{eq:CNR}. Generally using larger $n_{\omega} \and f_{\min}$ improves both the error and CNR.} \label{fig:praramFreq}
\end{figure}

To study the rest of the important parameters for PWE, we consider the double-push data with $6.49$\,mm inclusion as a representative example.
First, we investigate the effect of the number of frequency $n_{\omega}$ and minimum frequency $f_{\min}$ on CNR \eqref{eq:CNR} and normalized error, defined as
$ \text{err} = {\norm{ \bbmu - \bbmu^{\text{nom}} }} / { \norm{ \bbmu^{\text{nom}} } } , $
%\begin{equation} \label{eq:err}
%\text{err} = \frac{\norm{ \bbmu - \bbmu^{\text{nom}} }}{ \norm{ \bbmu^{\text{nom}} } } ,
%\end{equation}
%
where $\bbmu$ is the output of PWE Algorithm \ref{alg:PWE} and $\bbmu^{\text{nom}}$ is the corresponding nominal vector of values. Fig. \ref{fig:praramFreq} shows the results for three different frequency numbers $n_{\omega}$ and a range of minimum frequency $f_{\min}$ values.
As discussed in Section \ref{sec:paramSelec}, increasing $n_{\omega}$ and $f_{\min}$ generally leads to better reconstructions. However, there is a value of $f_{\min}$ above which the selected frequencies mostly contain noise and are not as informative; this is the reason for upward trend in Fig. \ref{fig:praramFreq_l2err} for large values of $f_{\min}$.
This upward trend starts earlier for larger $n_{\omega}$ since given a large value of $f_{\min}$, using more frequencies, leads to earlier inclusion of higher frequency noisy data.

% The minipage struct used to combine the frequency parameters (above) and window size (below) analyses to save space:
%
%\begin{figure*}
%%
%\begin{minipage}{0.66\textwidth}
%	\centering
%	\begin{subfigure}[b]{0.47\textwidth}
%		\includegraphics[width=\textwidth]{TypeIV/6p49/freqParamStudy_l2err.eps}
%	\caption{} \label{fig:praramFreq_l2err}
%	\end{subfigure}
%	\quad
%	\begin{subfigure}[b]{0.485\textwidth}
%		\includegraphics[width=\textwidth]{TypeIV/6p49/freqParamStudy_CNR.eps}
%	\caption{} \label{fig:praramFreq_CNR}
%	\end{subfigure}
%	\caption{(a) Fig. \ref{fig:praramFreq_l2err} plots the normalized $\ell_2$-error over the ROI as a function of the frequency lower-bound $f_{\text{min}}$ for three different number of dominant frequencies $n_{\omega}$. (b) Fig. \ref{fig:praramFreq_CNR} shows the similar plot for the CNR \eqref{eq:CNR}. Generally using larger $n_{\omega} \and f_{\min}$ improves both the error and CNR.} \label{fig:praramFreq}
%\end{minipage}
%%
%\hspace{1mm}
%%
%\begin{minipage}{0.34\textwidth}
%	\centering
%	\includegraphics[width=0.95\textwidth]{TypeIV/6p49/windowParamStudy.eps}
%	\caption{Absolute errors of the average background and inclusion shear moduli as a function of the window size $w$ corresponding to the double-push data of Section \ref{sec:stateArt_comp} with inclusion diameter of $6.49$\,mm.} \label{fig:paramWinSize}
%\end{minipage}
%%
%\end{figure*}

\begin{figure}[t!]
  \centering
    \includegraphics[width=0.45\textwidth]{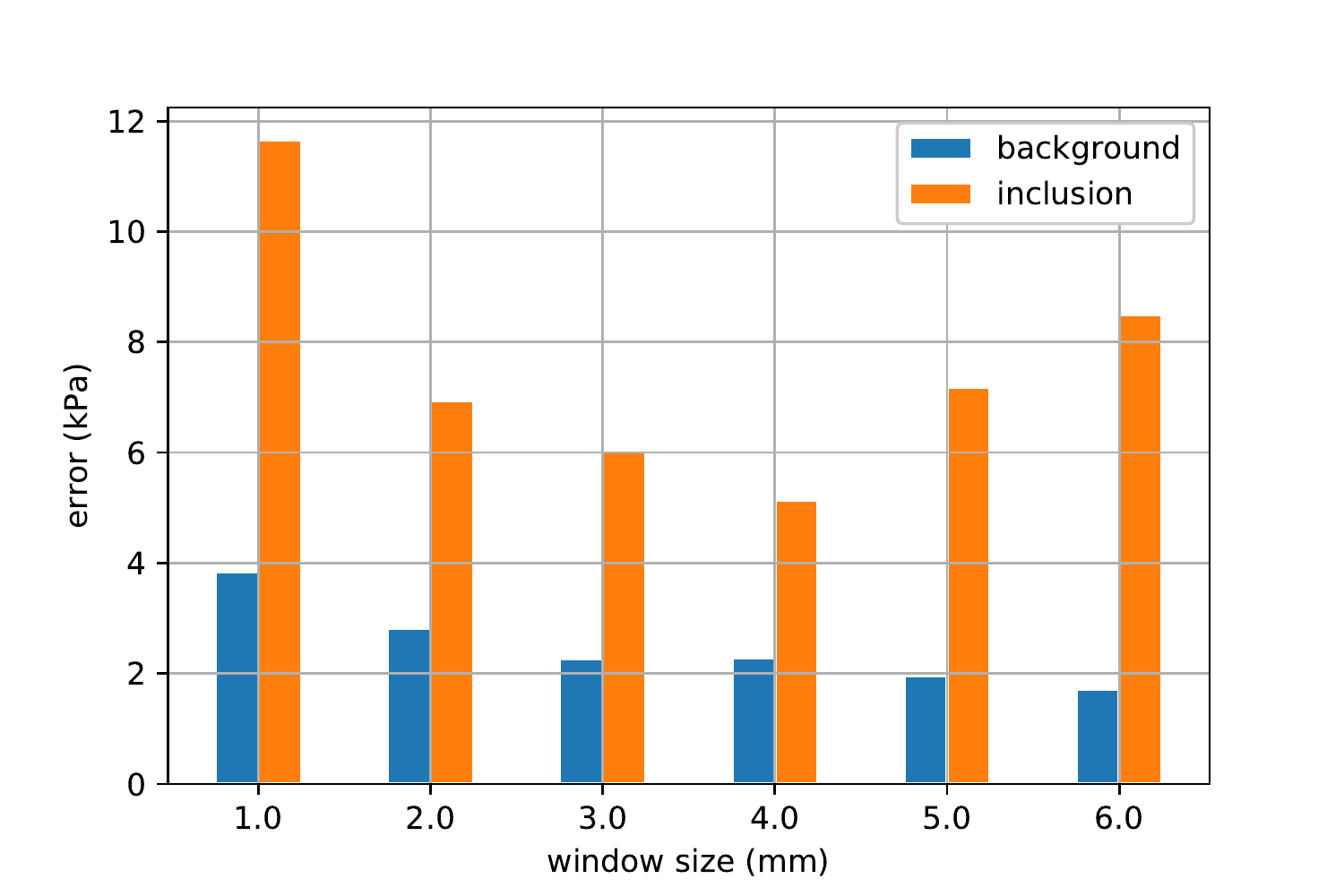}
  \caption{Absolute errors of the average background and inclusion shear moduli as a function of the window size $w$ for the double-push data of Section \ref{sec:stateArt_comp} with inclusion diameter of $6.49$\,mm.} \label{fig:paramWinSize}
\end{figure}

Finally, we study the effect of window size $w$. Fig. \ref{fig:paramWinSize} shows a bar plot similar to Fig. \ref{fig:paramBasisNum}, where we compare the average estimates in the background and inclusion to the nominal values.
As discussed in Section \ref{sec:paramSelec}, overly small values of $w$, compared to the wavelength corresponding to $f_{\min}$, cannot resolve the waves and result in poor reconstructions. On the other hand, large values of $w$ result in over-smoothed, poor reconstructions. This can be seen here for the estimation of the inclusion shear modulus.
Note that among the parameters studied in this section, the error values are more sensitive to window size $w$ and minimum frequency $f_{\min}$. Since for a given dataset, the range of dominant frequencies $[f_{\min}, f_{\max}]$ is determined, to obtain reasonable reconstructions, we need to adjust $w$; the lower $f_{\min}$, the larger $w$ needs to be  for accurate reconstruction; see Section \ref{sec:paramSelec}.
%We used $w$ = 2\,mm for the digital phantom of Section \ref{sec:digital-phantom} with high frequency content, w = 7\,mm for multi-push data of Section \ref{sec:multi-push} with low frequency content, and w = 3.7\,mm for the double-push data of Section \ref{sec:stateArt_comp}.

% ------------------------------------------------------------------------------------------------------------------------------ %
\section{Discussion} \label{sec:discussion}
% paragraph theme: PWE parameters
%
The PWE Algorithm \ref{alg:PWE} depends on a number of parameters. Among those, the number of dominant frequencies and plane wave basis functions can generally be fixed, as in Section \ref{sec:exp}. The frequency range $[f_{\min}, f_{\max}]$ is dictated by the shear wave data and should be selected by inspecting the Fourier spectrum and the knowledge of $f_{\min}$ enables the selection of window size $w$, as discussed in Section \ref{sec:paramSelec}. Ultimately, the only parameter that needs to be tuned in practice is the regularization parameter $\tau$ for which we outlined the L-curve approach in Section \ref{sec:paramSelec}. Moreover, unlike the reconstructions in Section \ref{sec:exp} which are based on data obtained from acquisitions with different settings, in practice the PWE algorithm will be used on an ultrasound system with consistent acquisition settings and noise characteristics. Under these circumstances, all of the parameters including $\tau$ can often be pre-selected. For instance, observe the consistency among parameters used for double-push reconstructions of Section \ref{sec:stateArt_comp}. Finally, the FSC method \cite{FSCSWSC2014SMZU} used for validation in Section \ref{sec:exp}, requires selecting at least as many parameters. These include the propagation directions, window and patch sizes, and parameters of the directional and radial filters including temporal and spatial frequency ranges, and power and order; see \ref{app:directionalFilter}.

% paragraph theme: extensions of PWE
%
As we discussed in Section \ref{sec:paramStudy}, when the plane wave directions in \eqref{eq:basisFun} happen to align with the propagation direction, PWE can reconstruct the shear modulus field with a few basis functions; see Fig. \ref{fig:paramBasisNum_double}. This means that we can exploit the prior knowledge of propagation directions if available, by using a non-uniform distribution on the directions instead of the uniform distribution in \eqref{eq:dirDisc}.
PWE can also be utilized with prior filtering and compounding, similar to the FSC method. 
%
% paragraph theme: multiple experiments
%
When multiple sets of data from independent experiments are available, instead of compounding, we can also (i) extend the optimization objective in \eqref{eq:elastProb} to include another summation over these sets of data, or (ii) superpose the data and process them at once as a single multi-push data.

The computational cost of the PWE Algorithm \ref{alg:PWE} depends on the basis number $n_b$, the measurement number $m$, the number of frequencies $n_{\omega}$, method used to solve \eqref{eq:PWEsol}, and the number of subdomains $n_s$.
From \eqref{eq:closed-form}, it can be seen that the dependence on the number of bases and measurements is $O(m \, n_b^2 + n_b^3)$ in the worst case while from \eqref{eq:PWEsol}, dependence on $n_{\omega}$ is linear. Assuming we use a simple discretization of the feasible wave-speed range $[c_{\min}, c_{\max}]$ with $n_c$ points to approximately solve \eqref{eq:PWEsol}, dependence on $n_c$ is also linear. Thus, the worst case computational cost of solving \eqref{eq:elastProb} for a homogeneous subdomain is bounded by $O(n_b^2 \, (m + n_b) \, n_{\omega} \, n_c)$.
From Algorithm \ref{alg:PWE}, observe that there is an explicit loop over the $n_s$ subdomains. Thus, the worst case computational cost of the PWE algorithm is bounded by $O(n_b^2 \, (m + n_b) \, n_{\omega} \, n_c \, n_s)$.
Nevertheless, the optimal coefficients for different frequencies can be calculated independently and in parallel using \eqref{eq:closed-form}. Moreover, the solution for each subdomain is independent and can be parallelized. Thus, the effective cost is only bounded by $O(n_b^2 \, (m + n_b) \, n_c)$. Note that the window size $w$ affects the computational cost through the number of measurements $m$ and dependence on the regularization parameter $\tau$ is negligible.
%
%using constant dominant frequencies across windows
The reconstructions reported in this paper typically require less than a minute on a desktop computer with an \textsc{Intel} Core i$9$-$3.10$\,GHz processor and $128$\,GB of memory, using our initial implementation of PWE. This computation time could however be considerably improved due the highly parallelizable nature of PWE and particularly, by utilizing GPUs to solve the linear system in \eqref{eq:closed-form}.

% paragraph theme: other applications
%
The proposed PWE method has potential in other elastography approaches particularly in those that use vibration for excitation resulting in complicated motion fields that cannot easily be directionally decoupled. This includes magnetic resonance elastography \cite{MREDVPA1995MLRGM}, vibration-based ultrasound elastography \cite{EVMUSE2014ZSMK}, passive ultrasound elastography \cite{TREWSS2008CBBN,QSEICEWSS2015BCBN}, and optical coherence elastography \cite{OCTECWBP2020LKU,FPVOCEH2020LKU}.

Finally, processing \textit{in vivo} patient data poses new challenges that we plan to investigate. This includes the viscoelastic nature of the soft tissue, as opposed to the elastic assumption made in this paper, and significantly higher noise levels due to physiological movement, severe inhomogeneity of the soft tissue, and dissipation caused by viscosity. As we demonstrated in Section \ref{sec:multi-push}, PWE seems to be more robust to noise than FSC and has potential for even more improvement when applied to \textit{in vivo} data with low SNR. It is known that in addition to the shear modulus, the shear viscosity of soft tissue also has diagnostic value \cite{VPDBM2018KDGBM}. PWE can be extended to estimate the shear viscosity by considering a complex modulus in wave equation \eqref{eq:scalarWE} and conducting a 2D search instead of the line search in \eqref{eq:PWEsol}.

% ------------------------------------------------------------------------------------------------------------------------------ %
\section{Conclusion} \label{sec:concl}
We proposed PWE, a novel ultrasound SWE approach that unlike commonly used techniques, can handle multiple waves with arbitrary incident angles at once and does not rely on directionality of the propagation or the prior knowledge of the propagation direction. %Assuming an isotropic, incompressible, linear-elastic medium, the PWE method utilizes the scalar wave equation to model the propagation of shear waves. Particularly, it uses a plane wave expansion of the solution to the wave equation and formulates a nonlinear $\ell_2$-regularized least-squares optimization problem that can be solved very efficiently.
We demonstrated through various phantom studies that PWE can reconstruct the shear modulus field with an accuracy comparable to state-of-the-art and provide feedback on the reconstruction. When the prior knowledge of the inclusion geometry was available, we obtained more efficient and accurate reconstructions.

% ------------------------------------------------------------------------------------------------------------------------------ %

\ack
This work is supported in part by the National Science Foundation under grant CNS \#1837499.

% ------------------------------------------------------------------------------------------------------------------------------ %

%\ifCLASSOPTIONcaptionsoff
%  \newpage
%\fi

\section*{References}
\bibliographystyle{dcu}\bibliography{MyBibliography}

% ------------------------------------------------------------------------------------------------------------------------------ %
%\newpage
\setcounter{page}{1}

\renewcommand{\thefigure}{A\arabic{figure}}
\setcounter{figure}{0}

\renewcommand{\thetable}{A.\Roman{table}}
\setcounter{table}{0}

\appendix

\section{Connection to Directional Filter} \label{app:directionalFilter}
\begin{figure}[t!]
  \centering
    \includegraphics[width=0.35\textwidth]{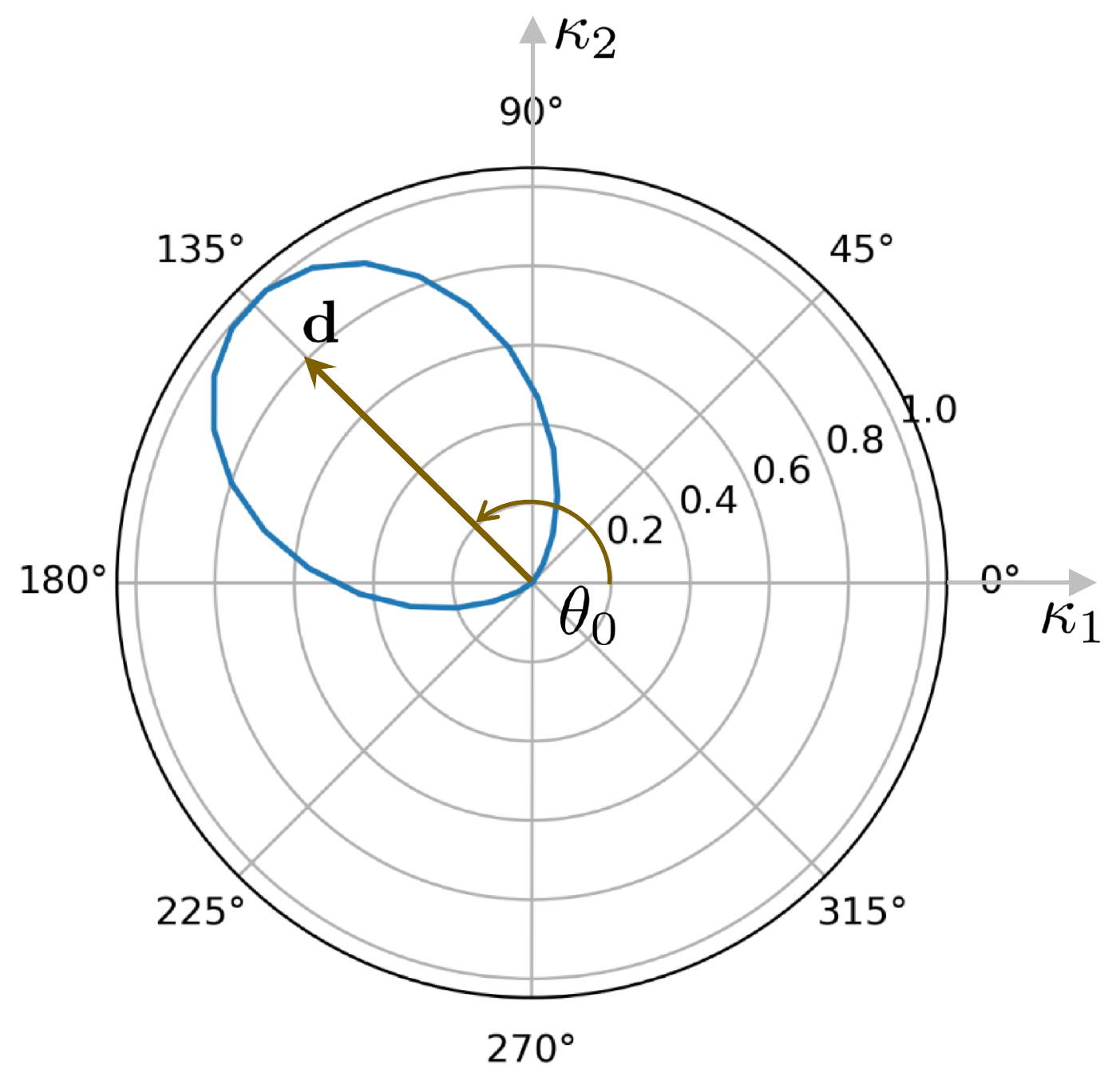}
  \caption{Polar plot of the directional filter $\max[0, \cos(\theta-\theta_0)]^2$ in the $\kappa_1-\kappa_2$ plane for $\theta_0 = 135^o$, where $\bbd \in \reals^2$ denotes the unit direction vector corresponding to the propagation direction $\theta_0$. The filter peaks at $\theta_0$ and quickly decays for directions far from $\theta_0$. Particularly, it equals zero for any direction with $\abs{\theta-\theta_0}>90^o$, meaning that reflections are mostly filtered.} \label{fig:directionalFilter}
\end{figure}

%
% paragraph theme:directional filter
%
In Section \ref{sec:intro}, we discussed the importance of directional filtering to time-of-flight (ToF) methods. Here we discuss the connection between directional filtering and the PWE method. The directional filter operates on the 3D Fourier transformed signal in the spatial frequency domain. Let $\kappa_1 \and \kappa_2$ denote the components of spatial frequency (wavenumber). Given a propagation direction $\theta_0$, the directional filter is defined in the $\kappa_1-\kappa_2$ plane as $\max[0, \cos(\theta-\theta_0)]^p$, where $\theta$ denotes the angle in polar coordinates and the power $p$ is a parameter to be chosen; in \cite{SDFIIMR2003MLKE}, $2 \leq p \leq 3$. Fig. \ref{fig:directionalFilter} depicts the directional filter for $\theta_0 = 135^o$.
This filter is applied across all positive temporal frequencies $\omega > 0$. For negative frequencies, the direction needs to be reversed since those waves travel backward in time.
%
% paragraph theme: radial filter
%
Often, a radial component is also added to improve the SNR by eliminating oscillations with unrealistically high wavenumbers. In \cite{SDFIIMR2003MLKE}, this radial component is a bandpass Butterworth filter. As we discussed in Section \ref{sec:exp}, both the directional and radial components are essential for ToF methods.

% paragraph theme: connection to directional filter
%
To see the connection between the PWE method and the directional filter, observe that given a frequency $\omega$ and for each angle $\theta_0$, there exists a plane wave that travels in the direction $\bbd = [\cos \theta_0, \sin \theta_0]$. Because ToF technique relies on directional propagation, we need to manually decompose the shear wave into its directional components specified by angles $\theta_0$, process each component separately, and then combine them through compounding. The PWE method on the other hand, searches for dominant directions to capture the shear wave in its entirety (including the reflections and refractions) and to simultaneously compute the wave-speed that best describes the observed data at once. This removes the need for the arbitrary compounding (averaging) step\footnote{One could for instance argue for selecting the point-wise maximum of the shear modulus fields instead of averaging. There is no particular advantage to compounding via averaging.}and instead selects the shear modulus value considering all data together.
To improve the nonlinearity of the PWE optimization problem and at the expense of increasing the dimension of the problem being solved, we explicitly discretized the plane wave directions $\theta_0$ in \eqref{eq:dirDisc} to obtain a finite set of $n_b$ plane wave basis functions given by \eqref{eq:basisFun}. The dominance of each direction $\bbd_j$ in capturing the shear wave is then determined by the magnitude of the corresponding basis coefficient $a_j$.
Note that the plane waves in expansion \eqref{eq:basisExpansion} are fundamental solutions of the scalar wave equation \eqref{eq:scalarWE} and form a complete set of basis functions meaning that by increasing $n_b$, we can approximate the solution to the wave equation as closely as we desire \cite{ODHWFEHP2001CK}.

\section{Nonlinearity of Objective Function} \label{app:homogeneous}
\begin{figure}[t!]
  \centering
    \includegraphics[width=0.4\textwidth]{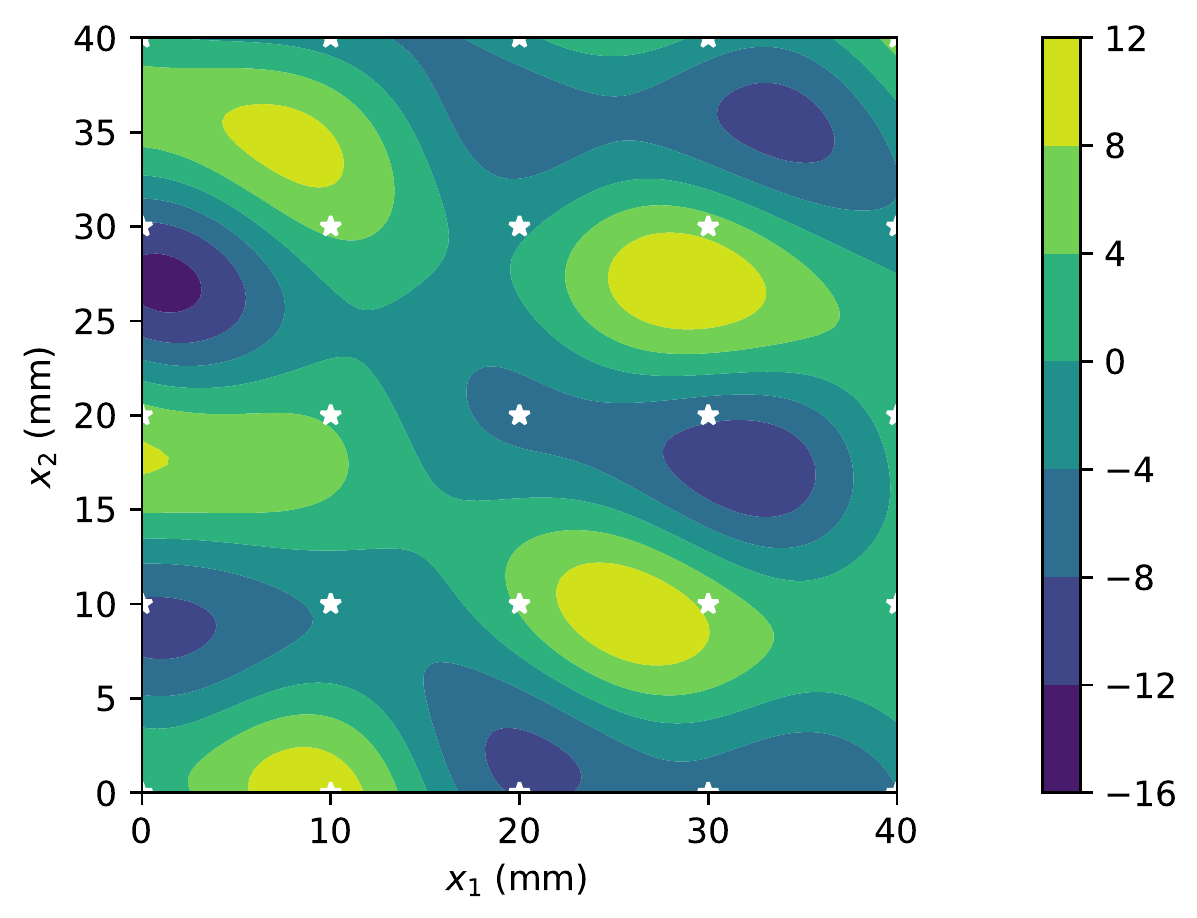}
  \caption{The real component of the fabricated data (units are arbitrary) within a homogeneous subdomain $\Omega = [0, 40] \times [0, 40]$\,mm$^2$ with wave-speed of $c = 5$\,m/s at frequency $\omega = 600\pi$\,rad/s, composed of $5$ plane waves propagating at uniformly spaced angles; the wave traveling at angle $72^o$ is dominant. The white stars indicate the grid of $5 \times 5$ measurements used for elastography.} \label{fig:real_disp}
\end{figure}

In this appendix, we consider more closely the ability of the PWE method to recover dominant propagation directions and the importance of this ability for reconstructing the desired shear modulus field. We also take a closer look at the shape (nonlinearity) of the objective function in \eqref{eq:PWEsol} and how the presence of noise and absence of true propagation directions affect it.
For simplicity, we use simulated data in a homogeneous medium with shear modulus of $\mu$ = 25\,kPa, amounting to a wave-speed of $c$ = 5\,m/s. We directly fabricate a frequency-domain displacement field using \eqref{eq:basisExpansion} composed of $5$ plane waves propagating with uniform angular spacing of $72^o$ at frequency $\omega$ = 600$\pi$\,rad/s. Fig. \ref{fig:real_disp} shows the real component of the field along with a grid of $m = 5 \times 5$ measurements used for elastography.
We choose the coefficients in \eqref{eq:basisExpansion} such that the dominant propagation directions are ordered as $[72^o, -72^o, 0^o, -144^o, 144^o]$.

In Table \ref{table:directionRecovery} we study the effect of including the true propagation directions and noise on recovering the dominant propagation directions.
\begin{table*}
\renewcommand{\arraystretch}{1.3}
\caption{Effect of excluding true propagation directions and noise on recovering the dominant propagation directions using PWE for the fabricated frequency-domain displacement data.}
\begin{center}
\footnotesize
\begin{tabular}{|c||c|c|}
\hline
true directions	& 	noiseless measurements			& 	noisy measurements	 			\\ \hline\hline
included		& 	$[72^o, -72^o, 0^o, -144^o, 144^o]$	& 	$[72^o, -72^o, 0^o, 144^o, -126^o]$	 \\ \hline
excluded		& 	$[69^o, -34^o, 34^o, -69^o, 0^o]$	& 	$[69^o, -34^o, 120^o, -69^o, 34^o]$	 \\ \hline
\end{tabular}
\end{center}
\label{table:directionRecovery}
\end{table*}
To include the true directions, we use $n_b = 20$ whereas to exclude them, we set $n_b = 21$. Furthermore, we use additive Gaussian noise resulting in SNR $= 19.92$\,dB. In the noiseless case, we set the regularization parameter to $\tau = 10^{-10}$ whereas in the noisy cases, we use $\tau = 10^{-2}$.
Observe that when true directions are included, in the absence of noise the true directions are exactly recovered up to their dominancy order. On the other hand, in the presence of noise the last dominant direction is not in the top five selected directions. When the true directions are excluded, the less dominant directions $-144^o \and 144^o$ are not closely approximated in the top five propagation directions although plane waves close to these directions might still have high coefficients. The reason for this behavior is that $\ell_2$-regularization is known to result in many small coefficients. Using a regularization term in \eqref{eq:elastProb} that enhances sparsity helps with better recovering the dominant directions at the expense of higher computational cost but as we show next, exactly recovering these dominant directions is not necessary for accurate shear modulus estimation.

\begin{figure*}
	\centering
	\begin{subfigure}[b]{0.4\textwidth}
		\includegraphics[width=\textwidth]{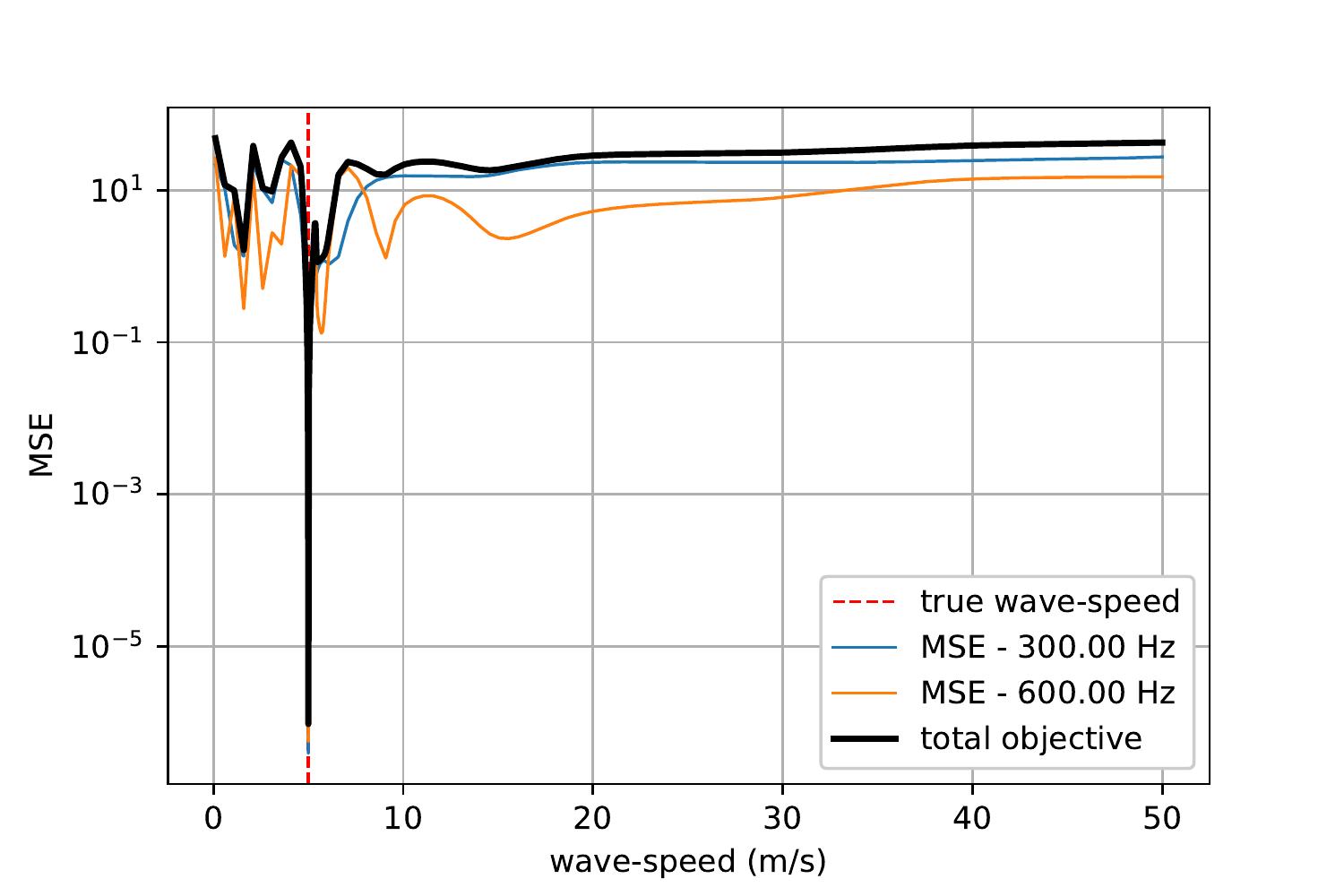}
	\caption{} \label{fig:included_noiseless}
	\end{subfigure}
	\quad
	\begin{subfigure}[b]{0.4\textwidth}
		\includegraphics[width=\textwidth]{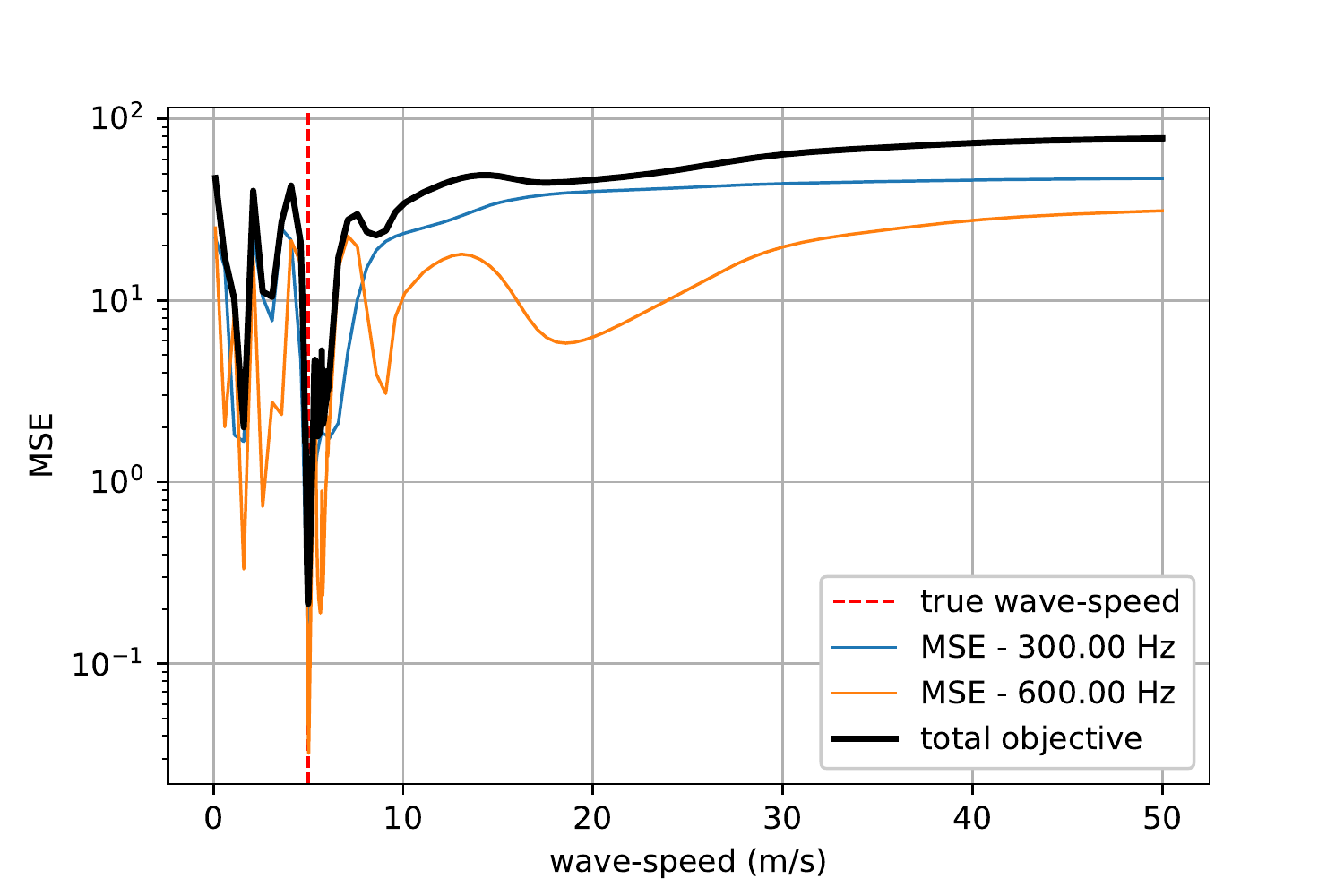}
	\caption{} \label{fig:included_noisy}
	\end{subfigure} \\
	\begin{subfigure}[b]{0.4\textwidth}
		\includegraphics[width=\textwidth]{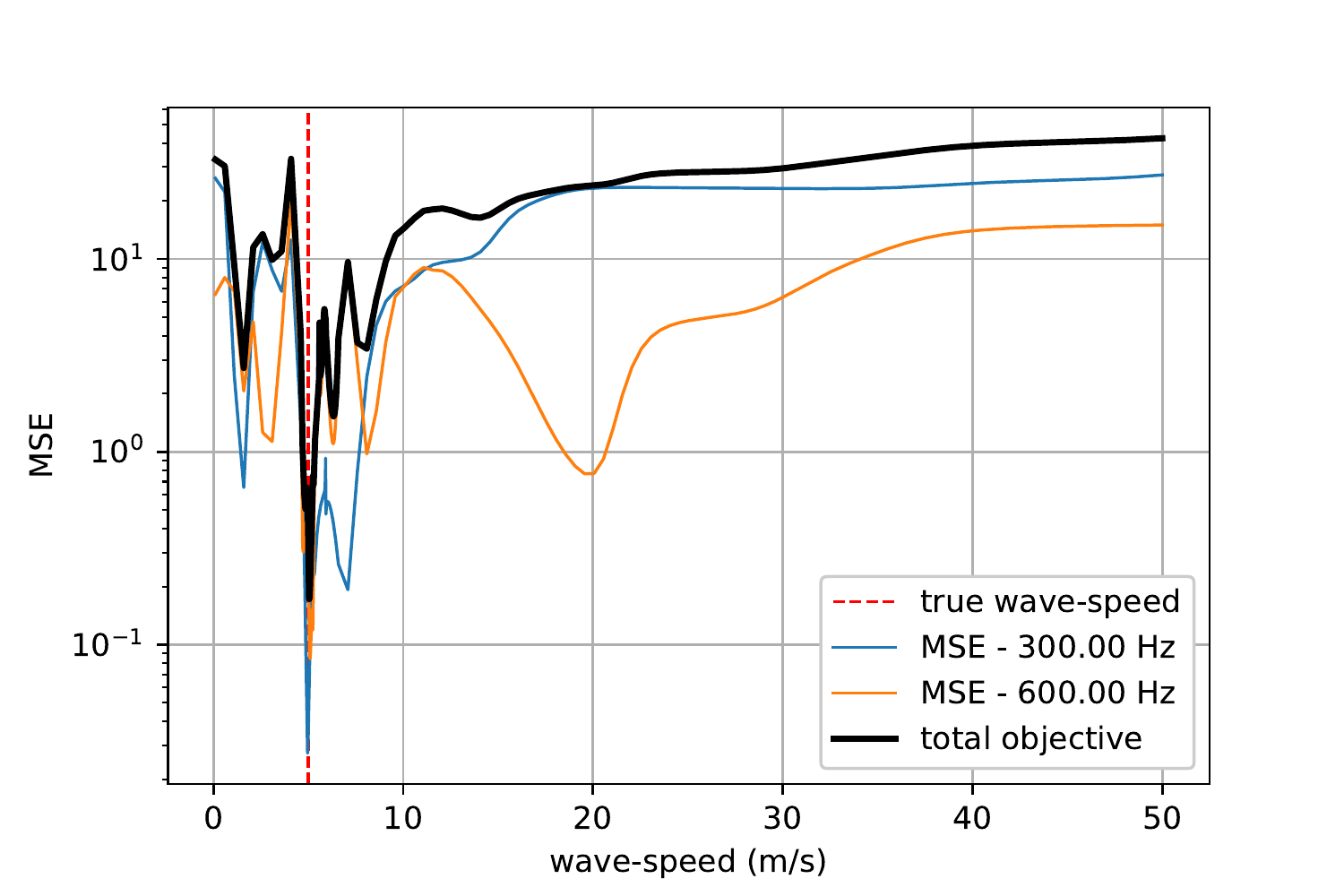}
	\caption{} \label{fig:excluded_noiseless}
	\end{subfigure}
	\quad
	\begin{subfigure}[b]{0.4\textwidth}
		\includegraphics[width=\textwidth]{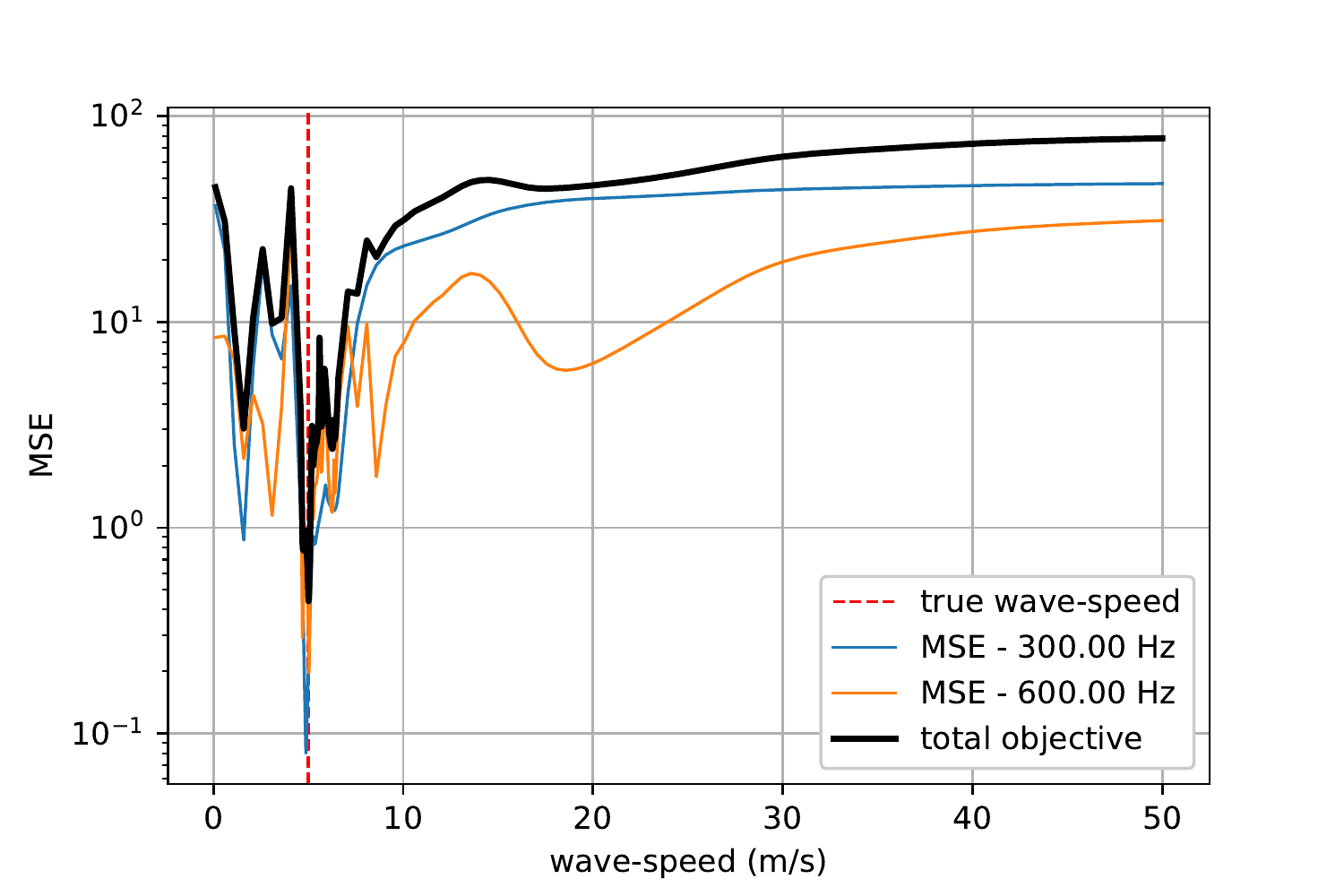}
	\caption{} \label{fig:excluded_noisy}
	\end{subfigure}
\caption{Objective function in \eqref{eq:PWEsol} as a function of the constant wave-speed $c$ (homogeneous medium) for the fabricated data with 5 plane waves at two frequencies $\omega = 600\pi$\,rad/s and $1200\pi$\,rad/s. (a) Fig. \ref{fig:included_noiseless} shows the objective function for the case of noiseless data when the true propagation directions are included in the bases \eqref{eq:basisExpansion}. (b) Fig. \ref{fig:included_noisy} shows the similar plot for noisy data. Plots in the second row exclude the true propagation directions. (c,d) Fig. \ref{fig:excluded_noiseless} corresponds to noiseless data whereas Fig. \ref{fig:excluded_noisy} depicts the objective for noisy data.}	\label{fig:homogeneous-obj}
\end{figure*}

To further elaborate on the last comment, we study the performance of PWE using displacement data corresponding to two frequencies $\omega = 600\pi$\,rad/s and $1200\pi$\,rad/s with SNR $= 19.92$\,dB and SNR $= 21.84$\,dB for noisy measurements. Fig. \ref{fig:homogeneous-obj} depicts the objective function in \eqref{eq:PWEsol} for the four different cases of including or excluding true directions and noiseless or noisy data.
Observe that the objective in all cases is extremely non-smooth. Also notice that the individual frequencies can have multiple local minima, some of which might have a smaller objective value than the true wave-speed. Nevertheless, the total objective function summed over the two frequencies often has its global minimum close to the true wave-speed. This shows the importance of using large number of frequencies $n_{\omega}$ to ensure that enough information is available for reconstruction. Finally, notice that when the true directions are excluded from reconstruction or data are noisy, the global minimum becomes less prominent.
Table \ref{table:homogeneous-l2-err} reports the estimated constant shear modulus value at each case.
\begin{table}[t]
\renewcommand{\arraystretch}{1.3}
\caption{The estimated shear modulus value (kPa) using the PWE method for the homogeneous medium with the ground-truth value of 25\,kPa.}
\begin{center}
\footnotesize
\begin{tabular}{|c||c|c|}
\hline
true directions	& 	noiseless measurements	& 	noisy measurements	 	\\ \hline\hline
included		& 	$25.00$				& 	$24.90$	 			\\ \hline
excluded		& 	$25.50$				& 	$25.10$				\\ \hline
\end{tabular}
\end{center}
\label{table:homogeneous-l2-err}
\end{table}
Note that although when the true directions are excluded, the dominant directions might not be properly identified as we observed in Table \ref{table:directionRecovery}, the PWE method still succeeds in approximating the true shear modulus values.

%\section{Additional Experiments} \label{app:exp2}
%\input{files/exp2}

\end{document}